\documentclass[a4paper,10pt]{article}   	
\usepackage{geometry}                		
\usepackage[dvipsnames]{xcolor}
\usepackage{graphicx}				
\usepackage{amsmath}
\usepackage{amsfonts}
\usepackage{amssymb}

\usepackage{mathrsfs}

\usepackage{gensymb} 
\usepackage{pdfpages} 
\usepackage{mathtools}

\usepackage{tensor}
\usepackage[cal=boondox]{mathalfa}

\usepackage[T1]{fontenc}
\usepackage{enumitem}

\usepackage{etoolbox}
\patchcmd{\thebibliography}{\chapter*}{\section*}{}{}

\makeatletter
\@addtoreset{footnote}{section}
\makeatother 

\usepackage[bottom]{footmisc}

\usepackage{braket}
\usepackage{slashed}

\allowdisplaybreaks 

\usepackage{appendix}

\usepackage{hyperref}
\usepackage{cite}

\usepackage{authblk}

\newcommand{\I}{\mathrm{i}}
\newcommand{\D}{\mathrm{D}}
\newcommand{ \Biggbr }[1]{\bigg[ #1 \bigg] }

\usepackage[normalem]{ulem}

\title{Gauge Fixing and the Semiclassical Interpretation of Quantum Cosmology}
\author[]{Leonardo Chataignier\thanks{E-mail: lcmr@thp.uni-koeln.de}}
\affil[]{\small\textit{Institute for Theoretical Physics, University of Cologne}, \\ \textit{Z\"{u}lpicher Stra\ss e 77, 50937 K\"{o}ln, Germany}}

\date{}							

\begin{document}
\maketitle

\begin{abstract}\noindent
We make a critical review of the semiclassical interpretation of quantum cosmology and emphasise that it is not necessary to consider that a concept of time emerges only when the gravitational field is (semi)classical. We show that the usual results of the semiclassical interpretation, and its generalisation known as the Born-Oppenheimer approach to quantum cosmology, can be obtained by gauge fixing, both at the classical and quantum levels. By `gauge fixing' we mean a particular choice of the time coordinate, which determines the arbitrary Lagrange multiplier that appears in Hamilton's equations. In the quantum theory, we adopt a tentative definition of the (Klein-Gordon) inner product, which is positive definite for solutions of the quantum constraint equation found via an iterative procedure that corresponds to a weak coupling expansion in powers of the inverse Planck mass. We conclude that the wave function should be interpreted as a state vector for both gravitational and matter degrees of freedom, the dynamics of which is unitary with respect to the chosen inner product and time variable.
\end{abstract}
\pagebreak

\section{Introduction}
In canonical general relativity, the bulk Hamiltonian is constrained to vanish~\cite{ADM:1962}. This constraint is related to the symmetry of the theory (`general covariance'), which is enforced by the Bergmann-Komar group in phase space~\cite{Bergmann:1972,Pons:2010}. Upon quantisation, one may promote the constraints to operators that annihilate the wave functional, which is equivalent to requiring that physical states are invariant under the action of the symmetry group. In the absence of boundary terms, this implies that physical states are annihilated by the Hamiltonian. Such states are, therefore, independent of the choice of spacetime coordinates and, in particular, independent of coordinate time. This time independence seems to imply that the wave functional is static and there is no dynamics. This is the so-called `problem of time' in canonical quantum gravity. There are many approaches to understanding and solving this problem (see, e.g. \cite{Kuchar:1991,Isham:1992,Anderson:book} and references therein).

In this paper, we will examine and reinterpret one such approach, known as the semiclassical interpretation of quantum gravity (see, e.g. \cite{Kuchar:1991,Isham:1992} for a review), which proposes that the notion of time emerges if the gravitational wave functional is semiclassical, i.e. if it can be approximated by its Wentzel-Kramers-Brillouin (WKB) counterpart. In this case, the first approximation to the phase of the WKB wave functional is a solution to the Einstein-Hamilton-Jacobi equations~\cite{Gerlach:1969}. This solution defines a congruence of classical gravitational trajectories and a standard of time with respect to which quantum matter evolves according to the (functional) time-dependent Schr\"{o}dinger equation (TDSE)~\cite{LapRuba:1979}. Thus, one is able to derive quantum field theory on a classical gravitational background from the quantum constraint equations for the composite system of gravitational and matter degrees of freedom. If one proceeds to higher orders in the semiclassical expansion, usually performed as a formal expansion in powers of the inverse Planck mass~\cite{Banks:1984,BFS:1984,Singh:1990,Kiefer:1991}, it is possible to compute corrections to the TDSE~\cite{Singh:1990,Kiefer:1991,Kim:1995-1,Barvinsky:1997}.

In this approach, the concept of time is taken to be inherently semiclassical and it cannot be defined when the gravitational field is fully quantum. This was argued by Banks~\cite{Banks:1984}, who followed an earlier argument of DeWitt~\cite{DeWitt:1967} that time should be a phenomenological concept in a covariant theory.
The
view expressed by Banks was shared by many authors in subsequent works in quantum gravity and cosmology~\cite{Halliwell:1984,Brout:1987-1,Brout:1987-2,Brout:1987-3,Brout:1987-4,Vilenkin:1988,PadSingh:1990-1,PadSingh:1990-2,Kiefer:1993-1,Kiefer:1993-2,Kiefer:2009T}, as well as in articles regarding the non-relativistic quantum mechanics of closed, isolated systems~\cite{Englert:1989,Briggs:2000-1,Briggs:2000-2}.

In the present article, we take a different view, motivated by the fact that the choice of time coordinate in canonical general relativity is analogous to a choice of gauge in canonical Yang-Mills theories. More precisely, both general relativity and Yang-Mills theories are constrained systems and, thus, the canonical field equations contain arbitrary Lagrange multipliers. In the Yang-Mills case, the multipliers are fixed by a gauge condition on the vector potential, whereas in the case of general relativity, they are fixed by a choice of spacetime coordinates, i.e. a `coordinate condition'. In analogy to the Yang-Mills case, we refer to the choice of spacetime coordinates as `gauge fixing'. This terminology is unrelated to gauge theories of gravitation, such as Poincar\'{e} gauge theory~\cite{Hehl:2013}.

As it is possible to fix the gauge both in the classical and quantum versions of Yang-Mills theories, we assume that the same is true for (quantum) general relativity. Although it might be indeed meaningless to discuss the interpretation of clock readings in a fully quantum regime, we assume that there is in principle no inconsistency in \emph{parametrising} the dynamics with respect to a given choice of time coordinate also in the quantum theory.
We will
thus
argue that it is unnecessary to relegate the concept of time to the (semi)classical level and that the usual results obtained in the semiclassical interpretation of quantum gravity coincide with a particular gauge fixing of the theory, both at the classical and quantum levels. 

It is not unexpected that the results of the semiclassical approach should coincide with a particular choice of gauge. 
Indeed, the emergent semiclassical time \emph{is} a time coordinate associated with the background geometry defined from the phase of the semiclassical gravitational wave functional. 
Once this coordinate is defined, it may be used to parametrise the dynamics not only of the background geometry, but also of the composite system of gravitational and matter degrees of freedom.
\emph{Classically},
time evolution is only defined once a choice of coordinates has been made, as this corresponds to fixing the arbitrary multipliers that appear in Hamilton's equations. Therefore, any notion of time (even if it is ``emergent'') must correspond to a particular fixation of the multipliers. 
We will provide closed-form expressions for the Lagrange multiplier (the `lapse') as well as the classical reduced gauge-fixed Hamiltonian associated with the choice of time given by the phase of the semiclassical gravitational wave functional. To the best of our knowledge, such closed-form expressions for this choice of time have not been derived before.

In~\cite{Parentani:1998}, Parentani showed that a time-dependent Hamilton-Jacobi equation (TDHJE) for matter fields could be derived from the Einstein-Hamilton-Jacobi equations in a way analogous to the derivation of the TDSE from the quantum constraints in the standard semiclassical approach. He emphasised that such a derivation amounts to a \emph{background field approximation}, as the notion of time in both classical and quantum cases is defined from the (classical) gravitational background. The higher order corrections to the TDHJE and TDSE thus depend on the \emph{choice} of this background. More recently, Briggs~\cite{Briggs:2015} has independently derived the TDHJE from the time-independent Hamilton-Jacobi equation. It remained unclear, however, whether the derivation of the approximate TDHJE corresponds to a choice of gauge by fixing the arbitrary multipliers in the equations of motion.
As already mentioned,
we will show that this is indeed the case.

More generally, one can fix the coordinates to be given by functions of the canonical variables. This `canonical gauge' choice is in line with DeWitt's view~\cite{DeWitt:1967} that, in a covariant theory, the contents of the universe itself should be used to define the coordinates and, in particular, time. In this way, the evolution of physical quantities is described through the correlation of their configuration with the trajectory of a quantity chosen to be the `clock'~\cite{Parentani:1997-1,Parentani:1997-2}. Moreover, it is important to note that gauge fixing is not merely a mathematical convenience, since different choices of coordinates may also be associated with different reference frames and observers. This marks a difference between the \emph{external} symmetry of general relativity and the \emph{internal} symmetries of Yang-Mills theories, for which different gauge choices are unobservable.

Regarding the quantum theory, we take the position that the independence of the wave functional on the choice of coordinates does not preclude its dynamical and probabilistic interpretation. Rather, the invariance of physical states implies that the quantum \emph{dynamics} is the same for any choice of spacetime coordinates. To fix the gauge in the canonical quantum theory, we proceed in analogy to the classical Hamilton-Jacobi theory. 
At the classical level, the Einstein-Hamilton-Jacobi equations are gauge-independent. However, if we choose suitable functions of the canonical variables (e.g., the Weyl scalars or matter scalar fields) to define the spacetime coordinates, the solutions to the Einstein-Hamilton-Jacobi equations may be interpreted dynamically, since their variation in the chosen time coordinate will be encoded in their dependence on the canonical variables. If we take the view that the same is true for the quantum constraint equations, the time-dependence of the wave functional will be encoded in its dependence on the configuration or momentum variables in the appropriate representation.

The difficulty in establishing such a quantum theory resides in defining the inner product on the space of physical states and assessing whether the theory is unitary with respect to different choices of the time parameter. There have been various proposals for selecting such a physical inner product and constructing the physical Hilbert space (see, for instance,~\cite{Marolf:1995-1,Marolf:1995-2} and references therein). This is a subtle issue that is outside the scope of this paper. We will adopt a tentative definition of the inner product, with respect to which we will interpret the usual results of the semiclassical approach as a particular gauge fixing. Thus, the quantum theory here presented is provisional. Our goal is \emph{not} to provide a definitive solution to the problem of time, but rather to reinterpret the standard semiclassical approach and to emphasise how the emergent semiclassical time is related to the freedom of choosing coordinates in general relativity: it corresponds to a \emph{particular} class of canonical gauge choices and there are more general choices of time coordinate that could be employed.

The semiclassical interpretation can be generalised to what is often referred to as the Born-Oppenheimer (BO) approach to quantum gravity and cosmology~\cite{Brout:1987-1,Brout:1987-2,Brout:1987-3,Brout:1987-4,Kamenshchik:2013,Kamenshchik:2014,Kamenshchik:2017,Kamenshchik:2018,Balbinot:1990,Anderson:2006vs-1,Anderson:2006vs-2,Anderson:2011wq,Anderson:2013vaa,Anderson:2013yza}, as it was inspired by the BO approximation in molecular physics~\cite{BO:1927,Cederbaum:2008}. We will use the terms `semiclassical interpretation', `semiclassical approach' and `BO approach' interchangeably. Since our focus is on the problem of time (in its simplest manisfestation) and its solution given by the semiclassical approach, we will not be concerned with field-theoretic issues, such as regularisation of the quantum constraint equations. In the main body of the paper, we will restrict ourselves to finite-dimensional models with a single constraint, which are useful for analysing homogeneous and isotropic cosmologies.
We include an appendix with the formal generalisation of the results of the paper to the field-theoretic case.

Finally, we mention a recent paper in which Kamenshchik et al.~\cite{Kamenshchik:2018} compared the results of the BO approach and the gauge-fixing approach for a simple minisuperspace model and obtained similar results for both methods. The formalism presented here is complementary to their work. We show that the BO approach is simply a particular choice of gauge for a general homogeneous, isotropic minisuperspace model with non-vanishing potential.

The paper is organised as follows. In section~\ref{sec:overview-BO}, we give a critical overview of the standard semiclassical interpretation and the BO approach to the problem of time. We then describe the particular gauge fixing with which it coincides at the classical level in section~\ref{sec:classical}. There, we show how the reduced gauge-fixed Hamiltonian can be approximated by the `corrected' Hamiltonian used in the standard semiclassical approach. In section~\ref{sec:quantum}, we analyse the corresponding quantum theory, in particular the question of unitarity in perturbation theory. The archetypical example of the relativistic particle is analysed in section~\ref{sec:example}. Finally, in section~\ref{sec:conclusions}, we summarise our results and present our conclusions. We include three appendices. Appendices~\ref{app:formulas-Qx-transf} and~\ref{app:conservation-innerprod} contain formulae that are needed in sections~\ref{sec:classical} and~\ref{sec:quantum}. Appendix~\ref{app:field-theory} formally extends the results of the paper to field theory. We work in units in which $c = \hbar = 1$.

\section{The Born-Oppenheimer Approach to the Problem of Time}\label{sec:overview-BO}
In this section, we make a critical review of the BO approach to the problem of time in preparation for sections~\ref{sec:classical} and~\ref{sec:quantum}, where we show that the standard results of the BO approach can be obtained by a particular choice of gauge.
\subsection{Overview}
The Hamiltonian constraint of quantum gravity, referred to as the Wheeler-DeWitt (WDW) equation, does not depend on a time variable and it is thus analogous to the time-independent Schr\"{o}dinger equation (TISE) of quantum theory. If some of the variables on which the TISE or the WDW equation depends can be treated semiclassically, one can define a time-variable from the phase of the semiclassical part of the wave function. In this way, time emerges from a timeless \emph{quantum} equation (TISE or WDW) in a \emph{semiclassical} regime. The basis for such a semiclassical interpretation of time was laid in Mott's work~\cite{Mott:1929,Mott:1931} on $\alpha$-particle tracks. In~\cite{Mott:1929,Mott:1931}, Mott analysed the TISE for the composite system of atoms and $\alpha$ particles and showed that, by treating the high-energy $\alpha$-particle wave function semiclassically, one can derive a TDSE for the atoms. The time parameter is defined from the phase of the $\alpha$-particle wave function. Mott's derivation of the TDSE from the TISE was further analysed in \cite{Barbour:1994-2,Briggs:2000-1,Briggs:2000-2} and inspired an application to quantum cosmology in \cite{Halliwell:2000}.

The semiclassical regime in which time emerges can be understood in the context of the BO approximation~\cite{Brout:1987-1,Brout:1987-2,Brout:1987-3,Brout:1987-4,Kamenshchik:2013,Kamenshchik:2014,Kamenshchik:2017,Kamenshchik:2018}, which is frequently employed in molecular physics~\cite{BO:1927,Cederbaum:2008}. There, one is interested in computing molecular spectra by analysing the quantum dynamics of a system of heavy nuclei and light electrons. In many cases, one can make a WKB approximation to the nuclear wave function and consider an adiabatic approximation in which the electronic wave function follows the semiclassical dynamics of the nuclei. The combination of the WKB expansion for the heavy nuclei and the adiabatic approximation comprises the BO approximation. This can be generalised to any composite system, composed of a `heavy' (or `slowly-varying') part and a `light' (sub)system \cite{Briggs:2000-1,Briggs:2000-2,Briggs:2015}. In the BO approximation, the evolution of the `light' system follows adiabatically the semiclassical dynamics of the `heavy' part. In the BO approach to the problem of time, the time parameter is defined from the phase of the semiclassical wave function of the `heavy' sector and is sometimes referred to as `WKB time'~\cite{Zeh:1987}.

Englert~\cite{Englert:1989} and subsequently Briggs and Rost~\cite{Briggs:2000-1,Briggs:2000-2} took the position that the (non-relativistic) quantum mechanics of closed, isolated systems should be fundamentally timeless and thus described by the TISE. They suggested that the BO approximation (or a generalisation thereof) should be used to derive the TDSE from the TISE in a procedure analogous to Mott's derivation. In this way, Briggs and Rost emphasise that the TDSE would be only an approximation and a mixed classical-quantum equation, since the time parameter is defined in the limit in which the heavy sector becomes classical. Thus, the TISE would be promoted to the fundamental equation of quantum mechanics. This approach was further pursued by Arce \cite{Arce:2012}.

In the context of quantum gravity, Lapchinsky and Rubakov~\cite{LapRuba:1979} derived the functional TDSE for matter fields propagating in a fixed, vacuum gravitational background by treating the gravitational field semiclassically in the quantum constraint equations. Thus, the gravitational variables served as the `heavy' sector for the `light' matter fields. Essentially the same procedure was followed independently by Banks~\cite{Banks:1984} and Banks, Fischler and Susskind~\cite{BFS:1984}. Although other separations are possible, the `heavy' variables usually coincide with the gravitational degrees of freedom, whereas the `light' sector consists of the matter variables.

In~\cite{BFS:1984,Banks:1984}, the semiclassical approximation for the gravitational sector was obtained by formally expanding the quantum constraint equations and their solution in powers of the inverse Planck mass. Such an expansion is valid when all energy scales are much smaller than the Planck scale (weak-coupling expansion) and it is analogous to what was done in Born and Oppenheimer's original paper~\cite{BO:1927}. In~\cite{BO:1927}, the average mass $M$ of the nuclei was considered to be much larger than the mass $m$ of electrons, such that it was possible to expand the Hamiltonian operator and energy eigenfunctions in powers of $\left(\frac{m}{M}\right)^{\frac{1}{4}}$.  To lowest order, one recovered the dynamics of electrons while the nuclei remained at fixed positions and the nuclear position variables were treated as parameters. Analogously, the gravitational variables enter as parameters in the functional TDSE for matter fields, which propagate in a fixed geometry to lowest order in the weak coupling expansion~\cite{BFS:1984,Banks:1984}.

It is
possible to include corrections to the (functional) TDSE by computing terms of higher orders in the inverse Planck mass. Such corrections have been computed in~\cite{Singh:1990,Kiefer:1991,Kim:1995-1,Barvinsky:1997}. In~\cite{Kiefer:1991}, it was concluded that the corrections include terms that violate unitarity in the matter sector. We shall reexamine this question in sections~\ref{sec:assess} and~\ref{sec:pt2} and find that the dynamics of the gravity-matter system is unitary with a suitable definition of the inner product.

In~\cite{LapRuba:1979,Banks:1984,BFS:1984}, the backreaction of quantum matter onto the classical gravitational background was not included. In a series of papers~\cite{Brout:1987-1,Brout:1987-2,Brout:1987-3,Brout:1987-4}, Brout and colleagues refined the method of Banks to include the effect of backreaction of matter in the form of a source term in the Einstein-Hamilton-Jacobi equations for the classical gravitational background. The source term consisted of the expectation value of the matter Hamiltonian (averaged only over matter degrees of freedom) and was also accompanied by Berry connection~\cite{MeadBerry:1979-1,MeadBerry:1979-2} terms, which is in line with the usual BO approximation used in molecular physics~\cite{Cederbaum:2008,Abedi:2010,Abedi:2012-1,Abedi:2012-2}. In~\cite{Bertoni:1996,Kamenshchik:2017,Kiefer:2018}, it was claimed that the inclusion of backreaction and Berry connection terms leads to a unitary description of the matter-sector dynamics to all orders in the weak-coupling expansion.

However, following an earlier paper by Hartle~\cite{Hartle:1987}, the authors Halliwell~\cite{Halliwell:1987}, D'Eath and Halliwell~\cite{DEath:1986} and Padmanabhan~\cite{Padmanabhan:1988} stressed that the backreaction terms were spurious, as they depend on the \emph{arbitrary} choice of phase for the gravitational wave function, which in turn is related to the freedom associated with the definition of the Berry connection. Halliwell~\cite{Halliwell:1987}, Singh and Padmanabhan~\cite{PadSingh:1990-1,PadSingh:1990-2} concluded that a semiclassical theory of gravity sourced by the expectation value of the matter Hamiltonian or, in a covariant setting, of the matter energy-momentum tensor is well defined only when the distribution of the matter Hamiltonian is peakead about its average value or the quantum corrections to the energy-momentum tensor of matter are small in comparison to the classical contribution. This arbitrariness related to the definition of backreaction terms casts doubt on the claim that the inclusion of backreaction guarantees unitarity in the matter sector. In what follows, we will see how this can be resolved.

\subsection{A Critical Assessment of the Method. Non-Relativistic Case}\label{sec:assess}
To clarify the conceptual points mentioned above and illustrate the BO approach to the problem of time, let us consider a non-relativistic example, analysed in a different way in~\cite{Briggs:2000-1,Briggs:2000-2,Kiefer:book,Briggs:2015}. We focus on a composite system of a `heavy' sector interacting with a `light' subsystem. The `heavy' sector is associated with a mass scale $M$ and degrees of freedom $Q_a$, $a = 1,...n$, whereas the `light' system is associated with a scale $m\ll M$ and degrees of freedom $q_{\mu}$, $\mu = 1,...,d$. The TISE reads
\begin{equation}\label{eq:TISE}
-\frac{1}{2M}\sum_{a = 1}^n\frac{\partial^2\Psi}{\partial Q_a^2}+V(Q)\Psi + \hat{H}_{\mathscr{S}}\left(Q;\frac{\partial}{\partial q_{\mu}},q_{\mu}\right)\Psi = E\Psi \ ,
\end{equation}
where $V$ is a potential term for the `heavy' sector and $\hat{H}_{\mathscr{S}}$ is the `light'-system Hamiltonian, which depends only parametrically on the `heavy' variables $Q$. In the traditional BO approach, one expands the wave function as the superposition
\begin{equation}\label{eq:BO-factorisation}
\Psi(Q;q) = \sum_{k}\chi_k(Q)\psi_k(Q;q) \ ,
\end{equation}
where $\psi_k$ form a complete system which is orthonormal with respect to the inner product taken only over the `light' variables $q$, i.e.
$\braket{\psi_k,\psi_l}_{\mathscr{S}}(Q) := \int\prod_{\mu = 1}^d\mathrm{d}q_{\mu}\ \bar{\psi}_k(Q;q)\psi_l(Q;q) \equiv \delta_{kl} \ .$
For example, one may choose $\psi_k$ to be the eigenstates of $\hat{H}_{\mathscr{S}}$ (if the spectrum is continuous, we replace $\sum_k\to\int\mathrm{d}k$ and $\delta_{kl}\to\delta(k-l)$). For simplicity, we can also rewrite (\ref{eq:BO-factorisation}) as
\begin{equation}\label{eq:exact-factorisation}
\Psi(Q;q) = \chi(Q)\sum_{k}\frac{\chi_k(Q)}{\chi(Q)}\psi_k(Q;q) \equiv \chi(Q)\psi(Q;q) \ .
\end{equation}
Such an \emph{exact factorisation} was considered in~\cite{Hunter:1975,Abedi:2010,Arce:2012,Abedi:2012-1,Abedi:2012-2,Cederbaum:2013,Briggs:2015} and it avoids the complication of having to consider the dynamics of each of the $\psi_k$ states. Evidently, this factorisation is ambiguous, as one can redefine each factor as follows:
\begin{equation}\label{eq:Berry-gauge-transf}
\chi(Q) = e^{\xi(Q)+\I\eta(Q)}\chi'(Q) \ , \ \psi(Q;q) = e^{-\xi(Q)-\I\eta(Q)}\psi'(Q;q) \ ,
\end{equation}
where $\xi$ and $\eta$ are smooth functions of the $Q$-variables. Under such a redefinition, the total state remains invariant, $\Psi = \chi\psi = \chi'\psi' = \Psi'$.

The usual procedure is to insert (\ref{eq:BO-factorisation}) or~(\ref{eq:exact-factorisation}) into (\ref{eq:TISE}), multiply the result by $\bar{\psi}_k$ or $\bar{\psi}$ and integrate over the $q$-variables to obtain an equation for $\chi_k$ or $\chi$~\cite{Brout:1987-1,Brout:1987-2,Brout:1987-3,Brout:1987-4,Bertoni:1996,Briggs:2000-1,Briggs:2000-2,Kiefer:book}. Such an equation involves the partial averages over the $q$-variables $\braket{\frac{\partial}{\partial Q_a}}_{\mathscr{S}}$, which are related to the Berry connection, and $\braket{\hat{H}_{\mathscr{S}}}_{\mathscr{S}}$, which is interpreted as a `backreaction' term. One then uses the equation for $\chi_k$ or $\chi$ to obtain an equation for $\psi_k$ or $\psi$, which will also involve partial averages over the $q$-variables. The result is a coupled non-linear system, which can be solved in an iterative, self-consistent way~\cite{Anderson:2013}. Here, we decide to take a slightly different but equivalent route. For convenience, we will work with the exact factorisation given in~(\ref{eq:exact-factorisation}). 
For any choice of $\chi$, we define
\begin{equation}\label{eq:nonrel-source}
\mathfrak{J}(Q):= \frac{1}{2M\chi(Q)}\sum_{a = 1}^n\frac{\partial^2\chi}{\partial Q_a^2}-V(Q)+E\equiv J(Q)+\I K(Q) \ ,
\end{equation}
where $J$ and $K$ are the real and imaginary parts of the `source' $\mathfrak{J}$. They can be written in terms of the amplitude and phase of $\chi$ as follows.
\begin{equation}
\begin{aligned}
J(Q) &=\frac{1}{2M}\sum_{a = 1}^n\left[\frac{1}{R}\frac{\partial^2R}{\partial Q_a^2}-\left(\frac{\partial\varphi}{\partial Q_a}\right)^2\right]-V(Q)+E \ , \\
K(Q) &= \frac{1}{2M}\sum_{a = 1}^n\left(\frac{2}{R}\frac{\partial R}{\partial Q_a}\frac{\partial\varphi}{\partial Q_a}+\frac{\partial^2\varphi}{\partial Q_a^2}\right) \ , 
\end{aligned}
\end{equation}
where we used the polar decomposition $\chi = R e^{i\varphi}$. We note that there is no loss of generality in choosing $\chi$ to be a complex wave function ($\varphi\neq0$), even if $\Psi$ is real, due to the freedom of redefining the states according to~(\ref{eq:Berry-gauge-transf}). If we redefine $\chi = e^{\xi+\I\eta}\chi'$~(cf.~(\ref{eq:Berry-gauge-transf})), the `source' $\mathfrak{J}$ changes accordingly,
\begin{equation}\label{eq:nonrel-Berry-gauge-transf-source}
\begin{aligned}
J(Q) &= J'(Q) +\frac{1}{2M}\sum_{a = 1}^n\left[\frac{2}{R'}\frac{\partial R'}{\partial Q_a}\frac{\partial\xi}{\partial Q_a}+\left(\frac{\partial\xi}{\partial Q_a}\right)^2+\frac{\partial^2\xi}{\partial Q_a^2}-2\frac{\partial\varphi'}{\partial Q_a}\frac{\partial\eta}{\partial Q_a}-\left(\frac{\partial\eta}{\partial Q_a}\right)^2\right] \ , \\
K(Q) &= K'(Q) + \frac{1}{M}\sum_{a = 1}^n\left(\frac{\partial\xi}{\partial Q_a}\frac{\partial\varphi'}{\partial Q_a}+\frac{1}{R'}\frac{\partial R'}{\partial Q_a}\frac{\partial\eta}{\partial Q_a}+\frac{\partial\xi}{\partial Q_a}\frac{\partial\eta}{\partial Q_a}+\frac{1}{2}\frac{\partial^2\eta}{\partial Q_a^2}\right) \ .
\end{aligned}
\end{equation}
We now insert~(\ref{eq:exact-factorisation}) into~(\ref{eq:TISE}) to obtain
\begin{equation}\label{eq:nonrel-pre-schrodinger}
\frac{\I}{M}\sum_{a = 1}^n\frac{\partial\varphi}{\partial Q_a}\frac{\partial\psi}{\partial Q_a} = \left(\hat{H}_{\mathscr{S}}-\mathfrak{J}\right)\psi-\frac{1}{M}\sum_{a = 1}^n\frac{\partial\log R}{\partial Q_a}\frac{\partial\psi}{\partial Q_a}-\frac{1}{2M}\sum_{a = 1}^n\frac{\partial^2\psi}{\partial Q_a^2} \ .
\end{equation}
If we define $\frac{\partial}{\partial t} := \frac{1}{M}\sum_{a = 1}^n\frac{\partial\varphi}{\partial Q_a}\frac{\partial}{\partial Q_a}$, then~(\ref{eq:nonrel-pre-schrodinger}) reads
\begin{equation}\label{eq:nonrel-pre-schrodinger-2}
\I\frac{\partial\psi}{\partial t} =  \left(\hat{H}_{\mathscr{S}}-\mathfrak{J}\right)\psi-\frac{1}{M}\sum_{a = 1}^n\frac{\partial\log R}{\partial Q_a}\frac{\partial\psi}{\partial Q_a}-\frac{1}{2M}\sum_{a = 1}^n\frac{\partial^2\psi}{\partial Q_a^2} \ ,
\end{equation}
which resembles a TDSE. Nevertheless, the presence of higher derivatives with respect to the $Q$-variables on the right-hand side makes it more akin to a Klein-Gordon equation. Traditionally, the `time' derivative in~(\ref{eq:nonrel-pre-schrodinger-2}) is defined only when $\chi$ (the `heavy'-sector wave function) is approximated by its WKB counterpart~\cite{LapRuba:1979,Banks:1984,BFS:1984}, such that $t$ is the `WKB-time'. However, we stress that this is not necessary. Indeed, we have defined $t$ from the general phase\footnote{It is worth mentioning that there have been suggestions in the literature to define time not from the phase $\varphi$, but rather from the state $\chi$ itself, e.g. as $\frac{\partial}{\partial t} := \frac{-i\hbar}{M\chi}\sum_{a = 1}^n\frac{\partial\chi}{\partial Q_a}\frac{\partial}{\partial Q_a}$. Such a notion of ``quantum time'' was considered in~\cite{Briggs:2015,Schild:2018}. However, since the choice of $\chi$ is arbitrary due to equation~(\ref{eq:Berry-gauge-transf}), the definition of $t$ from the phase $\varphi$ is sufficient. We also mention~\cite{Greensite:1989-1,Greensite:1989-2,Greensite:1989-3,Greensite:1989-4}, in which the phase of the total state $\Psi$ was used to define a time variable. Moreover, in the de Broglie-Bohm approach to quantum gravity~\cite{Nelson:2018}, time can also be defined from the phase of the total state $\Psi$.} of $\chi$ and used the exact polar decomposition $\chi = R e^{\I\varphi}$. Moreover, we note that $\varphi$ (and, hence, $t$) is freely specifiable due to the freedom of performing the redefinitions given in~(\ref{eq:Berry-gauge-transf}). Finally, let us mention that we can reinterpret~(\ref{eq:nonrel-source}) as a definition of $\chi$ given $\mathfrak{J}$, instead of as a definition of $\mathfrak{J}$ given $\chi$. In this way, equations~(\ref{eq:nonrel-source}) and~(\ref{eq:nonrel-pre-schrodinger-2}) can be seen as a coupled system for $\chi$ and $\psi$.
We will see in what follows how
this is related to
the treatment involving the Berry connection and backreaction terms.

If the terms proportional to $\frac{1}{M}$ can be neglected (e.g. by considering that all energy scales related to the `light'-sector are much smaller than $M$, or by performing a semiclassical expansion), then~(\ref{eq:nonrel-pre-schrodinger-2}) reduces to a TDSE
\begin{equation}\label{eq:nonrel-schrodinger}
\I\frac{\partial\psi}{\partial t} =  \left(\hat{H}_{\mathscr{S}}(t)-\mathfrak{J}\right)\psi+\mathcal{O}\left(\frac{1}{M}\right) \ .
\end{equation}
In the literature~\cite{Banks:1984,BFS:1984,Kiefer:1993-1,Kiefer:1993-2,Briggs:2000-1,Briggs:2000-2,Arce:2012,Briggs:2015}, equation~(\ref{eq:nonrel-schrodinger}) is often interpreted as the Schr\"{o}dinger equation for the `light' system alone, in which the `heavy' variables provide the clock which parametrises the evolution of the `light' degrees of freedom. The (real part of the) source term $J$ in~(\ref{eq:nonrel-schrodinger}) can be removed by a phase transformation of $\psi$~\cite{Bertoni:1996,Briggs:2015}. However, such a phase transformation corresponds to a redefinition given in~(\ref{eq:Berry-gauge-transf}), which would lead to a redefinition of $\varphi$ and $t$, unless one defines time only from a part of the phase of $\chi$ or in some other way. By taking into account the terms of order $\frac{1}{M}$, one obtains from~(\ref{eq:nonrel-pre-schrodinger-2}) a ``corrected'' Schr\"{o}dinger equation~\cite{Singh:1990,Kiefer:1991,Kim:1995-1,Barvinsky:1997}. We will see in what follows under what circumstances can one interpret such an equation as a dynamical equation for the `light' sector alone.

\subsubsection{Partial Averages. Berry Connection}
Given an operator $\hat{O}$, we define its `light'-sector partial average as
\begin{equation}\label{eq:nonrel-partial-average-def}
\braket{\hat{O}}_{\mathscr{S}}(Q) := \frac{\int\prod_{\mu}\mathrm{d}q_{\mu}\ \bar{\psi}(Q;q)\hat{O}\psi(Q;q)}{\int\prod_{\nu}\mathrm{d}q_{\nu}\ \bar{\psi}(Q;q)\psi(Q;q)} \ ,
\end{equation}
provided the integrals converge. The partial averages
\begin{equation}\label{eq:nonrel-partial-average-derivatives}
\left<\frac{\partial}{\partial Q_a}\right>_{\mathscr{S}}(Q) =: V_a(Q)+\I A_a(Q) \ ,
\end{equation}
where $A_a$ and $V_a$ are real, are of particular interest. The one-form with components $A_a$ will be referred to as ``Berry connection'', in analogy to the usual Berry connections that appear in adiabatic quantum mechanics \cite{MeadBerry:1979-1,MeadBerry:1979-2}. We can write $V_a$ and $A_a$ explicitly as
\begin{align}
&V_a(Q) :=\frac{1}{2}\frac{\partial}{\partial Q_a}\log\left(\int\prod_{\mu}\mathrm{d}q_{\mu}\ \bar{\psi}(Q;q)\psi(Q;q)\right) \ , \label{eq:nonrel-V}\\
&A_a(Q) :=-\frac{\I}{2}\left(\frac{\int\prod_{\mu}\mathrm{d}q_{\mu}\ \bar{\psi}(Q;q)\frac{\partial\psi(Q;q)}{\partial Q_a}}{\int\prod_{\nu}\mathrm{d}q_{\nu}\ \bar{\psi}(Q;q)\psi(Q;q)}-\frac{\int\prod_{\mu}\mathrm{d}q_{\mu}\ \psi(Q;q)\frac{\partial\bar{\psi}(Q;q)}{\partial Q_a}}{\int\prod_{\nu}\mathrm{d}q_{\nu}\ \bar{\psi}(Q;q)\psi(Q;q)}\right) \label{eq:nonrel-A} \ .
\end{align}
Under the redefinition given in~(\ref{eq:Berry-gauge-transf}), we obtain the transformation laws
\begin{equation}\label{eq:nonrel-Berry-gauge-transf-VA}
\begin{aligned}
V_a &= V'_a-\frac{\partial\xi}{\partial Q_a} \ ,\ A_a = A'_a-\frac{\partial\eta}{\partial Q_a} \ .
\end{aligned}
\end{equation}
We will also be interested in the partial average $\left<\frac{\partial^2}{\partial Q_a^2}\right>_{\mathscr{S}}$. Under the redefinition given in~(\ref{eq:Berry-gauge-transf}), we find
\begin{equation}\label{eq:nonrel-Berry-gauge-transf-nabla2}
\begin{aligned}
\mathfrak{Re}\left<\frac{\partial^2}{\partial Q_a^2}\right>_{\mathscr{S}} &= \mathfrak{Re}\left<\frac{\partial^2}{\partial Q_a^2}\right>_{\mathscr{S}}^{\prime}-\frac{\partial^2\xi}{\partial Q_a^2}+\left(\frac{\partial\xi}{\partial Q_a}\right)^2-\left(\frac{\partial\eta}{\partial Q_a}\right)^2+2\frac{\partial\eta}{\partial Q_a}A'_a-2\frac{\partial\xi}{\partial Q_a}V'_a\ , \\
\mathfrak{Im}\left<\frac{\partial^2}{\partial Q_a^2}\right>_{\mathscr{S}} &=\mathfrak{Im}\left<\frac{\partial^2}{\partial Q_a^2}\right>_{\mathscr{S}}^{\prime}-\frac{\partial^2\eta}{\partial Q_a^2}+2\left(\frac{\partial\xi}{\partial Q_a}\frac{\partial\eta}{\partial Q_a}-\frac{\partial\xi}{\partial Q_a}A'_a-\frac{\partial\eta}{\partial Q_a}V'_a\right)\ .
\end{aligned}
\end{equation}
By multiplying~(\ref{eq:nonrel-pre-schrodinger}) by $\bar{\psi}$ and integrating over the $q$-variables, we find\footnote{We assume that the `light'-sector Hamiltonian $\hat{H}_{\mathscr{S}}(Q;\hat{p},q)$ is self-adjoint with respect to the inner product taken only over the $q$ variables, such that $\braket{\hat{H}_{\mathscr{S}}}_{\mathscr{S}}$ is a real function of the remaining `heavy' variables $Q$.}
\begin{equation}\label{eq:nonrel-source-Berry}
\begin{aligned}
-\mathfrak{J} 
&= \frac{1}{M\chi}\sum_{a = 1}^n\frac{\partial\chi}{\partial Q_a}\left<\frac{\partial}{\partial Q_a}\right>_{\mathscr{S}}-\braket{\hat{H}_{\mathscr{S}}}_{\mathscr{S}}+\frac{1}{2M}\sum_{a = 1}^n\left<\frac{\partial^2}{\partial Q_a^2}\right>_{\mathscr{S}} \ .
\end{aligned}
\end{equation}
The real and imaginary parts of~(\ref{eq:nonrel-source-Berry}) form a coupled system for $V_a$ and $A_a$, which can be used to eliminate two of the $2n$ real components of the partial average $\left<\frac{\partial}{\partial Q_a}\right>_{\mathscr{S}}$. By using the transformation laws of $\chi$ and $\mathfrak{J}$ given in equations~(\ref{eq:Berry-gauge-transf}) and~(\ref{eq:nonrel-Berry-gauge-transf-source}) and using~(\ref{eq:nonrel-Berry-gauge-transf-VA}) and~(\ref{eq:nonrel-Berry-gauge-transf-nabla2}), one may verify that~(\ref{eq:nonrel-source-Berry}) is invariant under the state redefinitions given in~(\ref{eq:Berry-gauge-transf}), as it should be.

\subsubsection{Backreaction}
Let us now relate~(\ref{eq:nonrel-source}) and~(\ref{eq:nonrel-pre-schrodinger}) to the non-linear system of equations with backreaction terms which was analysed in~\cite{Brout:1987-1,Brout:1987-2,Brout:1987-3,Brout:1987-4}. We first define the `covariant' derivatives~\cite{Brout:1987-1,Brout:1987-2,Brout:1987-3,Brout:1987-4,Venturi:1990,Bertoni:1996}
\begin{equation}\label{eq:nonrel-covariant-derivative-Berry}
\D_a^{\pm}:= \frac{\partial}{\partial Q_a}\pm\left<\frac{\partial}{\partial Q_a}\right>_{\mathscr{S}} \ ,
\end{equation}
which, under the state redefinition given in~(\ref{eq:Berry-gauge-transf}), transform as follows.
\begin{equation}\label{eq:nonrel-covariant-transf}
\begin{aligned}
\D^+_a\chi &=e^{\xi+\I\eta}\frac{\partial \chi'}{\partial Q^a}+e^{\xi+\I\eta}\left<\frac{\partial}{\partial Q^a}\right>_{\mathscr{S}}^{\prime}\chi' = e^{\xi+\I\eta}\D^{\prime+}_a\chi' \ , \\
\D^-_a\psi &=e^{-\xi-\I\eta}\frac{\partial \psi'}{\partial Q^a}-e^{-\xi-\I\eta}\left<\frac{\partial}{\partial Q^a}\right>_{\mathscr{S}}^{\prime}\psi' = e^{-\xi-\I\eta}\D^{\prime-}_a\psi' \ .
\end{aligned}
\end{equation}
We now substitute~(\ref{eq:nonrel-source-Berry}) in equations~(\ref{eq:nonrel-source}) and~(\ref{eq:nonrel-pre-schrodinger}) to find the system
\begin{align}
&-\frac{1}{2M}\sum_{a = 1}^n\left[\left(\D_a^{+}\right)^2+\left<\left(\D_a^-\right)^2\right>_{\mathscr{S}}\right]\chi+V\chi = \left(E-\braket{\hat{H}_{\mathscr{S}}}_{\mathscr{S}}\right)\chi \ , \label{eq:nonrel-heavy-Berry}\\
&-\frac{1}{M\chi}\sum_{a = 1}^n\D_a^+\chi\D_a^-\psi-\frac{1}{2M}\left[\left(\D_a^-\right)^2-\left<\left(\D_a^-\right)^2\right>_{\mathscr{S}}\right]\psi + \left(\hat{H}_{\mathscr{S}}-\braket{\hat{H}_{\mathscr{S}}}_{\mathscr{S}}\right)\psi = 0 \ . \label{eq:nonrel-light-Berry}
\end{align}
Due to~(\ref{eq:nonrel-covariant-transf}), one can immediately verify that~(\ref{eq:nonrel-heavy-Berry}) and~(\ref{eq:nonrel-light-Berry}) are invariant under the state redefinitions given in~(\ref{eq:Berry-gauge-transf}). Although equations~(\ref{eq:nonrel-heavy-Berry}) and~(\ref{eq:nonrel-light-Berry}) form a non-linear system due to the presence of the partial averages, they are equivalent to~(\ref{eq:nonrel-source}) and~(\ref{eq:nonrel-pre-schrodinger}), respectively, which form a linear system if $\mathfrak{J}$ is considered as an independent function.

Equations~(\ref{eq:nonrel-heavy-Berry}) and~(\ref{eq:nonrel-light-Berry}) were analysed in~\cite{Brout:1987-1,Brout:1987-2,Brout:1987-3,Brout:1987-4,Venturi:1990,Bertoni:1996,Kamenshchik:2017} as equations incorporating the non-linear effects of the backreaction of the `light' sector (which usually corresponds to matter in quantum cosmology) onto the `heavy' sector (normally gravity in quantum cosmology). Backreaction is here understood as the collection of terms involving the `light'-sector partial averages, in particular the term $\braket{\hat{H}_{\mathscr{S}}}_{\mathscr{S}}$. The presence of the partial averages in~(\ref{eq:nonrel-heavy-Berry}) and~(\ref{eq:nonrel-light-Berry}), especially of the Berry connection terms, is in line with the usual BO approximation performed in molecular physics~\cite{Cederbaum:2008,Abedi:2010,Abedi:2012-1,Abedi:2012-2}. 
In~\cite{Brout:1987-1,Brout:1987-2,Brout:1987-3,Brout:1987-4,Venturi:1990,Bertoni:1996},~(\ref{eq:nonrel-heavy-Berry}) was interpreted as an equation determining the heavy-sector wave function $\chi$. From the above construction, we see that this is equivalent to interpreting~(\ref{eq:nonrel-source}) as defining $\chi$ given $\mathfrak{J}$. Nonetheless, as we have already noted, it is also possible to interpret~(\ref{eq:nonrel-source}) as defining $\mathfrak{J}$ given $\chi$ and to consider $\chi$ as arbitrary due to~(\ref{eq:Berry-gauge-transf}).

Halliwell~\cite{Halliwell:1987} and Padmanabhan~\cite{Padmanabhan:1988} already stressed that the arbitrariness of $\chi$ implies that the backreaction terms in~(\ref{eq:nonrel-heavy-Berry}) and~(\ref{eq:nonrel-light-Berry}) are also arbitrary. This can be understood from the fact that, although $\braket{\hat{H}_{\mathscr{S}}(Q;\hat{p},q)}_{\mathscr{S}}$ is invariant under the state redefinitions of~(\ref{eq:Berry-gauge-transf}), the Berry connection terms which appear implicitly in~(\ref{eq:nonrel-heavy-Berry}) and~(\ref{eq:nonrel-light-Berry}) are not~(cf.~(\ref{eq:nonrel-Berry-gauge-transf-VA})). We can interpret a choice of $\chi$ (or $\mathfrak{J}$), as a particular fixation of (some components of) the partial averages $\left<\frac{\partial}{\partial Q_a}\right>_{\mathscr{S}}$ via~(\ref{eq:nonrel-source-Berry}). Thus, the arbitrariness of the partial-average terms is equivalent to the arbitrariness of $\chi$ (or $\mathfrak{J}$).

Furthermore, the physical meaning of different choices of $\chi$ is most clearly seen from~(\ref{eq:nonrel-pre-schrodinger-2}). Since we define the time variable from the phase $\varphi$, which is a particular function of the heavy variables $Q$, changing $\varphi$ via~(\ref{eq:Berry-gauge-transf}) corresponds to changing what one means by time. Equivalently, a transformation of the Berry connection $A_a$~(cf.~(\ref{eq:nonrel-Berry-gauge-transf-VA})) entails a redefinition of the time variable in the BO approach to the problem of time. This has not been emphasised in the literature so far.
In sections~\ref{sec:classical} and~\ref{sec:quantum}, we will analyse how this is related to a gauge choice of the time coordinate. In section~\ref{sec:example}, we examine the question of backreaction and partial averages for the archetypical example of the relativistic particle.

\subsubsection{`Light'-Sector Unitarity}
In the BO approach of~\cite{Brout:1987-1,Brout:1987-2,Brout:1987-3,Brout:1987-4,Venturi:1990,Bertoni:1996,Singh:1990,Kiefer:1991,Kiefer:2018}, the question of whether the dynamics of the `light'-sector is unitary arose due to the interpretation of $\psi$ as the `light'-sector wave function and of~(\ref{eq:nonrel-pre-schrodinger-2}) as a ``corrected'' Schr\"{o}dinger equation for the `light' sector\footnote{See also~\cite{Kim:1995-2,Massar:1998}.
}. By `light'-sector unitarity, we mean the condition
\begin{equation}\notag
\frac{\partial}{\partial t}\int\prod_{\mu}\mathrm{d}q_{\mu}\ \bar{\psi}(Q;q)\psi(Q;q) = 0 \ ,
\end{equation}
which is equivalent to
\begin{equation}\label{eq:nonrel-light-unitarity}
0 = \mathfrak{Re}\left<\frac{\partial}{\partial t}\right>_{\mathscr{S}} = \frac{1}{M}\sum_{a = 1}^n\frac{\partial\varphi}{\partial Q_a}\mathfrak{Re}\left<\frac{\partial}{\partial Q_a}\right>_{\mathscr{S}} = \frac{1}{M}\sum_{a = 1}^n\frac{\partial\varphi}{\partial Q_a}V_a \ .
\end{equation}
As $V_a$ is not necessarily zero, equation~(\ref{eq:nonrel-light-unitarity}) can only be enforced by a particular choice of $\chi$. Indeed, we see from~(\ref{eq:nonrel-V}) that $V_a = 0$ if and only if $\braket{\psi,\psi}_{\mathscr{S}} = \int\prod_{\mu}\mathrm{d}q_{\mu}\ \bar{\psi}(Q;q)\psi(Q;q)$ is a constant. For a general factorisation given in~(\ref{eq:exact-factorisation}), this will not be the case and $\braket{\psi,\psi}_{\mathscr{S}}$ will be a function of the heavy variables $Q$. Nevertheless, we are free to perform a state redefinition as in~(\ref{eq:Berry-gauge-transf}) and fix $\xi$ as follows. Starting from an arbitrary initial factorisation of the total state $\Psi(Q;q) = \chi'(Q)\psi'(Q;q)$, we demand
\begin{align*}
1 = \braket{\psi,\psi}_{\mathscr{S}}(Q) = \int\prod_{\mu}\mathrm{d}q_{\mu}\ \bar{\psi}(Q;q)\psi(Q;q) = e^{-2\xi} \int\prod_{\mu}\mathrm{d}q_{\mu}\ \bar{\psi}'(Q;q)\psi'(Q;q) \ ,
\end{align*}
which implies $\xi(Q) = \frac{1}{2}\log\left(\int\prod_{\mu}\mathrm{d}q_{\mu}\ \bar{\psi}'(Q;q)\psi'(Q;q)\right)$. For this choice of $\xi(Q)$, we obtain (cf.~(\ref{eq:nonrel-V}) and~(\ref{eq:nonrel-Berry-gauge-transf-VA}))
\begin{equation}\notag
\begin{aligned}
V_a &= V'_a -\frac{\partial\xi}{\partial Q^a} = 0 \ , \\
\psi(Q;q) &= \frac{\psi'(Q;q)}{\sqrt{\braket{\psi',\psi'}_{\mathscr{S}}(Q)}} \ .
\end{aligned}
\end{equation}
Equation~(\ref{eq:nonrel-light-unitarity}) does not follow from the equations with `backreaction' terms. Indeed, by multiplying~(\ref{eq:nonrel-light-Berry}) by $\bar{\psi}$ and integrating over $q_{\mu}$, one obtains the trivial result $0 = 0$ and no information is gained on the value of $V_a$. Thus, even with the inclusion of backreaction and Berry connection terms (which are arbitrary), one still needs to enforce the `light'-sector unitarity by a choice of $\chi$. This was emphasised, in a somewhat different way, in~\cite{Kiefer:2018}.

\subsubsection{Marginal and Conditional Wave Functions}
Let us assume that we are able to choose a factorisation $\Psi = \chi\psi_{\mathrm{n}}$ in which `light'-sector unitarity holds. We obtain
\begin{align*}
\braket{\psi_{\mathrm{n}},\psi_{\mathrm{n}}}_{\mathscr{S}}(Q) &= \int\prod_{\mu}\mathrm{d}q_{\mu}\ \bar{\psi}_{\mathrm{n}}(Q;q)\psi_{\mathrm{n}}(Q;q) = 1 \ ,\\
\braket{\chi,\chi}_{\mathscr{E}} &= \int\prod_{a}\mathrm{d}Q_{a}\ \bar{\chi}(Q)\chi(Q) = c_{\mathscr{E}} \ ,
\end{align*}
where $c_{\mathscr{E}}$ is a finite constant if $\chi$ is normalisable. We then define the normalised state $\chi_{\mathrm{n}} = \frac{1}{\sqrt{c_{\mathscr{E}}}}\chi$, such that $\braket{\chi_{\mathrm{n}},\chi_{\mathrm{n}}}_{\mathscr{E}} = 1$. The total state $\Psi_{\mathrm{n}} = \chi_{\mathrm{n}}\psi_{\mathrm{n}} = \frac{1}{\sqrt{c_{\mathscr{E}}}}\Psi$ is a solution to~(\ref{eq:TISE}) and is normalised\footnote{If the spectrum of the total Hamiltonian in~(\ref{eq:TISE}) is continuous, the eigenstate $\Psi_{\mathrm{n}}$ will not be normalisable with respect to the standard inner product. This is the usual situation in quantum cosmology, where one needs to consider alternative definitions for the inner product in the space of physical states. In section~\ref{sec:innerprod}, we will consider the Klein-Gordon inner product for simplicity.} with respect to the inner product over all variables,
\begin{align*}
\braket{\Psi_{\mathrm{n}},\Psi_{\mathrm{n}}} &=  \int\prod_{a}\mathrm{d}Q_{a}\prod_{\mu}\mathrm{d}q_{\mu}\ \bar{\Psi}_{\mathrm{n}}\Psi_{\mathrm{n}} = \int\prod_{a}\mathrm{d}Q_{a}\ \bar{\chi}_{\mathrm{n}}\chi_{\mathrm{n}}\braket{\psi_{\mathrm{n}},\psi_{\mathrm{n}}}_{\mathscr{S}}(Q)\\
&= \braket{\chi_{\mathrm{n}},\chi_{\mathrm{n}}}_{\mathscr{E}} = 1 \ . 
\end{align*}
In this case, it is possible to interpret $|\Psi_{\mathrm{n}}(Q;q)|^2$ as a \emph{joint} probability density for the `light' system \emph{and} the `heavy' sector to be in the $(Q,q)$ configuration, whereas $\chi_{\mathrm{n}}(Q)$ and $\psi_{\mathrm{n}}(Q;q)$ are interpreted as a \emph{marginal wave function} and a \emph{conditional wave function}, respectively. Indeed, $|\chi_{\mathrm{n}}|^2$ can be seen as the marginal probability density for the `heavy' variables to be in the $Q$ configuration regardless of the configuration of the `light' variables. Analogously, $|\psi_{\mathrm{n}}|^2$ is then interpreted as the conditional probability density to find the `light' system in the $q$ configuration given that the `heavy' sector is in the $Q$ configuration. In the context of molecular physics and the BO approximation, this interpretation was used in~\cite{Hunter:1975,Abedi:2010,Abedi:2012-1,Abedi:2012-2,Cederbaum:2013}. Such an interpretation was also adopted by Arce~\cite{Arce:2012} in the context of the problem of time in (non-relativistic) quantum mechanics. In~\cite{Arce:2012}, Arce referred to the partial averages defined in~(\ref{eq:nonrel-partial-average-def}) as \emph{conditional expectation values}. Such an interpretation is only possible when the `light'-sector unitarity is enforced~\cite{Abedi:2012-1,Abedi:2012-2}. If one is able to enforce light-sector unitarity together with a normalisable $\chi$ at the exact level of~(\ref{eq:nonrel-pre-schrodinger-2}), then one may interpret~(\ref{eq:nonrel-pre-schrodinger-2}) as a ``corrected'' Schr\"{o}dinger equation for the conditional `light'-sector wave function $\psi(Q;q)$.

Alternatively, one may \emph{choose} $\chi(Q) = e^{i\varphi(Q)}$ and interpret $\psi(Q;q)$ as the wave function for the \emph{full} system comprised of `heavy' \emph{and} `light' degrees of freedom. In this way, the factorisation $\Psi = \chi\psi = e^{i\varphi}\psi$ is merely a phase transformation of the full system. This is the interpretation that we will adopt in this paper (except in section~\ref{sec:quantum-matter-classical-gravity}), which does not require that `light'-sector unitarity be enforced. We recall that this phase transformation is needed in order to rewrite the constraint equation~(\ref{eq:TISE}) in the form of~(\ref{eq:nonrel-pre-schrodinger-2}), which leads to the TDSE~(\ref{eq:nonrel-schrodinger}) at lowest order in an expansion in powers of $\frac{1}{M}$.

At the non-perturbative level (without resorting to such an expansion), the choice of factorisation $\Psi = e^{i\varphi}\psi$ evades some of the problems mentioned by Kucha\v{r} in his critique of the semiclassical interpretation~\cite{Kuchar:1991,Isham:1992}. Indeed, all the states in the Hilbert space can be transformed according to the same phase factor $e^{\I \varphi}$. The time variable is defined from only one congruence of trajectories associated with $\varphi$ and we need not consider how time `emerges' if the state is a superposition of factors such as $\Psi = e^{\I\varphi_1}\psi_1+e^{\I\varphi_2}\psi_2$ (or more generally~(\ref{eq:BO-factorisation})), since this state can be rewritten as $\Psi = e^{\I\varphi}\left[e^{\I(\varphi_1-\varphi)}\psi_1+e^{\I(\varphi_2-\varphi)}\psi_2\right]\equiv e^{\I\varphi}\Psi_{\varphi}$ for any choice of $\varphi$. Thus, it is not necessary to consider the interference of states with different phase pre-factors and there is no `superposition or interference problem' in the \emph{definition} (choice) of the (phase) time variable. However, the general superposition $\Psi_{\varphi} = e^{\I(\varphi_1-\varphi)}\psi_1+e^{\I(\varphi_2-\varphi)}\psi_2$ may not admit an expansion in powers of $\frac{1}{M}$ and, in this case, it is necessary to invoke the \emph{decoherence}~\cite{Kiefer:1993-0,Kiefer:1993-1,Kiefer:1993-2,Kiefer:2009T} of $\Psi$ into separate factors $e^{\I\varphi_i}\psi_i$ which can be treated perturbatively and thus lead to the Schr\"{o}dinger equation~(\ref{eq:nonrel-schrodinger}) at lowest order. In general, decoherence is relevant to the study of the classical limit of (a subset of) the quantised variables~\cite{Kiefer:1987,Halliwell:1989}.

\section{The BO Approach as a Choice of Gauge. Classical Theory}\label{sec:classical}
We now illustrate how the results of the BO approach to the problem of time discussed in the previous section can be obtained by a choice of gauge in the classical theory.
We focus on cosmological minisuperspace models and consider for simplicity that the `heavy' sector coincides with the gravitational sector, while the `light' variables are given by the matter degrees of freedom, although more general separations are possible~\cite{Vilenkin:1988,Kiefer:1993-1,Kiefer:1993-2}.

The gravitational-sector configuration space is endowed with local coordinates $Q^a, a = 1,...,n$ and an indefinite metric ${\bf G}$. Indices $a,b,...$ are lowered and raised with the components $G_{ab}$ and $G^{ab}$ of the metric and its inverse, respectively. We choose local coordinates $q^{\mu}, \mu = 1,...,d$ for the matter-sector configuration space with positive-definite metric ${\bf h}(Q)$. Indices $\mu,\nu,...$ are raised and lowered with $h^{\mu\nu}$ and $h_{\mu\nu}$, respectively. We consider the action functional
\begin{equation}\label{eq:action}
S = \int\mathrm{d}t\ \left(P_a\dot{Q}^a+p_{\mu}\dot{q}^{\mu}-NH\right) \ ,
\end{equation}
where summation over repeated indices is implied. A dot over a variable indicates differentiation with respect to the parameter $t$. The lapse $N$ is taken to be an arbitrary multiplier\footnote{We have eliminated the momentum conjugate to the lapse, $p_N$, by solving the primary constraint $p_N = 0$.} and the Hamiltonian constraint is
\begin{equation}\label{eq:hamiltonian-constraint}
\begin{aligned}
H &= H_g(Q,P)+H_m(Q;p,q) = 0 \ , \\
H_g(Q,P)  &= \frac{1}{2M}G^{ab}(Q)P_aP_b+MV(Q) \ ,\\
H_m(Q;p,q) &= \frac{1}{2}h^{\mu\nu}(Q;q)p_{\mu}p_{\nu}+V_m(Q;q) \ ,
\end{aligned}
\end{equation}
where $M = \frac{1}{32\pi G}$ and $G$ is Newton's constant. 
This is a time-reparametrisation invariant system. We assume that the gravitational potential term $V(Q)$ is non-vanishing, which is achieved if, for example, the cosmological constant term is not zero.
The matter degrees of freedom $p_{\mu},q^{\mu}$ can be assumed to be associated with a typical mass scale $m\ll\sqrt{M}$.
The matter-sector Hamiltonian $H_m$ depends only parametrically on the $Q$ coordinates and we assume $V_m$ is a smooth, non-negative real function. The equations of motion of this system read 
\begin{equation}\label{eq:eom}
\begin{aligned}
\dot{Q}^a &= \frac{N}{M}G^{ab}(Q)P_b \ , \\
\dot{P}_a &= -N\left(\frac{1}{2M}\frac{\partial G^{cd}}{\partial Q^a}P_cP_d+M\frac{\partial V}{\partial Q^a}\right)-N\frac{\partial H_m}{\partial Q^a} \ ,\\
\dot{q}^{\mu} &= N\frac{\partial H_m}{\partial p_{\mu}} \ , \\
\dot{p}_{\mu} &= -N\frac{\partial H_m}{\partial q^{\mu}} \ , \\
H &= H_g+H_m = 0 \ .
\end{aligned}
\end{equation}
One may solve these equations after making a choice of lapse. Alternatively, we can perform a canonical transformation, with generating function $F = W(Q, P';q,p') - Q^{\prime a} P'_a-q^{\prime\mu}p'_{\mu}$, such that the momenta are substituted by the gradient components of $W$, $P_a = \frac{\partial W}{\partial Q^a}$ and $p_{\mu} = \frac{\partial W}{\partial q^{\mu}}$, and the constraint equation becomes the differential equation
\begin{equation}\label{eq:HJ-general}
\frac{1}{2M}G^{ab}(Q)\frac{\partial W}{\partial Q^a}\frac{\partial W}{\partial Q^b}+MV(Q)+H_m\left(Q;\frac{\partial W}{\partial q},q\right) = 0 \ ,
\end{equation}
which will be referred to simply as the Hamilton-Jacobi equation, while $W$ will be called the Hamilton characteristic function.
Given a solution of~(\ref{eq:HJ-general}), one may pass to the new canonical frame described by the variables $Q', P', q', p'$, with respect to which the dynamics is trivial. Equivalently, one may still work with the old coordinates $Q$ and $q$, with respect to which the dynamics is described by the reduced set of equations
\begin{equation}\label{eq:eom-HJ}
\begin{aligned}
\dot{Q}^a &= \frac{N}{M}G^{ab}(Q)\frac{\partial W}{\partial Q^b} \ ,\\
\dot{q}^{\mu} &= Nh^{\mu\nu}(Q;q)\frac{\partial W}{\partial q^{\nu}} \ ,
\end{aligned}
\end{equation}
for a given choice of lapse.

\subsection{Canonical Variables Adapted to a Choice of Background}
The presence of the matter-sector Hamiltonian $H_m(Q;p,q)$ affects the dynamics of the gravitational sector. If $H_m(Q;p,q)$ can be approximated (in a sense which will be discussed in section~\ref{sec:classical-pt}) by a function $J(Q)$ solely of the gravitational configuration variables, we can consider that the dynamics of the gravitational field is approximately dictated by the Hamilton-Jacobi equation (cf.~(\ref{eq:HJ-general})) for the gravitational system in the presence of a source
\begin{equation}\label{eq:source}
\frac{1}{2M}G^{ab}(Q)\frac{\partial \varphi}{\partial Q^a}\frac{\partial \varphi}{\partial Q^b}+MV(Q) = -J(Q) \ .
\end{equation}
At this stage, however, we can consider $J(Q)$ as arbitrary\footnote{Parentanti~\cite{Parentani:1998} and Briggs~\cite{Briggs:2015} have considered similar definitions of an arbitrary source.}.
The solution $\varphi$ will be referred to as the background Hamilton function and it is analogous to the phase used in~(\ref{eq:nonrel-pre-schrodinger}). It is convenient to define the quantities
\begin{equation}\label{eq:Phi-source}
\Phi_a(Q) = \frac{\partial\varphi}{\partial Q^a} \ , 
\end{equation}
which will be called background momenta. Equations~(\ref{eq:source}) and~(\ref{eq:Phi-source}) imply the background momenta are normalised to
\begin{equation}\label{eq:normalisation-Phi}
G^{ab}(Q)\Phi_a\Phi_b = -2M(J(Q)+MV(Q)) \ .
\end{equation}
As in~(\ref{eq:nonrel-source}), we note that~(\ref{eq:source}) may be regarded either as a definition of $\varphi$ given $J$ or as a definition of $J$ given $\varphi$.
If we change the background Hamilton function, $\varphi(Q)= \varphi'(Q) + \eta(Q)$, we may change the background momenta and source accordingly,
\begin{equation}\label{eq:Phi-source-transf}
\begin{aligned}
\Phi_a(Q) = \Phi'_a +\frac{\partial\eta}{\partial Q^a} \ , \ J(Q) = J'(Q)-\frac{1}{M}G^{ab}(Q)\Phi'_a\frac{\partial\eta}{\partial Q^b}-\frac{1}{2M}G^{ab}(Q)\frac{\partial\eta}{\partial Q^a}\frac{\partial\eta}{\partial Q^b} \ .
\end{aligned}
\end{equation}
If $\varphi$ is chosen to be a non-constant function of the $Q$ coordinates, then the background momenta $\Phi_a$ will be non-trivial. In this case, we assume that one may define a holonomic vector-field basis in the tangent bundle composed of the
vector fields $\left\{\mathbf{B}_1 = \frac{\mathcal{N}}{M}\boldsymbol\Phi, \mathbf{B}_{i}\right\}, i = 2,...n$, where $\boldsymbol\Phi = G^{ab}\Phi_a\frac{\partial}{\partial Q^b}$ and $\mathcal{N}\equiv\mathcal{N}(Q)$ is an arbitrary normalisation function, which can be interpreted as a `background lapse'. The calculations are somewhat simplified if we assume that $\mathbf{B}_i$ are orthogonal to $\mathbf{B}_1$\footnote{In the most general case, for a given $\varphi$, it will not be possible to find a coordinate basis with this property and the metric components $\tilde{G}_{1i}$ in~(\ref{eq:normalisation-basis-vectors}) will differ from zero. In this case, there will be additional contributions from such components to the formulae we present. However, such additional terms do not offer a substantial modification of the formalism and should not qualitatively alter the conclusions regarding the choice of gauge and unitarity of the quantum theory discussed in sections~\ref{sec:gauge-fixing-classical} and~\ref{sec:innerprod}, respectively.}. The basis vectors are then normalised as follows (cf.~(\ref{eq:normalisation-Phi})):
\begin{equation}\label{eq:normalisation-basis-vectors}
\begin{aligned}
G_{ab}B^a_1 B^b_1 &= -2\mathcal{N}^2(Q)\left(\frac{J(Q)}{M}+V(Q)\right) = \tilde{G}_{11} \ , \\
G_{ab}B^a_1 B^b_i &= 0 = \tilde{G}_{1i} \ , \\
G_{ab}B^a_iB^b_j &= \tilde{G}_{ij}\equiv g_{ij} \ .
\end{aligned}
\end{equation}
We then define new coordinates $x = (x^1, x^i)$ via the integral curves of the basis fields,
\begin{equation}\label{eq:coord-transf-Qx}
\begin{aligned}
B^a_1 &= \frac{\mathcal{N}}{M}G^{ab}\Phi_b = \frac{\partial Q^a}{\partial x^1} \ , \\
B^a_{i} &= \frac{\partial Q^a}{\partial x^i} \ .
\end{aligned}
\end{equation}
In this way, the gravitational-sector configuration space is foliated by surfaces of constant $x^1$, on which $g_{ij} = \tilde{G}_{ij}$ is the induced metric. We will denote its inverse by $g^{ij}$. The first of equations~(\ref{eq:coord-transf-Qx}) can be interpreted as a (fictitious) `background' equation of motion for $Q^a$, which depends on the background momenta $\Phi_a$ and the background lapse $\mathcal{N}$ and for which $x^1(Q)$ plays the role of a `background time' function (compare the first of equations~(\ref{eq:coord-transf-Qx}) to the first of equations~(\ref{eq:eom-HJ})). We also obtain the useful identities
\begin{equation}\label{eq:x-gradient-varphi}
\begin{aligned}
\frac{\partial\varphi}{\partial x^1} &= \frac{\mathcal{N}}{M}G^{ab}\Phi_a\Phi_b = -2\mathcal{N}(J+MV) \ , \\
\frac{\partial\varphi}{\partial x^i} &= B^a_i\Phi_a = 0 \ .
\end{aligned}
\end{equation}
In appendix~\ref{app:formulas-Qx-transf}, we collect formulae related to the change of coordinates given in~(\ref{eq:coord-transf-Qx}).

The change of coordinates $Q\mapsto x$ induces a canonical transformation. The momenta conjugate to the $x$ coordinates read
\begin{equation}
\begin{aligned}
\tilde{P}_1 &=B^a_1P_a = \frac{\mathcal{N}}{M}G^{ab}\Phi_aP_b = \frac{\mathcal{N}}{M}G^{ab}\Phi_a\frac{\partial W}{\partial Q^b} = \frac{\partial W}{\partial x^1} \ , \\
\tilde{P}_i &= B^a_i P_a = B^a_i\frac{\partial W}{\partial Q^a} = \frac{\partial W}{\partial x^i}  \ .
\end{aligned}
\end{equation}
We can now use the above variables adapted to the background Hamilton function $\varphi$ to rewrite the Hamiltonian constraint of the full theory, in which $H_m(Q;p,q)$ is present. In terms of the new canonical variables, eq.~(\ref{eq:hamiltonian-constraint}) reads
\begin{equation}\label{eq:hamiltonian-constraint-tilde}
H = \frac{1}{2M}\tilde{G}^{11}(x)\left(\tilde{P}_1\right)^2+\frac{1}{2M}g^{ij}(x)\tilde{P}_i\tilde{P}_j+MV(x)+H_m(x;p,q) = 0 \ .
\end{equation}

\subsection{Gauge Fixing. Reduced Phase Space}\label{sec:gauge-fixing-classical}
Due to time-reparametrisation invariance, we are free to choose a parametrisation in which the following gauge fixing condition holds,
\begin{equation}
\tau(Q(t)) = t \ ,
\end{equation}
where $\tau$ is some smooth function of the $Q$ coordinates\footnote{This is an \emph{intrinsic} choice of time parametrisation, i.e., it only depends on the configuration space variables. It is also possible to choose \emph{extrinsic} coordinate conditions, which include a dependence on the momenta~\cite{Kuchar:1991}. Any given parametrisation may be restricted to a given region of phase space and it may not be possible to extend it to the whole space. In the field-theoretic case~(see appendix~\ref{app:field-theory}), intrinsic gauge choices will not be spacetime scalars, as they will be defined solely from the spatial three-metric. This leads to the so-called `spacetime problem'~\cite{Kuchar:1991}.}. Such a choice of time parametrisation leads to the following equation for the lapse
\begin{equation}\label{eq:general-lapse}
\frac{1}{N} = \frac{1}{M}G^{ab}(Q)P_a\frac{\partial\tau}{\partial Q^b} = \frac{1}{M}G^{ab}(Q)\frac{\partial W}{\partial Q^a}\frac{\partial\tau}{\partial Q^b} \ .
\end{equation}
We can choose\footnote{This is the choice that will lead to `WKB time' in the quantum theory. Other parametrisations are, in principle, possible.} the parametrisation in which the `background time' function $x^1$ defines the time coordinate, i.e.,
\begin{equation}
\tau(Q(t)) = x^1(Q(t)) = t \ ,
\end{equation}
which leads to the fixation of the lapse
\begin{equation}\label{eq:fixed-lapse-P}
\frac{1}{N} = \frac{1}{M}G^{ab}(Q)\frac{\partial x^1}{\partial Q^a}P_b = \frac{1}{M}\tilde{G}^{11}(x)\tilde{P}_1 = -\frac{\tilde{P}_1}{2\mathcal{N}^2(x)\left(J(x)+MV(x)\right)} \ .
\end{equation}
The momentum conjugate to $\tau = x^1$ is $P_{\tau} = \tilde{P}_1$. We can now solve the Hamiltonian constraint of~(\ref{eq:hamiltonian-constraint-tilde}) for $P_{\tau} = \tilde{P}_1$ to obtain
\begin{equation}\label{eq:reduced-hamiltonian-tilde}
-P_{\tau} = \pm 2M\left|\mathcal{N}\right|\left[\left(\frac{J}{M}+V\right)\left(V+\frac{1}{M}H_m+\frac{1}{2M^2}g^{ij}\tilde{P}_i\tilde{P}_j\right)\right]^{\frac{1}{2}} \equiv H^{\pm}_{\text{red}} \ ,
\end{equation}
where we used the fact that $\tilde{G}^{11}(x) = \left[\tilde{G}_{11}(x)\right]^{-1}$ and~(\ref{eq:normalisation-basis-vectors}). The function $H^{\pm}_{\text{red}}$ is the \emph{reduced} Hamiltonian for the gauge-fixed system. The corresponding reduced phase-space action is
\begin{equation}\label{eq:reduced-action}
S_{\text{red}} = \int\mathrm{d}\tau\ \left(\tilde{P}_i\frac{\partial x^i}{\partial\tau}+p_{\mu}\frac{\partial q^{\mu}}{\partial\tau}-H^{\pm}_{\text{red}}\right) \ .
\end{equation}
The reduced phase space is thus comprised of the degrees of freedom $\tilde{P}_i, x^i, p_{\mu}, q^{\mu}$.

\subsection{Background Transformations}
In the presence of matter, the Hamilton characteristic function $W(Q,P'; q,p')$ will in general not coincide with the background Hamilton function $\varphi(Q)$. Let their difference be $S = W-\varphi$. If $S$ is not trivial, we may still interpret it as the Hamilton characteristic function for the system described by the action given in~(\ref{eq:action}), provided we perform the canonical transformation\footnote{Note that this transformation is unrelated to time-reparametrisation invariance.}
\begin{equation}\label{eq:background-gauge}
\begin{aligned}
&Q^a \mapsto Q^a \ , \ P_a = \frac{\partial W}{\partial Q^a}\mapsto\Pi_a = P_a - \frac{\partial\varphi}{\partial Q^a} = \frac{\partial S}{\partial Q^a} \ ,\\
&q^{\mu} \mapsto q^{\mu} \ , \ p_{\mu} = \frac{\partial W}{\partial q^{\mu}}\mapsto p_{\mu} = \frac{\partial S}{\partial q^{\mu}} \ .
\end{aligned}
\end{equation}
If we change the background Hamilton function, $\varphi= \varphi' + \eta$, $S$ transforms as $S = S'-\eta$ so as to leave $W = S+\varphi$ invariant. Equivalently, the $\Pi$-momenta transform as $\Pi_a = \Pi'_a -\frac{\partial\eta}{\partial Q^a}$ so as to keep $P_a = \Pi_a + \Phi_a$ invariant. The matter-sector momenta $p_{\mu}$ are also invariant, since the background Hamilton function does not depend on matter degrees of freedom. Such a change of background Hamilton function amounts to performing a new canonical transformation of the same type as the one given in~(\ref{eq:background-gauge}). We will refer to such canonical transformations as background transformations. They will be useful in perturbation theory. The invariance of $P_a, p_{\mu}$ implies that the constraint equation is invariant under background transformations and independent of the choice of $\varphi$, as it should be.

If we now change to the $x$-coordinates, we find
\begin{equation}\label{eq:tildePi}
\begin{aligned}
\tilde{\Pi}_1 &= \tilde{P}_1 - \frac{\partial\varphi}{\partial x^1} = \tilde{P}_1 + 2\mathcal{N}(x)\left(J(x)+MV(x)\right) \ , \\
\tilde{\Pi}_i &= \tilde{P}_i \ ,
\end{aligned}
\end{equation}
where we used~(\ref{eq:x-gradient-varphi}). Using~(\ref{eq:normalisation-basis-vectors}) and~(\ref{eq:tildePi}) and the fact that $\tilde{G}^{11} = \left(\tilde{G}_{11}\right)^{-1}$, we may write the constraint equation for an arbitrary choice of $\varphi$ in the $x$-coordinates as
\begin{equation}\label{eq:hamiltonian-constraint-arbitrary-varphi}
H = \frac{\tilde{\Pi}_1}{\mathcal{N}}-\frac{\left(\tilde{\Pi}_1\right)^2}{4\mathcal{N}^2\left(J+MV\right)}+\frac{1}{2M}g^{ij}(x)\tilde{\Pi}_i\tilde{\Pi}_j+H_m-J \ .
\end{equation}
Let us now choose the parametrisation $\tau(Q(t)) = x^1(Q(t)) = t$. Due to~(\ref{eq:x-gradient-varphi}) and~(\ref{eq:tildePi}), we find
\begin{align}
\frac{\partial\varphi}{\partial x^1} &\equiv \frac{\mathrm{d}\varphi}{\mathrm{d}\tau} \ , \label{eq:total-tau}\\
P_{\tau} &= \tilde{\Pi}_1+\frac{\mathrm{d}\varphi}{\mathrm{d}\tau} \label{eq:Ptau-total-tau} \ .
\end{align}
We may rewrite the gauge-fixed lapse of~(\ref{eq:fixed-lapse-P}) as
\begin{equation}\label{eq:fixed-lapse-Pi}
N = \frac{\mathcal{N}}{1-\frac{\tilde{\Pi}_1}{2\mathcal{N}\left(J+MV\right)}} \ ,
\end{equation}
where we used~(\ref{eq:tildePi}). Using~(\ref{eq:Ptau-total-tau}), we can also rewrite~(\ref{eq:reduced-hamiltonian-tilde}) as
\begin{equation}\label{eq:reduced-hamiltonian-Pi}
-\tilde{\Pi}_1 =-2\mathcal{N}\left(J+MV\right) \pm 2M\left|\mathcal{N}\right|\left[\left(\frac{J}{M}+V\right)\left(V+\frac{1}{M}H_m+\frac{1}{2M^2}g^{ij}\tilde{\Pi}_i\tilde{\Pi}_j\right)\right]^{\frac{1}{2}}\equiv H^{\prime\pm}_{\text{red}} \ ,
\end{equation}
which is the solution of the transformed constraint equation~(\ref{eq:hamiltonian-constraint-arbitrary-varphi}). The function $H^{\prime\pm}_{\text{red}}$ differs from $H^{\pm}_{\text{red}}$ by a total $\tau$ derivative and is the reduced Hamiltonian for the gauge-fixed system comprised of the degrees of freedom $\tilde{\Pi}_i, x^i, p_{\mu}, q^{\mu}$, with the associated reduced action (cf.~(\ref{eq:reduced-action}))
\begin{equation}\label{eq:reduced-action-2}
S'_{\text{red}} = \int\mathrm{d}\tau\ \left(\tilde{\Pi}_i\frac{\partial x^i}{\partial\tau}+p_{\mu}\frac{\partial q^{\mu}}{\partial\tau}-H^{\prime\pm}_{\text{red}}\right) = S_{\text{red}}-\int\mathrm{d}\tau\frac{\mathrm{d}\varphi}{\mathrm{d}\tau} \ .
\end{equation}
Therefore, the reduced canonical theories described by~(\ref{eq:reduced-action}) and~(\ref{eq:reduced-action-2}) are equivalent.

\subsection{Perturbation Theory}\label{sec:classical-pt}
\subsubsection{Expansion of the Reduced Hamiltonian}
If we assume that all energy scales involved are much smaller than the heavy scale $\sqrt{M}$, it is possible to develop a formal perturbative expansion of the above reduced Hamiltonian in powers of $\frac{1}{M}$, which corresponds to a weak coupling expansion. We consider that the normalisation function $\mathcal{N}$ is independent of $M$, and the source $J$ can be expanded as
\begin{equation}\label{eq:J-expansion}
J(Q) = \sum_{k = 0}^{\infty}\frac{J^{(k)}(Q)}{M^k} \ .
\end{equation}
We further assume that $J$ is chosen such that the inequality $|H_m-J|^2 \ll M$ holds on-shell. Equations~(\ref{eq:source}) and~(\ref{eq:J-expansion}) together imply that the background Hamilton function should be expanded as follows:
\begin{equation}\label{eq:phi-expansion}
\varphi(Q) = M\varphi^{(-1)}(Q)+\sum_{k = 0}^{\infty}\frac{\varphi^{(k)}(Q)}{M^k} \ .
\end{equation}
The term proportional to $M$ is needed if the potential $V$ is non-zero~(cf.~(\ref{eq:source})) and we assume this is the case. We then expand the square-root in~(\ref{eq:reduced-hamiltonian-Pi}) to obtain
\begin{equation}\label{eq:reduced-Hamiltonian-expanded}
\begin{aligned}
H_{\text{red}}^{\prime\kappa} = &-2\mathcal{N}(J+MV)+\kappa\Biggbr{2M|\mathcal{N}V|+|\mathcal{N}|\sigma(H_m+J)\\
&-\frac{|\mathcal{N}|}{4M|V|}(H_m-J)^2+\frac{|\mathcal{N}|\sigma}{2M}g^{ij}\tilde{\Pi}_i\tilde{\Pi}_j}+\mathcal{O}\left(\frac{1}{M^2}\right) \ ,
\end{aligned}
\end{equation}
where $\kappa = \pm1$ labels the two solutions of the constraint equation and $\sigma = \mathrm{sgn}(V)$. We refrained from expanding the source $J$. Equation~(\ref{eq:reduced-Hamiltonian-expanded}) should be interpreted as the solution of the Hamiltonian constraint for the momentum $\tilde{\Pi}_1$, which is conjugate to the coordinate $x^1$ defined as in~(\ref{eq:coord-transf-Qx}), when the expansion of $\varphi$ given in~(\ref{eq:phi-expansion}) is truncated at order $\frac{1}{M}$. The reduced equations of motion read
\begin{equation}\label{eq:reduced-eom-expanded}
\begin{aligned}
\frac{\partial x^i}{\partial\tau} &= \frac{\kappa\sigma|\mathcal{N}|}{M}g^{ij}\tilde{\Pi}_j+\mathcal{O}\left(\frac{1}{M^2}\right) \ , \\
-\frac{\partial\tilde{\Pi}_i}{\partial\tau} &= -2\frac{\partial}{\partial x^i}(\mathcal{N}J+M\mathcal{N}V)+\kappa\frac{\partial}{\partial x^i}\left[2M|\mathcal{N}V|+|\mathcal{N}|\sigma\left(H_m+J\right)-\frac{|\mathcal{N}|}{4M|V|}(H_m-J)^2\right.\\
&\left.\ \ \ +\frac{|\mathcal{N}|\sigma}{2M}g^{ij}\tilde{\Pi}_i\tilde{\Pi}_j\right]+\mathcal{O}\left(\frac{1}{M^2}\right)  \ , \\
\frac{\partial q^{\mu}}{\partial \tau} &=\kappa\sigma|\mathcal{N}|\left[1-\frac{(H_m-J)}{2MV}\right]\frac{\partial H_m}{\partial p_{\mu}}+\mathcal{O}\left(\frac{1}{M^2}\right) \ , \\
-\frac{\partial p_{\mu}}{\partial\tau} &=\kappa\sigma|\mathcal{N}|\left[1-\frac{(H_m-J)}{2MV}\right]\frac{\partial H_m}{\partial q^{\mu}}+\mathcal{O}\left(\frac{1}{M^2}\right) \ .
\end{aligned}
\end{equation}
If we choose $\kappa = -\sigma\mathrm{sgn}(\mathcal{N}) = -\mathrm{sgn}(\mathcal{N}V)$, we find from~(\ref{eq:reduced-Hamiltonian-expanded}) and~(\ref{eq:reduced-eom-expanded}) that all on-shell gravitational-sector momenta (including the on-shell reduced Hamiltonian) exhibit poles in the perturbative parameter, which are the terms proportional to $M$, whereas $x^i(\tau), q^{\mu}(\tau)$ and $p_{\mu}(\tau)$ are analytic functions of $\frac{1}{M}$. In particular, we find the lowest-order equations for the gravitational sector
\begin{equation}\label{eq:grav-lowest-1}
\begin{aligned}
\frac{\partial x^i}{\partial\tau} &= -\mathcal{N}g^{ij}\left(\frac{\tilde{\Pi}_j}{M}\right) +\mathcal{O}\left(\frac{1}{M}\right) \ ,\\
\frac{\partial}{\partial\tau}\left(\frac{\tilde{\Pi}_i}{M}\right) &= 4\frac{\partial}{\partial x^i}(\mathcal{N}V)+\mathcal{O}\left(\frac{1}{M}\right)  \ ,
\end{aligned}
\end{equation}
which imply that, to lowest order, $\Pi_i(\tau)$ are proportional to $M$ and $x^i(\tau)$ are time-\emph{dependent} functions of order $M^0$. The gravitational-sector trajectory $x^i(\tau)$ given in~(\ref{eq:grav-lowest-1}) is independent of the matter-sector trajectory at this order.

Alternatively, if we choose $\kappa = +\sigma\mathrm{sgn}(\mathcal{N}) = +\mathrm{sgn}(\mathcal{N}V)$, we find that all dynamical solutions (including the on-shell reduced Hamiltonian) are analytic functions of $\frac{1}{M}$. The lowest-order gravitational-sector equations read
\begin{equation}\label{eq:grav-lowest-2}
\begin{aligned}
\frac{\partial x^i}{\partial\tau} &= 0+\mathcal{O}\left(\frac{1}{M}\right) \ ,\\
\frac{\partial \Pi_i}{\partial\tau} &= -\frac{\partial}{\partial x^i}\left[\mathcal{N}\left(H_m-J\right)\right]+\mathcal{O}\left(\frac{1}{M}\right)  \ ,
\end{aligned}
\end{equation}
which imply that, to lowest order, both $\Pi_i(\tau)$ and $x^i$ are of order $M^0$ and $x^i$ are constants\footnote{In~\cite{Vilenkin:1988,Kuchar:1991}, they were called `comoving' coordinates. In general, they will not be constants to higher orders in perturbation theory.}, i.e. $x^i(\tau) = x^i(0) + \mathcal{O}\left(\frac{1}{M}\right)$.  Thus, we see that both sign choices yield solutions $x^i(\tau)$ which are independent of the matter-sector dynamics at lowest order. Evidently, $\mathrm{sgn}(\mathcal{N}V)$ can vary in different regions of configuration space. Therefore, the above perturbative conclusions for a fixed choice of $\kappa$ only hold when the dynamical trajectory is restricted to a region of phase space in which $\mathrm{sgn}(\mathcal{N}V)$ is constant.

It is also useful to expand the gauge-fixed lapse given in~(\ref{eq:fixed-lapse-Pi}). Using~(\ref{eq:reduced-Hamiltonian-expanded}), we find
\begin{align*}
\frac{1}{N} &= \frac{1}{\mathcal{N}}-\frac{\tilde{\Pi}_1}{2\mathcal{N}^2\left(J+MV\right)} = \frac{1}{\mathcal{N}}+\frac{H_{\text{red}}^{\prime\kappa}}{2\mathcal{N}^2\left(J+MV\right)}\\
&= \frac{1}{\mathcal{N}}-\frac{1}{\mathcal{N}}+\frac{\kappa\sigma}{|\mathcal{N}|}-\frac{\kappa J}{M|\mathcal{N} V|}+\frac{\kappa(H_m+J)}{2M|\mathcal{N}V|}+\mathcal{O}\left(\frac{1}{M^2}\right)\\
&= \frac{\kappa\sigma}{|\mathcal{N}|}+\frac{\kappa(H_m-J)}{2M|\mathcal{N}V|}+\mathcal{O}\left(\frac{1}{M^2}\right) \ ,
\end{align*}
which yields
\begin{equation}\label{eq:lowest-order-lapse}
\begin{aligned}
N(\tau,x^i(\tau)) &= \kappa\sigma|\mathcal{N}(\tau,x^i(\tau))|\\
&-\frac{\kappa|\mathcal{N}(\tau,x^i(\tau))|}{2M|V(\tau,x^i(\tau))|}\left(H_m(\tau, x^i(\tau); p, q)-J(\tau,x^i(\tau))\right)+\mathcal{O}\left(\frac{1}{M^2}\right) \ .
\end{aligned}
\end{equation}

\subsubsection{Propagation of Matter in a Fixed Gravitational Background}
We have seen that for both choices of $\kappa$ the trajectory of the gravitational configuration variables is independent of the matter-sector dynamics (i.e. there is no backreaction from the matter sector onto the gravitational configuration variables) at the lowest order of the weak-coupling expansion. This implies that the clock defined from the `heavy' variables is not affected by the dynamics of the `light' variables at this order and thus provides an `external' notion of time for their evolution. The lowest-order equations of motion for the matter sector read (cf.~(\ref{eq:reduced-eom-expanded}))
\begin{equation}\label{eq:matter-fixed-background}
\begin{aligned}
\frac{\partial q^{\mu}}{\partial \tau} &=\kappa\sigma|\mathcal{N}(\tau, x^i(\tau))|\frac{\partial}{\partial p_{\mu}}H_m(\tau, x^i(\tau); p, q)+\mathcal{O}\left(\frac{1}{M}\right) \ , \\
\frac{\partial p_{\mu}}{\partial\tau} &=-\kappa\sigma|\mathcal{N}(\tau, x^i(\tau))|\frac{\partial}{\partial q^{\mu}}H_m(\tau, x^i(\tau);p,q)+\mathcal{O}\left(\frac{1}{M}\right) \ ,
\end{aligned}
\end{equation}
which are the equations of motion for matter propagating in the fixed gravitational background characterised by the `lapse' $\kappa\sigma|\mathcal{N}(\tau, x(\tau))|$ and the functions $\tau = x^1, x^i(\tau)$. Note that the `lapse' $\kappa\sigma|\mathcal{N}(\tau, x(\tau))|$ is simply the lowest order term in the expansion of the gauge-fixed lapse given in~(\ref{eq:fixed-lapse-Pi}), as shown in~(\ref{eq:lowest-order-lapse}).  Higher orders in $\frac{1}{M}$ represent corrections from the full, time-reparametrisation invariant theory to the description where the gravitational background is given by a fixed trajectory independent from the matter sector, i.e. the (classical) backreaction of matter is taken into account at higher orders. 

\subsubsection{Iterative Procedure}\label{sec:iterative-procedure}
The choice $\kappa=+\sigma\mathrm{sgn}(\mathcal{N})\equiv\kappa_{+}$ yields the simple expression for the reduced Hamiltonian 
\begin{equation}\label{eq:Hamiltonian-BO}
H_{\text{red}}^{\prime\kappa_+} = \mathcal{N}(H_m-J)-\frac{\mathcal{N}}{4MV}(H_m-J)^2+\frac{\mathcal{N}}{2M}g^{ij}\tilde{\Pi}_i\tilde{\Pi}_j+\mathcal{O}\left(\frac{1}{M^2}\right) \ .
\end{equation}
This expression may be obtained directly from the transformed Hamiltonian constraint given in~(\ref{eq:hamiltonian-constraint-arbitrary-varphi}) in a self-consistent, iterative fashion. We first rewrite~(\ref{eq:hamiltonian-constraint-arbitrary-varphi}) as
\begin{equation}\label{eq:constraint-iterative}
-\tilde{\Pi}_1 = \mathcal{N}(H_m-J)+\frac{\mathcal{N}}{2M}g^{ij}\tilde{\Pi}_i\tilde{\Pi}_j-\frac{\left(\tilde{\Pi}_1\right)^2}{4\mathcal{N}(J+MV)} \ .
\end{equation}
By neglecting terms of order $\frac{1}{M}$ in the equation above, we obtain
\begin{equation}\label{eq:Pi1-lowest-order}
-\tilde{\Pi}_1 = \mathcal{N}(H_m - J) \ ,
\end{equation}
which is the zeroth-order part of~(\ref{eq:Hamiltonian-BO}). We then substitute~(\ref{eq:Pi1-lowest-order}) in the right hand side of~(\ref{eq:constraint-iterative}), with the result
\begin{align*}
-\tilde{\Pi}_1 &= \mathcal{N}(H_m - J)+\frac{\mathcal{N}}{2M}g^{ij}\tilde{\Pi}_i\tilde{\Pi}_j-\frac{\mathcal{N}}{4(J+MV)}(H_m-J)^2\\
&= \mathcal{N}(H_m - J)+\frac{\mathcal{N}}{2M}g^{ij}\tilde{\Pi}_i\tilde{\Pi}_j-\frac{\mathcal{N}}{4MV}(H_m-J)^2+\mathcal{O}\left(\frac{1}{M^2}\right) \ ,
\end{align*}
which coincides with~(\ref{eq:Hamiltonian-BO}). This iterative solution is essentially the one found in the Born-Oppenheimer approach in the \emph{quantum} theory. Indeed, the term proportional to $(H_m-J)^2$ is one of the correction terms obtained by Kiefer and Singh in~\cite{Kiefer:1991} for a vacuum background ($J = 0$). Additional correction terms involving the time-derivatives of $H_m$ and $V$ were also found by Kiefer and Singh and we will see how they arise in the quantum theory in section~\ref{sec:quantum}. The term proportional to $g^{ij}\tilde{\Pi}_i\tilde{\Pi}_j$ was neglected in~\cite{Kiefer:1991}. Here we see that it arises naturally from the expansion of the reduced Hamiltonian, even at the classical level.

The terms of~(\ref{eq:Hamiltonian-BO}) and~(\ref{eq:constraint-iterative}) which are of order $\frac{1}{M}$ (and higher) comprise the gravitational kinetic term $\frac{1}{2M}g^{ij}\tilde{\Pi}_i\tilde{\Pi}_j-\frac{\left(\tilde{\Pi}_1\right)^2}{4\mathcal{N}^2(J+MV)}$. Such terms were referred to in \cite{Briggs:2015} as ``corrections'' to Hamilton's equations for the `light' subsystem (here, the matter sector), given that its interaction with a `heavy' sector (here, the gravitational sector) provides the notion of time. We see from the above construction that such an interpretation is not entirely adequate. Hamilton's equations~(\ref{eq:eom}) are \emph{not} corrected or altered in any way for the full time-reparametrisation invariant system. They follow, as usual, from the extremisation of the action. The so-called corrections for the `light' subsystem stem from a formal perturbative treatment of the reduced Hamiltonian obtained after a particular choice of time parametrisation has been made. It is not appropriate to interpret such corrections as modifications to the dynamics of matter alone, since we see from the equations of motion~(\ref{eq:reduced-eom-expanded}) that the matter-gravity system is coupled at order $\frac{1}{M}$, i.e. the gravitational trajectory depends on the matter-sector dynamics. Indeed, for the choice $\kappa=+\sigma\mathrm{sgn}(\mathcal{N})$ in the reduced Hamiltonian, the coordinates $x^i(\tau)$ are no longer constants (`comoving') at order $\frac{1}{M}$ and their evolution follows the lowest order momenta $\tilde{\Pi}_i$, which in turn depend on the matter-sector Hamiltonian $H_m$.

\subsubsection{Hamilton-Jacobi Theory}\label{sec:iterative-HJ}
We may rewrite~(\ref{eq:constraint-iterative}) as the Hamilton-Jacobi equation\footnote{Equation~(\ref{eq:constraint-iterative-HJ}) is a Hamilton-Jacobi equation for the full time-reparametrisation invariant system and, therefore, $S$ is the Hamilton characteristic function for the full system. Equivalently, when $x^1$ is regarded as a time parameter, $\tau = x^1$, $S$ can be interpreted as the (time-dependent) Hamilton principal function for the reduced system of matter degrees of freedom $q^{\mu}(\tau)$ \emph{and} gravitational degrees of freedom $x^i(\tau)$. In general, it will not be possible to interpret $S$ as a (time-dependent) Hamilton principal function for the matter degrees of freedom alone, since the dynamics of $x^i(\tau)$ is coupled to that of matter.}
\begin{equation}\label{eq:constraint-iterative-HJ}
-\frac{\partial S}{\partial\tau} = \mathcal{N}(x)\left[H_m\left(x;\frac{\partial S}{\partial q}, q\right)-J(x)\right]+\frac{\mathcal{N}(x)}{2M}g^{ij}(x)\frac{\partial S}{\partial x^i}\frac{\partial S}{\partial x^j}-\frac{\left(\frac{\partial S}{\partial\tau}\right)^2}{4\mathcal{N}(x)(J(x)+MV(x))} \ ,
\end{equation}
and solve it iteratively, as before. To lowest order, one finds the TDHJE
\begin{equation}\label{eq:HJ-lowest-order}
-\frac{\partial S}{\partial\tau} = \mathcal{N}(x)\left[H_m\left(x;\frac{\partial S}{\partial q}, q\right)-J(x)\right] \ ,
\end{equation}
which may be interpreted as the ordinary Hamilton-Jacobi equation associated to~(\ref{eq:matter-fixed-background}) when $\kappa=+\sigma\mathrm{sgn}(\mathcal{N})$ is chosen. Thus, we may interpret~(\ref{eq:HJ-lowest-order}) as the TDHJE for matter propagating in a fixed gravitational background, since $x = (\tau, x^i(0))$ to lowest order for this choice of $\kappa$ (which corresponds to the iterative procedure). In this case, the arbitrary source term $J$ may be removed by redefining $S(\tau,x^i,q)\mapsto S(\tau,x^i,q)+\int^{\tau}\mathcal{N}(\lambda,x^i)J(\lambda,x^i)\mathrm{d}\lambda$. At the next order we obtain the \emph{corrected} Hamilton-Jacobi equation
\begin{equation}\label{eq:HJ-BO}
\begin{aligned}
-\frac{\partial S}{\partial\tau} = &\mathcal{N}(x)\left[H_m\left(x;\frac{\partial S}{\partial q}, q\right)-J(x)\right]\\
&-\frac{\mathcal{N}(x)}{4MV}\left[H_m\left(x;\frac{\partial S}{\partial q}, q\right)-J(x)\right]^2+\frac{\mathcal{N}(x)}{2M}g^{ij}(x)\frac{\partial S}{\partial x^i}\frac{\partial S}{\partial x^j}+\mathcal{O}\left(\frac{1}{M^2}\right) \ ,
\end{aligned}
\end{equation}
which corresponds to~(\ref{eq:reduced-eom-expanded}) when $\kappa=+\sigma\mathrm{sgn}(\mathcal{N})$ is chosen. At this order, the dynamics of $x^i(\tau)$ is taken into account. Therefore, $S$ is most appropriately interpreted as the ($\tau$-dependent) Hamilton principal function for the reduced system comprised of the degrees of freedom $\tilde{\Pi}_i, x^i, p_{\mu}, q^{\mu}$ and not as the ``corrected'' Hamilton principal function of a system spanned by $p_{\mu}, q^{\mu}$ alone. In this point, we differ from Briggs~\cite{Briggs:2015}, who interpreted the equation $W = S + \varphi$ as a decomposition of the total Hamilton (characteristic) function into a Hamilton (characteristic) function $\varphi$ for the `heavy' sector and a Hamilton (principal) function $S$ for the `light' degrees of freedom. We
interpret the decomposition $W = S + \varphi$ as a canonical transformation.
We will now construct the quantum theory in analogy to the Hamilton-Jacobi theory.

\section{The BO Approach as a Choice of Gauge. Quantum Theory}\label{sec:quantum}
The main challenge in quantising the constrained system associated with the action given in~(\ref{eq:action}) is to define the Hilbert space of physical states. A state is defined to be `physical' if it is annihilated by the constraint operator\footnote{We will follow the route of `Dirac Quantisation', i.e. we will promote the Hamiltonian constraint to an operator which annhiliates physical states. We will \emph{not} directly quantise the reduced Hamiltonians given in~(\ref{eq:reduced-hamiltonian-tilde}) and~(\ref{eq:reduced-hamiltonian-Pi}) for the gauge-fixed system.}. In section~\ref{sec:innerprod}, we will choose a tentative definition of the inner product that is conserved with respect to our chosen time variable.
To determine the quantum version of the constraint equation~(\ref{eq:hamiltonian-constraint}), we adopt the Laplace-Beltrami factor ordering for both the gravitational and matter-sector Hamiltonians, which yields\footnote{It is also possible to add a term proportional to the Ricci scalar of the configuration space of gravitational and matter degrees of freedom. However, this is inessential for the method we present and we omit such term from the quantum constraint equation.}
\begin{align}
\hat{H}_g\Psi &= -\frac{1}{2M\sqrt{|Gh|}}\frac{\partial}{\partial Q^a}\left(\sqrt{|Gh|}G^{ab}\frac{\partial\Psi}{\partial Q^b}\right)+MV(Q)\Psi  \ , \label{eq:LB-constraint-Hg}\\
\hat{H}_m\Psi &= -\frac{1}{2\sqrt{h}}\frac{\partial}{\partial q^{\mu}}\left(\sqrt{h}h^{\mu\nu}\frac{\partial\Psi}{\partial q^{\nu}}\right)+V_m(Q;q)\Psi  \ . \label{eq:LB-constraint-Hm}
\end{align}
For simplicity of notation, we define the \emph{gravitational-sector} Laplace-Beltrami operator to be
\begin{equation}\label{eq:def-nabla2}
\nabla^2 = \frac{1}{\sqrt{|Gh|}}\frac{\partial}{\partial Q^a}\left(\sqrt{|Gh|}G^{ab}\frac{\partial}{\partial Q^b}\right) \ .
\end{equation}
The quantum constraint equation then reads
\begin{equation}\label{eq:wdw}
\hat{H}\Psi = \left(-\frac{1}{2M}\nabla^2+MV(Q)+\hat{H}_m(Q;\hat{p}, q)\right)\Psi(Q,q) = 0 \ ,
\end{equation}
which will be referred to as the Wheeler-DeWitt (WDW) equation. The factor ordering in~(\ref{eq:LB-constraint-Hg}),~(\ref{eq:LB-constraint-Hm}) and~(\ref{eq:def-nabla2}) was chosen so as to guarantee that the WDW equation is covariant under arbitrary coordinate transformations in the configuration space of both gravitational and matter degrees of freedom~\cite{Kuchar:1991,Isham:1992}.

\subsection{Quantum Background Transformations}\label{sec:quantum-background-transf}
In the classical theory, the Hamilton characteristic function $W$ can be decomposed into $W = S + \varphi$ for a given choice of background Hamilton function $\varphi$. Analogously, given a choice of (classical) $\varphi$, we consider the following phase transformation in the quantum theory
\begin{equation}\label{eq:phase-transf-1}
\begin{aligned}
\Psi(Q,q) &= e^{\I\varphi(Q)}\Psi_{\varphi}(Q,q) \ , \\
\hat{O}(\hat{P},Q;\hat{p},q) &= e^{\I\varphi(Q)}\hat{O}_{\varphi}(\hat{P}_{\varphi},Q;\hat{p},q)e^{-\I\varphi(Q)} \ ,
\end{aligned}
\end{equation}
for any (physical) state $\Psi$ and any operator $\hat{O}$. 
If we change the background Hamilton function, $\varphi = \varphi' + \eta$, the state $\Psi_{\varphi}$ and the operator $\hat{O}_{\varphi}$ change according to the \emph{quantum} background transformations
\begin{equation}\label{eq:background-transf-psi-O}
\begin{aligned}
\Psi_{\varphi} = e^{-\I\eta}\Psi_{\varphi'} \ , \ \hat{O}_{\varphi} = e^{-\I\eta}\hat{O}_{\varphi'}e^{\I\eta} \ ,
\end{aligned}
\end{equation}
so as to keep $\Psi$ and $\hat{O}$ invariant. 
We will fix $\varphi$ by demanding that it be a solution to the (classical) equation~(\ref{eq:source}), after a choice of source $J$ is made. As before, we define the (real-valued) background momenta to be $\Phi_a = \frac{\partial\varphi}{\partial Q^a}$. We now define
\begin{align}
\mathfrak{J}(Q,q)&:= e^{-\I\varphi(Q)}\left(\frac{1}{2M}\nabla^2-MV(Q)\right)e^{\I\varphi(Q)} =J(Q)+\I K(Q,q) \ , \label{eq:complex-source}\\
K(Q,q) &:=\frac{1}{2M}\nabla^2\varphi \label{eq:K}\ .
\end{align}
We note that $\mathfrak{J}$ is not the operator $-\hat{H}_{g,\varphi}$, but rather the complex function obtained by acting with $-\hat{H}_g$ on the complex exponential $e^{\I\varphi}$ and subsequently multiplying by the conjugate exponential. The dependence of $K$ on the $q$ coordinates comes from the Laplace-Beltrami factor ordering in~(\ref{eq:def-nabla2}). Under a quantum background transformation, we obtain
\begin{equation}\label{eq:background-gauge-transf-JK}
K(Q,q) = K'(Q,q)+\frac{1}{2M}\nabla^2\eta \ ,
\end{equation}
in addition to the transformation laws given in~(\ref{eq:Phi-source-transf}). From the full WDW equation~(\ref{eq:wdw}) and from~(\ref{eq:complex-source}) we obtain an equation for $\Psi_{\varphi}$ (cf.~(\ref{eq:nonrel-pre-schrodinger})),
\begin{equation}\label{eq:pre-schrodinger}
\I G^{ab}(Q)\frac{\Phi_a}{M}\frac{\partial\Psi_{\varphi}}{\partial Q^b} = \left(\hat{H}_m(Q; \hat{p}, q)-\mathfrak{J}(Q,q)\right)\Psi_{\varphi}-\frac{1}{2M}\nabla^2\Psi_{\varphi} \ .
\end{equation}
One may verify that this equation is invariant under quantum background transformations by using~(\ref{eq:Phi-source-transf}),~(\ref{eq:background-transf-psi-O}) and~(\ref{eq:background-gauge-transf-JK}). This is also understood from the fact that~(\ref{eq:pre-schrodinger}) is simply $\hat{H}_{\varphi}\Psi_{\varphi} = 0$. In~(\ref{eq:K-app0}) and~(\ref{eq:K-app}) of appendix~\ref{app:formulas-Qx-transf}, we find that the function $K$ may be rewritten in the $x$ coordinates as
\begin{equation}\label{eq:K-Laplace-Beltrami}
K(Q,q) = \frac{1}{2\sqrt{|\tilde{G}h|}}\frac{\partial}{\partial x^1}\left(\frac{\sqrt{|\tilde{G}h|}}{\mathcal{N}}\right)\ ,
\end{equation}
for the Laplace-Beltrami factor ordering. By performing the coordinate transformation given in~(\ref{eq:coord-transf-Qx}) (see also appendix~\ref{app:formulas-Qx-transf}), we can rewrite~(\ref{eq:pre-schrodinger}) as
\begin{equation}\label{eq:pre-schrodinger-2}
\begin{aligned}
\I\frac{\partial\Psi_{\varphi}}{\partial x^1} &= \mathcal{N}(x)\left(\hat{H}_m(x; \hat{p}, q)-\mathfrak{J}(x,q)\right)\Psi_{\varphi}\\
&-\frac{\mathcal{N}(x)}{2M\sqrt{h|\tilde{G}|}}\frac{\partial}{\partial x^i}\left(\sqrt{h|\tilde{G}|}g^{ij} \frac{\partial}{\partial x^j}\Psi_{\varphi}\right)-\frac{\mathcal{N}(x)}{2M\sqrt{h|\tilde{G}|}}\frac{\partial}{\partial x^1}\left(\sqrt{h|\tilde{G}|}\tilde{G}^{11}\frac{\partial}{\partial x^1}\Psi_{\varphi}\right) \ ,
\end{aligned}
\end{equation}
which is a quantisation of the corresponding classical equations
~(\ref{eq:constraint-iterative}) and~(\ref{eq:constraint-iterative-HJ}). The solution to~(\ref{eq:pre-schrodinger-2}) is the wave function of the gauge-fixed system comprised of the degrees of freedom $\hat{p}^{\mu}, \hat{q}^{\mu}, \hat{\Pi}_i, \hat{x}^i$, while $x^1$ plays the role of the time parameter. Therefore, the evolution of $\Psi_{\varphi}$ accounts for the (coupled) dynamics of \emph{both} matter and gravitational degrees of freedom, contrary to what is usually assumed in a Born-Oppenheimer context, which is that $\Psi_{\varphi}$ is only the matter-sector wave function. The interpretation of $\Psi_{\varphi}$ as the wave function for both gravitational and matter degrees of freedom is analogous to the interpretation of the Hamilton (characteristic) function $S = W-\varphi$ as describing the dynamics of the composite system in the classical theory. We will see in section~(\ref{sec:quantum-matter-classical-gravity}) how to recover the description of the dynamics of quantum matter in a fixed gravitational background.

It is worthwhile to mention the `complex structure problem'~\cite{Kuchar:1991,Isham:1992}, which in the formalism presented here can be understood as follows. The factor of $\I$ in~(\ref{eq:pre-schrodinger-2}) leads to the coupling of the real and imaginary parts of $\Psi_{\varphi}$. On the other hand, the WDW equation~(\ref{eq:wdw}) is real and no such coupling occurs for the real and imaginary parts of $\Psi$. In fact, we can take $\Psi$ to be real. The complex structure of~(\ref{eq:pre-schrodinger-2}) originates solely from the phase pre-factor in $\Psi = e^{\I\varphi}\Psi_{\varphi}$, which may seem \emph{ad hoc}. This has been criticised by Barbour~\cite{Barbour:1993} and Kucha\v{r}~\cite{Kuchar:1991} in the context of the WKB approximation. 
However, one may invoke decoherence~\cite{Kiefer:1993-0,Kiefer:1993-1,Kiefer:1993-2} to justify the appearance of the phase pre-factor. In the present formalism, we assume from the start that the Hilbert space is complex and, in this context, the phase transformation given in~(\ref{eq:phase-transf-1}) is the quantum analogue of the canonical transformation given in~(\ref{eq:background-gauge}) in the classical theory. Thus, the phase transformation $\Psi_{\varphi} = \Psi e^{-\I\varphi}$ can be employed without loss of generality, even if $\Psi$ is real.  

\subsection{Inner Product and Unitarity}\label{sec:innerprod}
Given a solution $\varphi$ to the (classical) equation~(\ref{eq:source}), we may define the coordinates $x$ adapted to the background Hamilton function $\varphi$ as in~(\ref{eq:coord-transf-Qx}) (and in appendix~\ref{app:formulas-Qx-transf}). Due to the Laplace-Beltrami factor ordering, we may change coordinates $Q\mapsto x$ in the WDW equation~(\ref{eq:wdw}) to obtain
\begin{equation}\label{eq:wdw-x}
\begin{aligned}
0 &= -\frac{1}{2M\sqrt{|\tilde{G}h|}}\frac{\partial}{\partial x^A}\left(\sqrt{|\tilde{G}h|}\tilde{G}^{AB}\frac{\partial\Psi}{\partial x^B}\right)+MV(x)\Psi\\
&-\frac{1}{2\sqrt{h}}\frac{\partial}{\partial q^{\mu}}\left(\sqrt{h}h^{\mu\nu}\frac{\partial\Psi}{\partial q^{\nu}}\right)+V_m(x;q)\Psi \ .
\end{aligned}
\end{equation}
Equation~(\ref{eq:wdw-x}) leads to the continuity equation
\begin{equation}\label{eq:KGcontinuity}
\frac{1}{\sqrt{|\tilde{G}h|}}\frac{\partial}{\partial x^A}\left(\sqrt{|\tilde{G}h|} j^A\right)+\frac{1}{\sqrt{h}}\frac{\partial}{\partial q^{\mu}}\left(\sqrt{h}j^{\mu}\right) =0  \ ,
\end{equation}
where the Klein-Gordon current is defined as~\cite{Vilenkin:1988,Kuchar:1991,Isham:1992}
\begin{equation}\label{eq:KG-current}
\begin{aligned}
j^A(x,q) &= \frac{\I f\tilde{G}^{AB}}{2M}\left(\bar{\Psi}_1\frac{\partial\Psi_2}{\partial x^B}-\Psi_2\frac{\partial\bar{\Psi}_1}{\partial x^B}\right)\ , \\
j^{\mu}(x,q) &= \frac{\I f h^{\mu\nu}}{2}\left(\bar{\Psi}_1\frac{\partial\Psi_2}{\partial q^{\nu}}-\Psi_2\frac{\partial\bar{\Psi}_1}{\partial q^{\nu}}\right)\ ,
\end{aligned}
\end{equation}
for any two solutions $\Psi_1, \Psi_2$ of the WDW equation~(\ref{eq:wdw-x}). The parameter $f$ is a real constant that will be fixed in what follows. Given $\varphi$, the continuity equation~(\ref{eq:KGcontinuity}) implies that the quantity\footnote{We note that we have assumed that $\tilde{G}^{1i} = 0$. In the most general case, such terms are non-vanishing and lead to the presence of derivatives with respect to $x^i$ in~(\ref{eq:KGprod}).}
\begin{equation}\label{eq:KGprod}
\begin{aligned}
(\Psi_1,\Psi_2)_{\text{KG}} &:= \int\prod_{i=2}^n\mathrm{d}x^i\sqrt{|\tilde{G}|}\prod_{\mu=1}^d\mathrm{d}q^{\mu}\sqrt{h}\ j^1(x,q)\\
&=\int\prod_i\mathrm{d}x^i\sqrt{|\tilde{G}|}\prod_{\mu}\mathrm{d}q^{\mu}\sqrt{h}\ \frac{\I f\tilde{G}^{11}}{2M}\left(\bar{\Psi}_1\frac{\partial\Psi_2}{\partial x^1}-\Psi_2\frac{\partial\bar{\Psi}_1}{\partial x^1}\right) 
\end{aligned}
\end{equation}
is conserved with respect to the $x^1$-coordinate,
\begin{equation}\label{eq:KG-conserved}
\frac{\partial}{\partial x^1}(\Psi_1,\Psi_2)_{\text{KG}} = 0 \ .
\end{equation}
The conserved charge given in~(\ref{eq:KGprod}) is the Klein-Gordon inner product. If $x^1$ is considered to be the time parameter, then~(\ref{eq:KG-conserved}) implies that the dynamics based on the Klein-Gordon inner product is unitary with respect to $x^1$-evolution. As is well-known, the Klein-Gordon inner product is indefinite. Nevertheless, we will see in sections~\ref{sec:pt1} and~\ref{sec:pt2} that this inner product is of a definite sign in the perturbative regime, i.e. for solutions of the WDW equation found via the iterative procedure. Thus, a probability interpretation is possible in perturbation theory. 

If we perform a quantum background transformation, $\Psi_1 = e^{\I\varphi}\Psi_{1,\varphi}, \Psi_2 = e^{\I\varphi}\Psi_{2,\varphi}$, the inner product in~(\ref{eq:KGprod}) can be rewritten as
\begin{equation}
\begin{aligned}
&(\Psi_1,\Psi_2)_{\text{KG}}\\
&=\int\prod_i\mathrm{d}x^i\sqrt{|\tilde{G}|}\prod_{\mu}\mathrm{d}q^{\mu}\sqrt{h}\ \frac{(-f)\tilde{G}^{11}}{M}\frac{\partial\varphi}{\partial x^1}\bar{\Psi}_{1,\varphi}\Psi_{2,\varphi} + (\Psi_{1,\varphi},\Psi_{2,\varphi})_{\text{KG}}
\end{aligned}
\end{equation}
Using the first of equations~(\ref{eq:x-gradient-varphi}), which implies $\frac{\partial\varphi}{\partial x^1} = \frac{M}{\mathcal{N}\tilde{G}^{11}}$, we obtain
\begin{equation}
\begin{aligned}
&(\Psi_1,\Psi_2)_{\text{KG}}=\int\prod_i\mathrm{d}x^i\sqrt{|\tilde{G}|}\prod_{\mu}\mathrm{d}q^{\mu}\sqrt{h}\ \frac{(-f)}{\mathcal{N}}\bar{\Psi}_{1,\varphi}\Psi_{2,\varphi} + (\Psi_{1,\varphi},\Psi_{2,\varphi})_{\text{KG}} \ ,
\end{aligned}
\end{equation}
which we can rewritte as
\begin{equation}\label{eq:BOprod}
\begin{aligned}
&(\Psi_1,\Psi_2)_{\text{KG}} = \int\prod_i\mathrm{d}x^i\sqrt{|\tilde{G}|}\prod_{\mu}\mathrm{d}q^{\mu}\sqrt{h}\\
&\times\bar{\Psi}_{1,\varphi}\left[\frac{(-f)}{\mathcal{N}}+\frac{\tilde{G}^{11}}{2M}\left(\I f\overset{\rightarrow}{\frac{\partial}{\partial x^1}}\right)-\left(\I f\overset{\leftarrow}{\frac{\partial}{\partial x^1}}\right)\frac{\tilde{G}^{11}}{2M}\right]\Psi_{2,\varphi} \ .
\end{aligned}
\end{equation}
This form of the Klein-Gordon inner product will be useful in perturbation theory.

\subsection{Perturbation Theory I}\label{sec:pt1}
As in the classical theory, if we restrict ourselves to a regime in which all energy scales are much smaller than the heavy scale $\sqrt{M}$, we can develop a formal perturbative expansion in powers of $\frac{1}{M}$. We assume that the states $\Psi_{\varphi}$ are analytic functions of $\frac{1}{M}$ and admit the formal expansion
\begin{equation}\label{eq:expansion-Psi-varphi}
\Psi_{\varphi} = \sum_{k = 0}^{\infty}\frac{1}{M^k}\Psi_{\varphi}^{(k)} \ .
\end{equation}
Together with~(\ref{eq:phi-expansion}), equation~(\ref{eq:expansion-Psi-varphi}) implies that the states $\Psi = e^{\I\varphi}\Psi_{\varphi}$ can be expanded as
\begin{equation}\label{eq:expansion-BO}
\Psi = e^{\I M \varphi^{(-1)}}\sum_{k = 0}^{\infty}\frac{1}{M^k}\xi^{(k)} \ ,
\end{equation}
where $\xi^{(k)}$ are coefficients which can be computed from the expansions in~(\ref{eq:phi-expansion}) and~(\ref{eq:expansion-Psi-varphi}). The expansion of~(\ref{eq:expansion-BO}) is the one usually performed in the semiclassical approach~\cite{Banks:1984,BFS:1984,Singh:1990,Kiefer:1991}.

We now set out to solve the constraint equation~(\ref{eq:pre-schrodinger-2}) in a self-consistent, iterative fashion in analogy to what was done in the Hamilton-Jacobi theory in section~\ref{sec:iterative-HJ}. Let us at first keep only terms to lowest order in $\frac{1}{M}$. The inner product in~(\ref{eq:BOprod}) becomes
\begin{equation}\label{eq:innerprod-order-0}
\begin{aligned}
&\left(\Psi_1, \Psi_2\right)_{\text{KG}}\\
& = (-f)\int\prod_i\mathrm{d}x^i\prod_{\mu}\mathrm{d}q^{\mu}\ \mathrm{sgn}(\mathcal{N})\sqrt{2|V(x)g(x)|h(x;q)}\ \bar{\Psi}_{1,\varphi}(x,q)\Psi_{2,\varphi}(x,q)+\mathcal{O}\left(\frac{1}{M}\right) \ .
\end{aligned}
\end{equation}
Let us restrict ourselves to a region of configuration space where $\mathrm{sgn}(\mathcal{N})$ is constant. Then, we see from~(\ref{eq:innerprod-order-0}) that the lowest-order approximation to the Klein-Gordon inner product is positive-definitive if we set $f = -\mathrm{sgn}(\mathcal{N})$. To lowest order, equation~(\ref{eq:pre-schrodinger-2}) reads
\begin{equation}\notag
\I\frac{\partial\Psi_{\varphi}}{\partial x^1} = \mathcal{N}(x)\left(\hat{H}_m(x; \hat{p}, q)-\mathfrak{J}(x,q)\right)\Psi_{\varphi}(x,q)+\mathcal{O}\left(\frac{1}{M}\right) \ ,
\end{equation}
which can be rewritten as
\begin{equation}\label{eq:schrodinger-order-0}
\begin{aligned}
\I\frac{\partial\Psi_{\varphi}}{\partial x^1} &= \mathcal{N}(x)\left(\hat{H}_m(x; \hat{p}, q)-J(x)\right)\Psi_{\varphi}-\I\left[\frac{\partial}{\partial x^1}\log\left(\left|2Vgh\right|^{\frac{1}{4}}\right)\right]\Psi_{\varphi}+\mathcal{O}\left(\frac{1}{M}\right) \ ,
\end{aligned}
\end{equation}
for constant $\mathrm{sgn}(\mathcal{N})$. Equation~(\ref{eq:schrodinger-order-0}) is a quantum version of the corresponding classical equations~(\ref{eq:Pi1-lowest-order}) and~(\ref{eq:HJ-lowest-order}). Evidently, the solutions of~(\ref{eq:schrodinger-order-0}) are only valid up to order $M^0$. The imaginary term on the right hand side corresponds to the truncation of $\I\mathcal{N}K$ to lowest order in $\frac{1}{M}$.
Such a term is present due to the dependence of measure given in~(\ref{eq:innerprod-order-0}) on the coordinate $x^1$, which plays the role of a time parameter\footnote{The object $\frac{\partial}{\partial x^1}+\frac{\partial}{\partial x^1}\log\left|2Vgh\right|^{\frac{1}{4}}$ is what DeWitt referred to as a `conservative time derivative'~\cite{DeWitt:1957}.} in the gauge-fixed theory. 
Indeed, if $\Psi_{1,\varphi}$ and $\Psi_{2,\varphi}$ are solutions of~(\ref{eq:schrodinger-order-0}), then the inner product given in~(\ref{eq:innerprod-order-0}) is conserved with respect to $x^1$ up to order $M^0$. Although this approximate conservation is a consequence of the exact equation~(\ref{eq:KG-conserved}), we find it instructive to explicitly verify that~(\ref{eq:innerprod-order-0}) is conserved using the solutions to~(\ref{eq:schrodinger-order-0}) and we have registered this computation in appendix~\ref{app:conservation-innerprod}, where we keep terms up to order $\frac{1}{M}$~(cf.~section~(\ref{sec:pt2})).

\subsection{Propagation of Quantum Matter in a Fixed Gravitational Background}\label{sec:quantum-matter-classical-gravity}

\subsubsection{Partial Ehrenfest Equations}
As before, we define the matter-sector partial averages of an operator $\hat{O}$ as
\begin{equation}\label{eq:matter-sector-partial-average}
\braket{\hat{O}}_m(x) := \frac{\int\prod_{\mu}\mathrm{d}q^{\mu}\sqrt{h}\ \bar{\Psi}_{\varphi}(x,q)\hat{O}\Psi_{\varphi}(x,q)}{\int\prod_{\mu}\mathrm{d}q^{\mu}\sqrt{h}\ \bar{\Psi}_{\varphi}(x,q)\Psi_{\varphi}(x,q)} \ ,
\end{equation}
provided the integrals converge. We note that $\braket{\hat{O}}_m(x)$ is a function of the remaining gravitational variables $x^1, x^i$ or $Q^a$. We find that self-adjoint matter-sector operators, $\hat{O}_m \equiv \hat{O}_m(x;\hat{p},\hat{q})$ obey the partial Ehrenfest equation
\begin{equation}\label{eq:partial-ehrenfest}
\begin{aligned}
&\frac{\partial}{\partial x^1}\braket{\hat{O}_m}_m(x)\\
&= \left<\frac{\partial\hat{O}_m}{\partial x^1}\right>_m+\I\mathcal{N}(x)\left<\left[\hat{H}_m,\hat{O}_m\right]\right>_m+\left<\left[\frac{\partial}{\partial x^1}\log\left|h\right|^{\frac{1}{4}},\hat{O}_m\right]\right>_m+\mathcal{O}\left(\frac{1}{M}\right) \ .
\end{aligned}
\end{equation}
We note that~(\ref{eq:partial-ehrenfest}) holds despite the fact that the dynamics of $\Psi_{\varphi}$ is not unitary in the matter-sector. Equation~(\ref{eq:partial-ehrenfest}) is the Ehrefent equation for a self-adjoint matter-sector operator defined in the gravitational background corresponding to the time parameter $x^1$, its associated lapse function $\mathcal{N}(x)$ and the `comoving' coordinates $x^i$.

\subsubsection{Matter-Sector Unitarity}
We can impose unitarity in the matter sector by considering the factorisation $\Psi_{\varphi}(x,q) = \chi(x)\psi(x,q)$, where we demand that the factor $\chi$ obeys the equation
\begin{equation}\label{eq:schrodinger-chi}
\I\frac{\partial\chi}{\partial x^1} = -\mathcal{N}(x)J(x)\chi(x)-\I\left[\frac{\partial}{\partial x^1}\log\left|2V(x)g(x)\right|^{\frac{1}{4}}\right]\chi(x)+\mathcal{O}\left(\frac{1}{M}\right) \ .
\end{equation}
By inserting $\Psi_{\varphi} = \chi\psi$ into~(\ref{eq:schrodinger-order-0}) and using~(\ref{eq:schrodinger-chi}), we obtain
\begin{equation}\label{eq:schrodinger-psi}
\I\frac{\partial\psi}{\partial x^1} = \mathcal{N}(x)\hat{H}_m(x; \hat{p}, q)\psi(x,q)-\I\left[\frac{\partial}{\partial x^1}\log\left|h\right|^{\frac{1}{4}}\right]\psi(x,q)+\mathcal{O}\left(\frac{1}{M}\right) \ .
\end{equation}
Using~(\ref{eq:schrodinger-chi}) and~(\ref{eq:schrodinger-psi}), one may explicitly verify that
\begin{equation}
\begin{aligned}
&\frac{\partial}{\partial x^1}\int\prod_{i}\mathrm{d}x^{i}\sqrt{2|Vg|}\ \bar{\chi}(x)\chi(x) = 0+\mathcal{O}\left(\frac{1}{M}\right) \ ,\\
&\frac{\partial}{\partial x^1}\int\prod_{\mu}\mathrm{d}q^{\mu}\sqrt{h}\ \bar{\psi}(x,q)\psi(x,q) = 0+\mathcal{O}\left(\frac{1}{M}\right) \ .
\end{aligned}
\end{equation}
Thus, unitarity is enforced separately in each sector. The solution to~(\ref{eq:schrodinger-chi}) is
\begin{equation}\label{eq:chi-solution}
\chi(x^1,x^i) = \left|2V(x^1,x^i)g(x^1,x^i)\right|^{-\frac{1}{4}}\gamma(x^i)\exp\left(\I\int^{x^1}\mathrm{d}\lambda\ \mathcal{N}(\lambda,x^i)J(\lambda,x^i)\right)+\mathcal{O}\left(\frac{1}{M}\right) \ ,
\end{equation}
where $J$ is understood as its lowest order approximation and $\gamma(x^i)$ is an arbitrary factor which satisfies the normalisation condition in the `flat' inner product $\int\prod_{i}\mathrm{d}x^i\ \bar{\gamma}(x^i)\gamma(x^i) = 1$. Similarly, we can normalise $\psi(x,q)$ in the matter-sector inner product. The dynamics of $\psi(x,q)$ is identical to the one usually studied in quantum theory in a fixed gravitational background.

The original total state $\Psi$, which is a solution to the WDW equation~(\ref{eq:wdw}), can thus be written as
\begin{equation}\label{eq:BO-fact-Cederbaum}
\Psi(x,q) = e^{\I\varphi(x)}\Psi_{\varphi}(x,q) = e^{\I\varphi(x)}\chi(x)\psi(x,q)+\mathcal{O}\left(\frac{1}{M}\right) \ ,
\end{equation}
which is just the BO exact factorisation for the total state $\Psi$~(cf.~(\ref{eq:exact-factorisation})). We note that we have defined the time variable from $\varphi$, which is only part of the phase of the `gravitational factor' $e^{\I\varphi(x)}\chi(x)$. The form of~(\ref{eq:BO-fact-Cederbaum}) was used in~\cite{Cederbaum:2008} as the BO ansatz for the total state of nuclei and electrons in the context of molecular physics.

\subsection{Perturbation Theory II}\label{sec:pt2}
Let us now continue with the iterative procedure and keep terms only up to order~$\frac{1}{M}$. Using the lowest-order equation~(\ref{eq:schrodinger-order-0}), we can rewrite the the inner product given in~(\ref{eq:BOprod}) as\footnote{The inner product given in~(\ref{eq:innerprod-order-m1}) is analogous to the one considered in~\cite{Lammerzahl:1995} for quantum optics in gravitational fields.}
\begin{align}\label{eq:innerprod-order-m1}
\left(\Psi_1, \Psi_2\right)_{\text{KG}} &= \int\prod_i\mathrm{d}x^i\prod_{\mu}\mathrm{d}q^{\mu}\ \bar{\Psi}_{1,\varphi}(x,q)\hat{\mathscr{M}}(x;\hat{p},q)\Psi_{2,\varphi}(x,q)+\mathcal{O}\left(\frac{1}{M^2}\right) \ , \\
\hat{\mathscr{M}}(x;\hat{p},q) &:= -f\mathrm{sgn}(\mathcal{N}(x))\sqrt{2|V(x)g(x)|h(x;q)}\left[1+\frac{1}{2MV(x)}\hat{H}_m(x;\hat{p},q)\right]\ , \label{eq:measure-M}
\end{align}
where we used the fact that $\hat{H}_m(x;\hat{p},q)$ is symmetric with respect to the matter-sector inner product. We now restrict ourselves to a region of configuration space where $\mathrm{sgn}(\mathcal{N})$ is constant and set $f = -\mathrm{sgn}(\mathcal{N})$. The potential $V(x)$ can have a positive or negative sign. Thus, the inner product given in~(\ref{eq:innerprod-order-m1}) is positive-definite if the following condition holds\footnote{We assume that the matter-sector Hamiltonian $\hat{H}_m(x;\hat{p},q)$ has non-negative eigenvalues.}
\begin{equation}\label{eq:condition-positive-definiteness}
\int\prod_i\mathrm{d}x^i\prod_{\mu}\mathrm{d}q^{\mu}\sqrt{2|Vg|h}\ \bar{\Psi}_{\varphi}\Psi_{\varphi}\gg\frac{1}{M}\int\prod_i\mathrm{d}x^i\prod_{\mu}\mathrm{d}q^{\mu}\sqrt{\frac{1}{2}\left|\frac{g}{V}\right|h}\ \bar{\Psi}_{\varphi}\hat{H}_m\Psi_{\varphi} \ .
\end{equation}
The inequality~(\ref{eq:condition-positive-definiteness}) should be satisfied in the regime of validity of perturbation theory.
To continue the iterative procedure, we use~(\ref{eq:schrodinger-order-0}) to eliminate the $x^1$-derivatives in the right-hand side of~(\ref{eq:pre-schrodinger-2}). After some algebra, we obtain
\begin{equation}\label{eq:schrodinger-order-1-V}
\begin{aligned}
&\I\frac{\partial\Psi_{\varphi}}{\partial x^1}+\I\hat{\Gamma}\Psi_{\varphi} = \mathcal{N}\left(\hat{H}_m-J\right)\Psi_{\varphi}-\frac{\mathcal{N}}{4MV}\left(\hat{H}_m-J\right)^2\Psi_{\varphi}\\
&-\frac{1}{2M\sqrt{2|Vgh|}}\frac{\partial}{\partial x^i}\left(\sqrt{2|Vgh|}\mathcal{N}g^{ij} \frac{\partial\Psi_{\varphi}}{\partial x^j}\right)+\frac{1}{M}\mathcal{V}\Psi_{\varphi}+\mathcal{O}\left(\frac{1}{M^2}\right) \ ,
\end{aligned}
\end{equation}
where we defined\footnote{We define the (explicit) time-derivative of an operator as $\left(\frac{\partial}{\partial x^1}\hat{O}\right)\psi := \frac{\partial}{\partial x^1}\left(\hat{O}\psi\right)-\hat{O}\frac{\partial\psi}{\partial x^1} = \left[\frac{\partial}{\partial x^1},\hat{O}\right]\psi$.}
\begin{align}
\hat{\Gamma}&:=\frac{\partial}{\partial x^1}\log\left|2Vgh\right|^{\frac{1}{4}}+\frac{1}{2M\sqrt{2|Vgh|}}\frac{\partial}{\partial x^1}\left(\sigma\sqrt{\frac{h}{2}\left|\frac{g}{V}\right|}\hat{H}_m\right)\notag\\
&-\frac{1}{4MV}\hat{H}_m\frac{\partial}{\partial x^1}\log\left|2Vgh\right|^{\frac{1}{4}}-\frac{1}{4MV}\left(\frac{\partial}{\partial x^1}\log\left|2Vgh\right|^{\frac{1}{4}}\right)\hat{H}_m \ ,\label{eq:Gamma-Unitarity}\\
\mathcal{V} &:= \frac{1}{32\mathcal{N}V|Vgh|}\left(\frac{\partial}{\partial x^1}\sqrt{2|Vgh|}\right)^2-\frac{1}{2\sqrt{2|Vgh|}}\frac{\partial}{\partial x^1}\left(\frac{1}{4\mathcal{N}V}\frac{\partial}{\partial x^1}\sqrt{2|Vgh|}\right) \ , 
\end{align}
and $\sigma = \mathrm{sgn}(V)$.
Equation~(\ref{eq:schrodinger-order-1-V}) is a quantum version of the corresponding classical equations~(\ref{eq:Hamiltonian-BO}) and~(\ref{eq:HJ-BO}). The term $\frac{1}{M}\mathcal{V}$ can be interpreted as a ``quantum correction'', which is present due to the fact that we quantised the constraint equation~(\ref{eq:hamiltonian-constraint-arbitrary-varphi}) rather than the reduced Hamiltonian given in~(\ref{eq:reduced-hamiltonian-Pi}) and subsequently adopted an iterative procedure to find the solution to the quantum constraint equation~(\ref{eq:pre-schrodinger-2}). We refrain from factorising $\Psi_{\varphi} = \chi\psi$ to enforce unitarity in the matter-sector by a suitable choice of $\chi$, since it is sufficient to interpret $\Psi_{\varphi}$ as the wave function for \emph{both} gravitational and matter degrees of freedom, which is a solution to~(\ref{eq:schrodinger-order-1-V}).

Equation~(\ref{eq:schrodinger-order-1-V}) was computed in~\cite{Kiefer:1991} for a vacuum background ($J = 0$) and the terms involving the derivatives with respect to the $x^i$ variables as well as the term proportional to $\mathcal{V}$ were absent\footnote{If $\tilde{G}^{1i} \neq 0$, there will also be additional terms in~(\ref{eq:Gamma-Unitarity}) involving derivatives with respect to the $x^i$ variables.}. Here, such terms arise from the iterative solution of the constraint equation~(\ref{eq:pre-schrodinger-2}). Moreover, the $x^1$-derivatives of the matter Hamiltonian $\hat{H}_m$ and of the potential $V$ present in the term $\I\hat{\Gamma}$ were interpreted in~\cite{Kiefer:1991} as unitarity-violating terms induced by gravity. One is led to this interpretation if one regards $\Psi_{\varphi}$ as the matter-sector wave function and~(\ref{eq:schrodinger-order-1-V}) as a ``corrected'' Schr\"{o}dinger equation for the matter-sector. However, as we have argued above, the state $\Psi_{\varphi}$ is most appropriately interpreted as the wave-function for the coupled system of \emph{both} gravity and matter comprised of the configurational degrees of freedom $x^i$ \emph{and} $q^{\mu}$ when $x^1$ is regarded as the time parameter. In this way, the inner product involves not only an integration over matter variables, but also over the $x^i$ variables (cf.~(\ref{eq:innerprod-order-m1})). Thus, rather than introducing a violation of unitarity, the term $\I\hat{\Gamma}$ is the factor which guarantees unitarity with respect to the inner product given in~(\ref{eq:innerprod-order-m1}). Indeed, we explicitly verify that unitarity holds due to this term in appendix~\ref{app:conservation-innerprod}.

\subsection{Backreaction}\label{sec:rel-backreaction}
We now show how the formalism of~\cite{Brout:1987-1,Brout:1987-2,Brout:1987-3,Brout:1987-4,Venturi:1990,Bertoni:1996,Kamenshchik:2017} can be recovered from the above construction. Following what was done in section~\ref{sec:assess}, we compute the equations with `backreaction' terms from~(\ref{eq:complex-source}) and~(\ref{eq:pre-schrodinger}). Upon taking the matter-sector partial average\footnote{The definition of the partial averages will be discussed in appendix~\ref{app:conservation-innerprod}.} of~(\ref{eq:pre-schrodinger}), we find
\begin{equation}\label{eq:pre-schrodinger-average}
-J(Q) = \I G^{ab}(Q)\frac{\Phi_a}{M}\left<\frac{\partial}{\partial Q^b}\right>_m-\braket{\hat{H}_m(Q; \hat{p}, q)}_m+\I\braket{K(Q,q)}_m+\frac{1}{2M}\braket{\nabla^2}_m \ .
\end{equation}
Using~(\ref{eq:coord-transf-Qx}) and~(\ref{eq:K-Laplace-Beltrami}), we can write
\begin{align*}
&\I G^{ab}(Q)\frac{\Phi_a}{M}\left<\frac{\partial}{\partial Q^b}\right>_m+\I\braket{K(Q,q)}_m = \I G^{ab}(Q)\frac{\Phi_a}{M}\left<\frac{\partial}{\partial Q^b}+\frac{\mathcal{N}}{2\sqrt{|\tilde{G}h|}}\frac{\partial}{\partial Q^b}\left(\frac{\sqrt{|\tilde{G}h|}}{\mathcal{N}}\right)\right>_m \ .
\end{align*}
If we now insert the real part of~(\ref{eq:pre-schrodinger-average}) into~(\ref{eq:source}), we obtain
\begin{equation}\label{eq:source-backreaction}
\begin{aligned}
&\frac{1}{2M}G^{ab}(Q)\left(\frac{\partial \varphi}{\partial Q^a}+A_a\right)\left(\frac{\partial \varphi}{\partial Q^b}+A_b\right)+MV(Q)\\
& = -\braket{\hat{H}_m(Q; \hat{p}, q)}_m+\frac{1}{2M}\left(\mathfrak{Re}\braket{\nabla^2}_m+G^{ab}A_aA_b\right) \ ,
\end{aligned}
\end{equation}
where we defined the `Berry connection' as $A_a = \mathfrak{Im}\left<\frac{\partial}{\partial Q^a}+\frac{\mathcal{N}}{2\sqrt{|\tilde{G}h|}}\frac{\partial}{\partial Q^a}\left(\frac{\sqrt{|\tilde{G}h|}}{\mathcal{N}}\right)\right>_m$~(cf.~(\ref{eq:nonrel-partial-average-derivatives})). Equation~(\ref{eq:source-backreaction}) can be seen as a Hamilton-Jacobi equation with quantum backreaction terms. Using the expansion of $\varphi$ given in~(\ref{eq:phi-expansion}), we may solve~(\ref{eq:source-backreaction}) at each order of $\frac{1}{M}$. The lowest orders read
\begin{align}
&\mathcal{O}\left(M^1\right) \ : \ \frac{1}{2}G^{ab}(Q)\frac{\partial \varphi^{(-1)}}{\partial Q^a}\frac{\partial \varphi^{(-1)}}{\partial Q^b}+V(Q) = 0 \ , \label{eq:source-backreaction-order-1}\\
&\mathcal{O}\left(M^0\right) \ : \ G^{ab}\frac{\partial\varphi^{(-1)}}{\partial Q^a}\frac{\partial\varphi^{(0)}}{\partial Q^b} = -G^{ab}\frac{\partial\varphi^{(-1)}}{\partial Q^a}A_b^{(0)} -\braket{\hat{H}_m(Q; \hat{p}, q)}_m^{(0)} \ ,\label{eq:source-backreaction-order-0}
\end{align}
where $\braket{\hat{O}}_m^{(0)}$ denotes the lowest order approximation to the partial average $\braket{\hat{O}}_m$, which is the matter-sector expectation value of the operator $\hat{O}$ taken with respect to a solution of~(\ref{eq:schrodinger-order-0}). Equation~(\ref{eq:source-backreaction-order-1}) is the vacuum Hamilton-Jacobi equation for the gravitational sector. Thus, we see that the effects of backreaction terms enter only in~(\ref{eq:source-backreaction-order-0}), at order $M^0$. This is consistent with the expansion of $J$ in~(\ref{eq:J-expansion}). The conclusion that backreaction effects are not found to lowest order was also reached in~\cite{Kiefer:1991,Kiefer:1993-1,Kiefer:1993-2,Kamenshchik:2017} in the context of the usual semiclassical (BO) interpretation of quantum gravity.

The background Hamilton function $\varphi$ in~(\ref{eq:source-backreaction}) is sourced by $\braket{\hat{H}_m(Q; \hat{p}, q)}_m$ \emph{and} the Berry connection terms. Since~(\ref{eq:source-backreaction}) is by construction equivalent to~(\ref{eq:source}), we see that the arbitrariness of the Berry connection terms corresponds to the freedom in choosing $J$. A given \emph{choice} of $J$ defines a background Hamilton function $\varphi$ and, thus, a background gravitational trajectory with respect to which the weak coupling expansion is to be performed.
This was also argued in a different way by Parentani in~\cite{Parentani:1998}, where he emphasised that this procedure corresponds to a \emph{background field approximation}. The classical (quantum) dynamics of the composite system of gravitational and matter degrees of freedom is encoded in the Hamilton (wave) function $W(Q,P'; q,p')$ ($\Psi(Q,q)$) and $W$ (the phase of $\Psi$) coincides with $\varphi$ only up to order $M^1$ in the weak coupling expansion~(cf.~(\ref{eq:source-backreaction-order-1})).

If we insert~(\ref{eq:pre-schrodinger-average}) back into~(\ref{eq:complex-source}) and~(\ref{eq:pre-schrodinger}), we obtain
\begin{align}
&-\frac{1}{2M}\left(\nabla^2+\braket{\nabla^2}_m\right)\chi_{\varphi}-\frac{G^{ab}}{M}\frac{\partial\chi_{\varphi}}{\partial Q^a}\left<\frac{\partial}{\partial Q^b}\right>_m+\I(K-\braket{K}_m)\chi_{\varphi}\notag\\
&+MV\chi_{\varphi}+\braket{\hat{H}_m}_m\chi_{\varphi} = 0 \ , \label{eq:heavy-Berry}\\
&\frac{G^{ab}}{M\chi_{\varphi}}\frac{\partial\chi_{\varphi}}{\partial Q^a}\left(\frac{\partial}{\partial Q^b}-\left<\frac{\partial}{\partial Q^b}\right>_m\right)\Psi_{\varphi}\notag\\
& = \left(\hat{H}_m-\braket{\hat{H}_m}_m\right)\Psi_{\varphi}-\I(K-\braket{K}_m)\Psi_{\varphi}-\frac{1}{2M}\left(\nabla^2-\braket{\nabla^2}_m\right)\Psi_{\varphi} \ , \label{eq:light-Berry}
\end{align}
where we defined $\chi_{\varphi}(Q) = e^{\I\varphi(Q)}$. Equations~(\ref{eq:heavy-Berry}) and~(\ref{eq:light-Berry}) are equivalent to~(\ref{eq:complex-source}) and~(\ref{eq:pre-schrodinger}), respectively, and are the analogues of the non-relativistic equations~(\ref{eq:nonrel-heavy-Berry}) and~(\ref{eq:nonrel-light-Berry}). We refrain from rewriting~(\ref{eq:heavy-Berry}) and~(\ref{eq:light-Berry}) in terms of `covariant' derivatives in analogy to~(\ref{eq:nonrel-covariant-derivative-Berry}).

Non-linear coupled equations such as~(\ref{eq:heavy-Berry}) and~(\ref{eq:light-Berry}) were used in~\cite{Brout:1987-1,Brout:1987-2,Brout:1987-3,Brout:1987-4,Venturi:1990,Bertoni:1996,Kamenshchik:2017} to describe the dynamics of the composite gravity-matter system and, in particular, $\Psi_{\varphi}$ was interpreted as the matter-sector wave function. We stress that such an interpretation requires the additional choice of matter-sector unitarity and the ensuing interpretation of $\chi_{\varphi}$ and $\Psi_{\varphi}$ as marginal and conditional wave functions, respectively. Alternatively, $\Psi_{\varphi}$ may be interpreted as the wave function for both gravitational and matter degrees of freedom, obtained from the solution $\Psi$ of the WDW equation by a phase transformation. `Semiclassical gravity' emerges from the BO approach to quantum gravity in the sense that~(\ref{eq:heavy-Berry}) is equivalent to~(\ref{eq:complex-source}), which leads to the Hamilton-Jacobi equation with backreaction terms~(cf.~(\ref{eq:source-backreaction})).

\section{A Simple Example: The Relativistic Particle}\label{sec:example}
As a simple example of the above formalism, let us consider the action for a massive relativistic particle in two spacetime dimensions,
\begin{equation}\label{eq:action-relpart}
\begin{aligned}
S &= \int\mathrm{d}t\ \left(p_{1}\dot{q}^1+p_{2}\dot{q}^2-NH\right) \ , \\
H &= -\frac{1}{c^2}(p_1)^2+(p_2)^2+m^2c^2 \ ,
\end{aligned}
\end{equation}
where we have restored factors of the speed of light $c$ and $m$ is the mass of the particle. We will develop an expansion in powers of $\frac{1}{c^2}$, such that $c$ is plays the role of the `heavy' scale $\sqrt{M}$ in the formalism we have presented. The `heavy' sector thus consists of the degrees of freedom $(p_1,q^1)$, while the `light' sector is comprised of the $(p_2, q^2)$ variables. This example was also considered in~\cite{Kiefer:1991} as an analogy to the semiclassical interpretation of quantum gravity. Here, we examine it to clarify the fact that the results of the semiclassical and BO approaches coincide with a particular choice of gauge both at the classical and quantum levels.

We will choose a gauge adapted to a given background Hamilton function $\varphi$. We take $\varphi(q) = mc^2 q^1$, which solves the `vacuum' ($J = 0$) Hamilton-Jacobi equation for the `heavy' sector,
\begin{align*}
-\frac{1}{c^2}\left(\frac{\partial\varphi}{\partial q^1}\right)^2+m^2c^2 = 0 \ .
\end{align*}
Let us now define the configuration-space coordinate adapted to this choice of $\varphi$. Let us set $\mathcal{N} = \frac{1}{2m}$. We then choose the basis vector~(cf.~(\ref{eq:coord-transf-Qx}))
\begin{equation}
\frac{\partial}{\partial x^1} =\frac{\mathcal{N}}{c^2}(-2)\frac{\partial\varphi}{\partial q^1}\frac{\partial}{\partial q^1}= -\frac{1}{mc^2}\frac{\partial\varphi}{\partial q^1}\frac{\partial}{\partial q^1} = -\frac{\partial}{\partial q^1} \ ,
\end{equation}
which leads to $x^1 = -q^1$, which is canonically conjugate to $\tilde{p}_1 = -p_1$. We now fix the gauge $x^1(t) = t$, which determines the lapse to be
\begin{equation}\label{eq:relpart-lapse}
N = -\frac{c^2}{2\tilde{p}_1} \ .
\end{equation}
Solving the constraint equation for $\tilde{p}_1$ yields the reduced Hamiltonian
\begin{equation}\label{eq:relpart-redhamiltonian}
-\tilde{p}_1 = \pm c^2\sqrt{\left(\frac{p_2}{c}\right)^2+m^2} \ .
\end{equation}
Inserting~(\ref{eq:relpart-redhamiltonian}) into~(\ref{eq:relpart-lapse}) yields~(cf.~(\ref{eq:lowest-order-lapse}))
\begin{equation}
N = \pm\frac{1}{2\sqrt{\left(\frac{p_2}{c}\right)^2+m^2}} = \pm\frac{1}{2m}+\mathcal{O}\left(\frac{1}{c^2}\right) \ .
\end{equation}
Upon performing the canonical transformation $\tilde{p}_1 = \tilde{\Pi}_1+\frac{\partial\varphi}{\partial x^1} = \tilde{\Pi}_1-mc^2$, we find
\begin{equation}
-\tilde{\Pi}_1 = -mc^2\pm mc^2\sqrt{1+\left(\frac{p_2}{mc}\right)^2} \ ,
\end{equation}
which is the solution to the transformed constraint equation~(cf.~(\ref{eq:hamiltonian-constraint-arbitrary-varphi}))
\begin{equation}\label{eq:relpart-constraint-pi}
2m\tilde{\Pi}_1-\frac{1}{c^2}\left(\tilde{\Pi}_1\right)^2+(p_2)^2 = 0 \ .
\end{equation}
The solution of~(\ref{eq:relpart-constraint-pi}) found in the iterative procedure is the one with the positive sign in front of the square root and reads
\begin{equation}
-\tilde{\Pi}_1 = \frac{1}{2m}(p_2)^2+\mathcal{O}\left(\frac{1}{c^2}\right) \ .
\end{equation}
In the quantum theory, we promote the variables to operators $\hat{q}^1\Psi = q^1\Psi, \hat{p}_1\Psi = -\I\frac{\partial\Psi}{\partial q^1}, \hat{q}^i\Psi = q^2\Psi, \hat{p}_2\Psi = -\I\frac{\partial\Psi}{\partial q^2}$. The quantum constraint equation reads
\begin{equation}\notag
\frac{1}{c^2}\frac{\partial^2\Psi}{\partial \left(x^1\right)^2}-\frac{\partial^2\Psi}{\partial \left(q^2\right)^2}+m^2c^2\Psi = 0 \ .
\end{equation}
The conserved Klein-Gordon inner product with a suitably chosen constant pre-factor is
\begin{equation}
(\Psi_1,\Psi_2)_{\text{KG}} = \int\mathrm{d}q^2\ \frac{\I}{2mc^2}\left(\bar{\Psi}_1\frac{\partial\Psi_2}{\partial x^1}- \Psi_2\frac{\partial\bar{\Psi}_1}{\partial x^1}\right) \ .
\end{equation}
We perform the phase transformation $\Psi = e^{\I\varphi}\Psi_{\varphi}$ to obtain the transformed quantum constraint equation
\begin{equation}\label{eq:relpart-pre-schrodinger}
\I\frac{\partial\Psi_{\varphi}}{\partial x^1} = -\frac{1}{2m}\frac{\partial^2\Psi_{\varphi}}{\partial \left(q^2\right)^2}+\frac{1}{2mc^2}\frac{\partial^2\Psi_{\varphi}}{\partial\left(x^1\right)^2} = -\frac{1}{2m}\frac{\partial^2\Psi_{\varphi}}{\partial \left(q^2\right)^2}+\mathcal{O}\left(\frac{1}{c^2}\right) \ .
\end{equation}
For solutions of~(\ref{eq:relpart-pre-schrodinger}) found in the iterative procedure, the transformed inner product reads
\begin{equation}
\begin{aligned}
(\Psi_1,\Psi_2)_{\text{KG}} &= \int\mathrm{d}q^2\ \left[\bar{\Psi}_{1,\varphi}\Psi_{2,\varphi}+\frac{\I}{2mc^2}\left(\bar{\Psi}_{1,\varphi}\frac{\partial\Psi_{2,\varphi}}{\partial x^1}- \Psi_{2,\varphi}\frac{\partial\bar{\Psi}_{1,\varphi}}{\partial x^1}\right)\right]\\
&=\int\mathrm{d}q^2\ \bar{\Psi}_{1,\varphi}\Psi_{2,\varphi}+\mathcal{O}\left(\frac{1}{c^2}\right) \ .
\end{aligned}
\end{equation}
Equation~(\ref{eq:heavy-Berry}) reads
\begin{equation}\notag
\frac{1}{c^2}\frac{\partial^2\chi_{\varphi}}{\partial \left(q^1\right)^2}+\frac{1}{c^2}\left<\frac{\partial^2}{\partial \left(q^1\right)^2}\right>_m\chi_{\varphi}+\frac{2}{c^2}\frac{\partial\chi_{\varphi}}{\partial q^1}\left<\frac{\partial}{\partial q^1}\right>_m+m^2c^2\chi_{\varphi}-\left<\frac{\partial^2}{\partial\left(q^2\right)^2}\right>_m\chi_{\varphi} = 0 \ .
\end{equation}
If we substitute $\chi_{\varphi} = e^{\I\varphi} = e^{\I mc^2 q^1}$ into the above equation, we find
\begin{equation}\notag
\I\left<\frac{\partial}{\partial q^1}\right>_m-\frac{1}{2m}\left<\frac{\partial^2}{\partial\left(q^2\right)^2}\right>_m +\frac{1}{2mc^2}\left<\frac{\partial^2}{\partial \left(q^1\right)^2}\right>_m= 0 \ ,
\end{equation}
which is simply the partial average of the constraint equation~(\ref{eq:relpart-pre-schrodinger}), since $x^1 = - q^1$. Therefore, $\chi_{\varphi} = e^{\I\varphi} = e^{\I mc^2 q^1}$ is a solution to~(\ref{eq:heavy-Berry}), which is considered to be the equation for the `heavy'-sector wave function in the BO approach of~\cite{Brout:1987-1,Brout:1987-2,Brout:1987-3,Brout:1987-4,Venturi:1990,Bertoni:1996,Kamenshchik:2017}. 

\section{Conclusions}\label{sec:conclusions}
The problem of time in canonical quantum gravity has many facets and has inspired various interpretations of the theory~\cite{Anderson:book,Kuchar:1991,Isham:1992}. In this paper, we have reinterpreted the usual results of the semiclassical interpretation of quantum cosmology based on the view that the independence of the wave function on the time parameter does not conceal the dynamical content of the quantum theory and that it is unnecessary to restrict some of the fields to the (semi)classical regime to recover a notion of dynamics. We have followed a conservative route, in which the interpretation of the quantum dynamics closely follows that of the canonical classical theory, which is far less controversial. The diffeomorphism-induced symmetry of the theory implies that the choice of time parameter is not unique.
At the classical level, this implies that Hamilton's equations can only be solved once the arbitrary Lagrange multipliers associated with the gauge symmetry are fixed. This corresponds to a choice of gauge, i.e., a choice of coordinate system. In a closed, isolated universe, it is natural to fix the gauge by functions of the canonical variables. In this way, the coordinates are fixed by the contents of the universe~\cite{DeWitt:1967}. In particular, the time variable is given by (the level sets of) a particular function of the canonical variables.

We took the position that the same is true in the quantum theory, which we constructed in analogy to the Hamilton-Jacobi theory. As in the classical case, the quantum dynamics must be understood with respect to a \emph{non-unique} choice of time as a function of configuration variables and momenta. Such a time function is measured by \emph{physical} clocks composed of the quantised variables themselves, as Singh~\cite{Singh:1990} noted for the case of `WKB time'. Indeed, the `WKB time' used in the semiclassical interpretation (and its generalisation known as the BO approach) is a function of the canonical variables and, thus, it must correspond to a particular gauge fixing. We have shown that this is the case and that the usual results of the BO approach are obtained by a particular class of gauge choices, which can be made both at the classical and quantum levels. The (WKB) time parameter $x^1$ is chosen from a congruence of (classical) trajectories associated with a background Hamilton function, which solves the Hamilton-Jacobi equation with arbitrary sources $J$. Its interpretation is that of the standard of time measured by clocks which travel along the background trajectories defined from the background Hamilton function.

At the quantum level, the chosen time parameter $x^1$ appears in the quantum constraint equation by a change of coordinates in configuration space and the usual BO factorisation of the wave function can be replaced by a phase transformation determined by the background Hamilton function. The ambiguity in the usual BO factorisation corresponds to the ambiguity in the choice of phase factor which, in turn, is related to the freedom to choose different time variables. Thus, there is no need to perform a WKB approximation to recover the \emph{concept} of time. The inner product and the dynamical interpretation of the theory here constructed, although provisional, are independent of the semiclassical limit. Nevertheless, some approximation method is needed to perform practical computations. In the formalism we have presented, the only approximation used was a weak-coupling expansion. The formalism is applicable not only to quantum cosmology, but also to the timeless non-relativistic quantum mechanics of closed systems, which was studied in~\cite{Englert:1989,Briggs:2000-1,Briggs:2000-2,Arce:2012,Briggs:2015}.

To reproduce the results of the usual semiclassical approach, the background Hamilton function is chosen such that a weak-coupling expansion is possible. Equivalently, the fields are separated into `heavy' and `light' variables, such that a perturbative expansion in the `light-to-heavy' ratio of mass scales can be performed. The background trajectories coincide with the trajectories of the `heavy' variables to lowest order in perturbation theory. In the case of general relativity, the `heavy' mass scale is the Planck scale. We have shown that, in the perturbative regime, one can expand the reduced Hamiltonian of the gauge-fixed classical theory to obtain some of the corrections found in the semiclassical approach to quantum cosmology. This shows that some of the ``corrections from quantum gravity'' found in~\cite{Singh:1990,Kiefer:1991} are, in fact, a result of the weak-coupling expansion of the gauge-fixed system and are present even in the classical theory. The correction terms can be obtained by an iterative solution of the constraint equation both at the classical and quantum levels. In the quantum theory with the Laplace-Beltrami factor ordering, additional correction terms are present which guarantee unitarity of the theory with respect to the (Klein-Gordon) inner product in the perturbative regime.

The arbitrary source $J$ approximates the backreaction of the `light'-sector Hamiltonian onto the `heavy'-sector dynamics in the perturbative regime. This is true both in the classical and quantum theories. In particular, we have shown that the arbitrary source $J$ is associated with the usual quantum backreaction terms, comprised of the expectation value of the `light'-sector Hamiltonian (averaged only over `light' variables) and Berry connection terms. We have shown that the ambiguity in the choice of $J$ is equivalent to the ambiguity in the definition of the Berry connection terms, which corresponds to the freedom to choose a particular background Hamilton function and its associated time parameter. 
We refer the reader to~\cite{Halliwell:1987,Halliwell:1989,Padmanabhan:1988} for a complementary discussion on the quantum backreaction terms. 

As we have seen, a TDSE appears as the lowest order approximation to the quantum constraint equation in the weak-coupling expansion. We have interpreted the solution of the quantum constraint equation as the wave function of the composite system of gravitational and matter degrees of freedom.
This interpretation is motivated by the fact that higher orders in the weak coupling expansion lead to a \emph{corrected} TDSE which includes the momenta of gravitational degrees of freedom (here denoted by $x^i$) and which, therefore, incorporates their dynamics. Nevertheless, we have argued that it is possible to factorise the wave function such that `light'-sector unitary is guaranteed and a marginal-conditional interpretation of the factors is warranted, as was advocated in~\cite{Arce:2012}. In this way, the conditional wave function describes the unitary dynamics of the `light' sector.

Such observations are important if one wants to analyse the phenomenology of the corrected TDSE. Several works in the literature~\cite{Kiefer:2011-1,Kiefer:2011-2,Bini:2013,Brizuela:2015-1,Brizuela:2015-2,Brizuela:2015-3,Kamenshchik:2013,Kamenshchik:2014,Kamenshchik:2016,Kamenshchik:2018rpw} have addressed this topic by applying the semiclassical or BO approaches to compute quantum gravitational corrections to the Cosmic Microwave Background (CMB) power spectrum of cosmological scalar and tensor perturbations. Kiefer and Kr\"{a}mer~\cite{Kiefer:2011-1,Kiefer:2011-2}, Bini et al.~\cite{Bini:2013} and Brizuela et al.~\cite{Brizuela:2015-1,Brizuela:2015-2,Brizuela:2015-3} used the corrected TDSE found in~\cite{Kiefer:1991} to compute the corrected power spectrum. Their method can be reinterpreted with the formalism presented in this paper as a particular gauge fixing and subsequent weak-coupling expansion. It is also worth mentioning that Kamenshchik et al.~\cite{Kamenshchik:2013,Kamenshchik:2014,Kamenshchik:2016} found corrections to the power spectrum by considering `non-adiabatic' effects related to the quantum backreaction and Berry connection terms.

The results of this paper lead us to the conclusion that time does not `emerge' only when a subset of the fields is (semi)classical, which is the central tenet of the semiclassical interpretation of quantum gravity. Rather, different notions of time are available in the full quantum theory and are associated with the different coordinate systems that one can employ. The semiclassical approach can be reinterpreted as a particular gauge fixing and the chosen time function $x^1$ is meaningful beyond the semiclassical level when interpreted as a combination of the quantised `heavy' variables. Thus, we expect that the semiclassical approach can be superseded by gauge-fixing methods in a more definitive version of the quantum theory. Moreover,
it would be desirable to generalise the inner product
that was employed in this paper
to guarantee positive-definiteness and unitarity beyond the perturbative regime and such that the interference between perturbative and non-perturbative solutions could be analysed. In addition to this, it would be useful to compare the quantum dynamics with respect to $x^1$ to the results associated with more general choices of the time variable (i.e., not given by the phase of the `heavy' part of the wave function). This will be left for future work.

\section*{Acknowledgements}
\addcontentsline{toc}{section}{Acknowledgements}
The author would like to thank David Brizuela, Claus Kiefer, Manuel Kr\"{a}mer, Ward Struyve and especially Branislav Nikoli\'{c} for useful discussions, and the Bonn-Cologne Graduate School of Physics and Astronomy for financial support.

\begin{appendix}
\section{Canonical Variables Adapted to a Choice of Background. Formulae}\label{app:formulas-Qx-transf}
In this appendix we collect useful formulae related to the change of coordinates given in~(\ref{eq:coord-transf-Qx}), which we repeat below,
\begin{equation}\label{eq:coord-transf-Qx-app}
\begin{aligned}
B^a_1 &= \frac{\mathcal{N}}{M}G^{ab}\Phi_b = \frac{\partial Q^a}{\partial x^1} \ , \\
B^a_{i} &= \frac{\partial Q^a}{\partial x^i} \ .
\end{aligned}
\end{equation}
The coordinates defined via~(\ref{eq:coord-transf-Qx-app}) induce the following transformation between basis vectors in the tangent space,
\begin{equation}\label{eq:vectors-xQ-app}
\begin{aligned}
\frac{\partial}{\partial x^1} &= \frac{\mathcal{N}}{M}G^{ab}\Phi_b\frac{\partial}{\partial Q^a} = B_1^a\frac{\partial}{\partial Q^a} \ , \\
\frac{\partial}{\partial x^i} &= B^a_i\frac{\partial}{\partial Q^a} \ .
\end{aligned}
\end{equation}
The metric tensor in the new coordinates has components (cf.~(\ref{eq:normalisation-basis-vectors}))
\begin{equation}
\begin{aligned}
\tilde{G}_{AB} &= G_{ab}\frac{\partial Q^a}{\partial x^A}\frac{\partial Q^b}{\partial x^B} = G_{ab}B^a_AB^b_B \ , \\
\tilde{G}_{11} &= G_{ab}B^a_1B^b_1 = -2\mathcal{N}^2\left(\frac{J}{M}+V\right) \ , \\
\tilde{G}_{1i} &= 0 \ , \\
\tilde{G}_{ij} &= G_{ab}B^a_iB^b_j\equiv g_{ij} \ ,
\end{aligned}
\end{equation}
while the inverse metric tensor has components $\tilde{G}^{11} =\left(\tilde{G}_{11}\right)^{-1}$, $\tilde{G}^{1i} = 0$ and $g^{ij} = \tilde{G}^{ij}$, such that $g^{ij}g_{jk} = \delta^i_k$. The determinants obey $\sqrt{|\tilde{G}|} = \sqrt{|G|}B$, where $B = \det B^a_A$.  The inverse of the jacobian $B^a_A$ is
\begin{equation}
\frac{\partial x^A}{\partial Q^a} = \left(B^{-1}\right)^A_{a} = \tilde{G}^{AB}G_{ab}B^b_B \ .
\end{equation}
The old basis vectors can thus be expressed in terms of the new basis as\footnote{The change of basis given in~(\ref{eq:vectors-Qx-app}) was performed in \cite{Kiefer:1991} in a perturbative context, where the terms proportional to $\mathbf{B}_i$ were neglected.}
\begin{equation}\label{eq:vectors-Qx-app}
\begin{aligned}
\frac{\partial}{\partial Q^a} &= \left(B^{-1}\right)^A_a\frac{\partial}{\partial x^A}=\tilde{G}^{11}G_{ab}B_1^b\frac{\partial}{\partial x^1}+\tilde{G}^{ij}G_{ab} B^b_i\frac{\partial}{\partial x^j}\\
&=-\frac{\Phi_a}{2\mathcal{N}\left(J+MV\right)}\frac{\partial}{\partial x^1}+g^{ij}G_{ab} B^b_i\frac{\partial}{\partial x^j} \ .
\end{aligned}
\end{equation}
By differentiating the first of equations~(\ref{eq:coord-transf-Qx-app}) with respect to $Q^b$, we obtain the useful identity
\begin{equation}\label{eq:derivative-Phi-app}
G^{ac}\frac{\partial\Phi_c}{\partial Q^b}  =\frac{M}{\mathcal{N}}\frac{\partial B_1^a}{\partial Q^b} -\frac{1}{\mathcal{N}}\frac{\partial}{\partial Q^b}\left(\mathcal{N}G^{ac}\right)\Phi_c \ .
\end{equation}
We also have the Hessian identities
\begin{equation}\label{eq:Hessian-app}
\begin{aligned}
\frac{\partial B^a_B}{\partial x^A} = \frac{\partial^2 Q^a}{\partial x^A\partial x^B} = \frac{\partial^2 Q^a}{\partial x^B\partial x^A} = \frac{\partial B^a_A}{\partial x^B} \ ,
\end{aligned}
\end{equation}
which lead to
\begin{equation}\label{eq:trace-derivative-app}
\begin{aligned}
\left(B^{-1}\right)^A_a\frac{\partial B^a_A}{\partial x^B} &= \left(B^{-1}\right)^A_a\frac{\partial B^a_B}{\partial x^A} = \frac{\partial B^a_B}{\partial Q^a} \ .
\end{aligned}
\end{equation}
Due to the identity given in~(\ref{eq:derivative-Phi-app}), we may rewrite the function $K$ defined in~(\ref{eq:K}) as follows.
\begin{equation}\label{eq:K-app0}
K(Q) = \frac{1}{2\mathcal{N}}\frac{\partial B_1^a}{\partial Q^a}-\frac{1}{2M\mathcal{N}}\frac{\partial}{\partial Q^a}\left(\mathcal{N}G^{ac}\right)\Phi_c+\frac{\Phi_a}{2M\sqrt{|Gh|}}\frac{\partial}{\partial Q^b}\left(\sqrt{|Gh|}G^{ab}\right) \ .
\end{equation}
Now using~(\ref{eq:coord-transf-Qx-app}),~(\ref{eq:vectors-Qx-app}) and~(\ref{eq:trace-derivative-app}), we can rewrite~(\ref{eq:K-app0}) in the $x$-coordinate system,
\begin{equation}\label{eq:K-app}
\begin{aligned}
K(Q) &= \frac{1}{2\mathcal{N}}\frac{\partial B_1^a}{\partial Q^a}-\frac{G^{ab}\Phi_a}{2M\mathcal{N}}\frac{\partial\mathcal{N}}{\partial Q^b}+\frac{G^{ab}\Phi_a}{2M}\frac{1}{\sqrt{|Gh|}}\frac{\partial\sqrt{|Gh|}}{\partial Q^b}\\
&= \frac{1}{2\mathcal{N}}\frac{\partial B_1^a}{\partial Q^a}+\frac{1}{2\sqrt{|Gh|}}\frac{\partial}{\partial x^1}\left(\frac{\sqrt{|Gh|}}{\mathcal{N}}\right)\\
&= \frac{1}{2\mathcal{N}}\frac{\partial B_1^a}{\partial Q^a}-\frac{1}{2\mathcal{N}}\left(B^{-1}\right)^A_a\frac{\partial B^a_A}{\partial x^1}+\frac{1}{2\sqrt{|\tilde{G}h|}}\frac{\partial}{\partial x^1}\left(\frac{\sqrt{|\tilde{G}h|}}{\mathcal{N}}\right)\\
&= \frac{1}{2\sqrt{|\tilde{G}h|}}\frac{\partial}{\partial x^1}\left(\frac{\sqrt{|\tilde{G}h|}}{\mathcal{N}}\right)\ .
\end{aligned}
\end{equation}

\section{Conservation of the Inner Product in Perturbation Theory. Definition of Partial Averages}\label{app:conservation-innerprod}
It is instructive to verify that the approximate inner product given in~(\ref{eq:innerprod-order-m1}) is conserved for solutions of~(\ref{eq:schrodinger-order-1-V}) up to order $\frac{1}{M}$. Evidently, this is a consequence of the exact equation~(\ref{eq:KG-conserved}). We assume $\mathrm{sgn}(\mathcal{N})$ is constant and set $f = -\mathrm{sgn}(\mathcal{N})$. We first note that, given the factor ordering of $\hat{H}_m$ in~(\ref{eq:LB-constraint-Hm}), we find that it obeys the symmetry condition
\begin{equation}\label{eq:symmetry-innerprod-1}
\int\prod_i\mathrm{d}x^i\sqrt{2|Vg|}\prod_{\mu}\mathrm{d}q^{\mu}\sqrt{h}\ \bar{\Psi}_{1}\hat{H}_m\Psi_{2} = \int\prod_i\mathrm{d}x^i\sqrt{2|Vg|}\prod_{\mu}\mathrm{d}q^{\mu}\sqrt{h}\ \overline{\left(\hat{H}_m\Psi_{1}\right)}\Psi_{2} \ .
\end{equation}
Secondly, recall that we define the derivative of an operator as $\left(\frac{\partial}{\partial x^1}\hat{O}\right)\psi := \frac{\partial}{\partial x^1}\left(\hat{O}\psi\right)-\hat{O}\frac{\partial\psi}{\partial x^1}$. Then, using the Laplace-Beltrami factor ordering of $\hat{H}_m$ given in~(\ref{eq:LB-constraint-Hm}) and the measure $\hat{\mathscr{M}}$ defined in~(\ref{eq:measure-M}), we define the operators
\begin{align}
&\frac{\partial}{\partial x^1}\left(\sigma\sqrt{\left|\frac{gh}{2V}\right|}\hat{H}_m\right):=-\frac{1}{2}\frac{\partial}{\partial q^{\mu}}\left[\frac{\partial}{\partial x^1}\left(\sigma\sqrt{\left|\frac{gh}{2V}\right|}h^{\mu\nu}\right)\frac{\partial}{\partial q^{\nu}}\right]+\frac{\partial}{\partial x^1}\left(\sigma\sqrt{\left|\frac{gh}{2V}\right|}V_m\right) \ , \label{eq:def-for-Gamma}\\
&\frac{\partial}{\partial x^1}\hat{\mathscr{M}} := \frac{\partial}{\partial x^1}\sqrt{2|Vgh|}+\frac{1}{M}\frac{\partial}{\partial x^1}\left(\sigma\sqrt{\left|\frac{gh}{2V}\right|}\hat{H}_m\right) \ , \\
&\hat{\mathfrak{H}} := \mathcal{N}\left(\hat{H}_m-J\right)-\frac{\mathcal{N}}{4MV}\left(\hat{H}_m-J\right)^2-\frac{1}{2M\sqrt{2|Vgh|}}\frac{\partial}{\partial x^i}\left(\sqrt{2|Vgh|}\mathcal{N}g^{ij} \frac{\partial}{\partial x^j}\right)\notag\\
&\ \ \ +\frac{1}{M}\mathcal{V} \ . \label{eq:hfrak}
\end{align}
Using~(\ref{eq:Gamma-Unitarity}),~(\ref{eq:def-for-Gamma}) and~(\ref{eq:hfrak}), it is possible to show that $\hat{\Gamma}$ and $\hat{\mathfrak{H}}$ obey the same symmetry condition as $\hat{H}_m$ given in~(\ref{eq:symmetry-innerprod-1}). With these definitions, it is straightforward to prove the identities
\begin{equation}\label{eq:Gamma-hfrak-identities}
\begin{aligned}
&\sqrt{2|Vg|h}\left[\hat{\Gamma}\left(1+\frac{1}{2MV}\hat{H}_m\right)+\left(1+\frac{1}{2MV}\hat{H}_m\right)\hat{\Gamma}\right] = \frac{\partial}{\partial x^1}\hat{\mathscr{M}}+\mathcal{O}\left(\frac{1}{M^2}\right) \ , \\
&\left[\hat{\mathfrak{H}},\frac{1}{2MV}\hat{H}_m\right] = \left[\mathcal{N}\left(\hat{H}_m-J\right),\frac{1}{2MV}\hat{H}_m\right]+\mathcal{O}\left(\frac{1}{M^2}\right) = 0+\mathcal{O}\left(\frac{1}{M^2}\right) \ .
\end{aligned}
\end{equation}
We can now rewrite~(\ref{eq:schrodinger-order-1-V}) as
\begin{equation}\label{eq:schrodinger-order-1-V-simple}
\frac{\partial\Psi_{\varphi}}{\partial x^1} =-\I\hat{\mathfrak{H}}\Psi_{\varphi}-\hat{\Gamma}\Psi_{\varphi}+\mathcal{O}\left(\frac{1}{M^2}\right) \ .
\end{equation}
Using~(\ref{eq:schrodinger-order-1-V-simple}), we obtain
\begin{align*}
&\frac{\partial}{\partial x^1}\int\prod_i\mathrm{d}x^i\prod_{\mu}\mathrm{d}q^{\mu}\ \bar{\Psi}_{1,\varphi}\hat{\mathscr{M}}\Psi_{2,\varphi}\\
& = \int\prod_i\mathrm{d}x^i\prod_{\mu}\mathrm{d}q^{\mu}\ \Biggbr{\I\left(\hat{\mathfrak{H}}\bar{\Psi}_{1,\varphi}\right)\hat{\mathscr{M}}\Psi_{2,\varphi}-\I\bar{\Psi}_{1,\varphi}\hat{\mathscr{M}}\hat{\mathfrak{H}}\Psi_{2,\varphi}\\
&-\left(\hat{\Gamma}\bar{\Psi}_{1,\varphi}\right)\hat{\mathscr{M}}\Psi_{2,\varphi}-\bar{\Psi}_{1,\varphi}\hat{\mathscr{M}}\hat{\Gamma}\Psi_{2,\varphi}+\bar{\Psi}_{1,\varphi}\left(\frac{\partial}{\partial x^1}\hat{\mathscr{M}}\right)\Psi_{2,\varphi}}+\mathcal{O}\left(\frac{1}{M^2}\right)\\
& = \int\prod_i\mathrm{d}x^i\sqrt{2|Vg|}\prod_{\mu}\mathrm{d}q^{\mu}\sqrt{h}\ \Biggbr{\I\left(\hat{\mathfrak{H}}\bar{\Psi}_{1,\varphi}\right)\left(1+\frac{1}{2MV}\hat{H}_m\right)\Psi_{2,\varphi}\\
&-\I\bar{\Psi}_{1,\varphi}\left(1+\frac{1}{2MV}\hat{H}_m\right)\hat{\mathfrak{H}}\Psi_{2,\varphi}-\left(\hat{\Gamma}\bar{\Psi}_{1,\varphi}\right)\left(1+\frac{1}{2MV}\hat{H}_m\right)\Psi_{2,\varphi}\\
&-\bar{\Psi}_{1,\varphi}\left(1+\frac{1}{2MV}\hat{H}_m\right)\hat{\Gamma}\Psi_{2,\varphi}+\frac{1}{\sqrt{2|Vg|h}}\bar{\Psi}_{1,\varphi}\left(\frac{\partial}{\partial x^1}\hat{\mathscr{M}}\right)\Psi_{2,\varphi}}+\mathcal{O}\left(\frac{1}{M^2}\right)\\
& = \int\prod_i\mathrm{d}x^i\prod_{\mu}\mathrm{d}q^{\mu}\ \left\{\I\bar{\Psi}_{1,\varphi}\sqrt{2|Vg|h}\left[\hat{\mathfrak{H}},\frac{1}{2MV}\hat{H}_m\right]\Psi_{2,\varphi}\right.\\
&\left.-\bar{\Psi}_{1,\varphi}\sqrt{2|Vg|h}\left[\hat{\Gamma}\left(1+\frac{1}{2MV}\hat{H}_m\right)+\left(1+\frac{1}{2MV}\hat{H}_m\right)\hat{\Gamma}\right]\Psi_{2,\varphi}\right.\\
&\left.+\bar{\Psi}_{1,\varphi}\left(\frac{\partial}{\partial x^1}\hat{\mathscr{M}}\right)\Psi_{2,\varphi}\right\}+\mathcal{O}\left(\frac{1}{M^2}\right) = 0+\mathcal{O}\left(\frac{1}{M^2}\right) \ ,
\end{align*}
where we used the symmetry condition given in~(\ref{eq:symmetry-innerprod-1}) for $\hat{\Gamma}$ and $\hat{\mathfrak{H}}$ and subsequently we applied~(\ref{eq:Gamma-hfrak-identities}). Thus, we arrive at the result
\begin{equation}\label{eq:unitarity-order-m1}
\frac{\partial}{\partial x^1}\int\prod_i\mathrm{d}x^i\prod_{\mu}\mathrm{d}q^{\mu}\ \bar{\Psi}_{1,\varphi}\hat{\mathscr{M}}\Psi_{2,\varphi} = 0+\mathcal{O}\left(\frac{1}{M^2}\right) \ ,
\end{equation}
which is consistent with the exact equation~(\ref{eq:KG-conserved}) and confirms that the term $\I\hat{\Gamma}$, which appears in~(\ref{eq:schrodinger-order-1-V}) and~(\ref{eq:schrodinger-order-1-V-simple}), is the term which guarantees that the dynamics is unitary to this order in perturbation theory.

A final comment about matter-sector partial averages is in order. In section~\ref{sec:rel-backreaction}, the matter-sector inner product was tacitly taken to be $\braket{\psi_1,\psi_2}_m = \int\prod_{\mu}\mathrm{d}q^{\mu}\sqrt{h}\ \bar{\psi}_1(x;q)\psi_2(x;q)$, which is associated with partial averages given in~(\ref{eq:matter-sector-partial-average}). However, due to~(\ref{eq:innerprod-order-m1}), this may be regarded as the lowest order approximation to the more general matter-sector inner product $\braket{\psi_1,\psi_2}_m = \int\prod_{\mu}\mathrm{d}q^{\mu}\ \bar{\psi}_1(x;q)\hat{\mathcal{M}}(x;\hat{p},q)\psi_2(x;q)$, where the measure $\hat{\mathcal{M}}$ has to be determined from the perturbative expansion of~(\ref{eq:BOprod}). In this case, the partial average of an operator is given by $\braket{\hat{O}}_m = \left(\int\prod_{\mu}\mathrm{d}q^{\mu}\ \bar{\psi}\hat{\mathcal{M}}\psi\right)^{-1}\int\prod_{\mu}\mathrm{d}q^{\mu}\ \bar{\psi}\hat{\mathcal{M}}\hat{O}\psi$.
At order $\frac{1}{M}$, we can take $\hat{\mathcal{M}}$ to be given by~(\ref{eq:measure-M}) (overall factors of $V(x)$ and $g(x)$ can be eliminated from the matter-sector inner product by suitable factorisations of the wave function) and, using~(\ref{eq:Gamma-Unitarity}), we can define $\hat{\Gamma} =: \frac{\mathcal{N}}{M}G^{ab}\Phi_a\hat{\Gamma}_b$. The operator $\hat{\Gamma}_a$ inherits its symmetry with respect to the lowest-order matter-sector inner product from the symmetry of $\hat{\Gamma}$. Using~(\ref{eq:Gamma-hfrak-identities}), one may verify that matter-sector unitarity is then equivalent to $\mathfrak{Re}\left<\frac{\partial}{\partial Q^a}+\hat{\Gamma}_a\right>_m = 0$  (cf.~(\ref{eq:nonrel-light-unitarity})) and the `Berry connection' can be defined as $A_a = \mathfrak{Im}\left<\frac{\partial}{\partial Q^a}+\hat{\Gamma}_a\right>_m$.

\section{Extension to Field Theory}\label{app:field-theory}
We now comment on how one can formally extend the formalism presented in this paper to the field-theoretic case. The canonical approach to General Relativity in $3+1$ spacetime dimensions involves a foliation of spacetime into a family of spacelike hypersurfaces $\Sigma_t$ defined as the level sets of some scalar function, $\tau(y) = t$. We parametrise each hypersuface by coordinates $y$, such that $(t,y)$ defines a coordinate system of spacetime. The spacetime line element is
\begin{equation}
\mathrm{d}s^2 = -N^2\mathrm{d}t^2 + Q_{ij}(\mathrm{d}y^i + N^i\mathrm{d}t)(\mathrm{d}y^j + N^j\mathrm{d}t) \ ,
\end{equation}
where $y^i$ are the coordinates used to parametrise each hypersurface, $Q_{ij}$ are the components of the induced metric on a given hypersurface, $N$ is the lapse function and $N^i$ is the shift vector. The covariant derivative compatible with the induced metric will be denoted by a semicolon, such that $Q_{ij;k} = 0$. The components of the inverse of the induced metric are denoted by $Q^{ij}$ and the determinant of the induced metric is written as $Q$. Latin indices are raised and lowered with the induced metric and its inverse. The canonical momentum conjugate to $Q_{ij}$ is $P^{ij}$. Matter fields and their conjugate momenta are generically written as $q$ and $p$, respectively. Neglecting boundary terms, we find the Hamiltonian for the gravity-matter system
\begin{align}
H &= \int_{\Sigma_t}\mathrm{d}^3y \ \left(N\mathcal{H}_{\perp}+N^i \mathcal{H}_i\right) \ , \\
\mathcal{H}_{\perp} &= \frac{1}{2M}G_{ijlm}P^{ij}P^{lm}-2M\sqrt{Q}(R-2\Lambda)+\mathcal{H}^m_{\perp}(Q; p,q)\ , \\
\mathcal{H}_i &=  -2Q_{ik}\tensor{P}{^{kl}_{;l}}+\mathcal{H}^m_{i}(Q;p,q)\ ,
\end{align}
where $M = \frac{1}{32\pi G}$, $R$ is the Ricci scalar of a given three-dimensional hypersurface, $\Lambda$ is a cosmological constant term, $\mathcal{H}^{m}_{\perp}$ and $\mathcal{H}^m_{i}$ are contributions from the matter-sector and 
\begin{equation}
G_{ijlm} = \frac{1}{2\sqrt{Q}}\left(Q_{il}Q_{jm}+Q_{im}Q_{jl}-Q_{ij}Q_{lm}\right)
\end{equation}
is the inverse DeWitt metric. The DeWitt metric reads
\begin{align}
G^{ijlm} &= \frac{\sqrt{Q}}{2}\left(Q^{il}Q^{jm}+Q^{im}Q^{jl}-2Q^{ij}Q^{lm}\right) \ , \\
G^{ijlm}G_{lmab} &= \frac{1}{2}\left(\delta^{i}_a\delta^j_b+\delta^i_b\delta^j_a\right) = \delta^i_{(a}\delta^j_{b)} \ .
\end{align}
We work only with closed three-manifolds such that no boundary terms are present in the Hamiltonian. The canonical momenta conjugate to the lapse and shift functions vanish and constitute primary constraints~(see, e.g.,~\cite{Pons:2010} and references therein), which we have already eliminated. The lapse and shift are thus multipliers. By varying the action with respect to $N$ and $N^i$, we obtain the constraints $\mathcal{H}_{\perp} = 0$ and $\mathcal{H}_i = 0$, which are referred to as the Hamiltonian and diffeomorphism (momentum) constraints, respectively. In the absence of boundary terms, the dynamics of the theory is entirely contained in these constraints. The Einstein-Hamilton-Jacobi equations are
\begin{equation}\label{eq:field-HJ}
\begin{aligned}
&\frac{1}{2M}G_{ijlm}\frac{\delta W}{\delta Q_{ij}}\frac{\delta W}{\delta Q_{lm}}+MV(Q)+\mathcal{H}^m_{\perp}\left(Q; \frac{\delta W}{\delta q},q\right) = 0 \ , \\
&-2Q_{ik}\left(\frac{\delta W}{\delta Q_{kl}}\right)_{;l}+\mathcal{H}^m_{i}\left(Q;\frac{\delta W}{\delta q},q\right) = 0 \ ,
\end{aligned}
\end{equation}
where $W$ is the Hamilton characteristic functional for the composite system of both gravitational and matter degrees of freedom and we defined the potential $V(Q) = -2\sqrt{Q}(R-2\Lambda)$. In analogy to the finite-dimensional model considered in this paper, we define the background Hamilton functional $\varphi[Q]$ as the solution to the equations
\begin{equation}\label{eq:field-varphi}
\begin{aligned}
&\frac{1}{2M}G_{ijlm}\frac{\delta\varphi}{\delta Q_{ij}}\frac{\delta\varphi}{\delta Q_{lm}}+MV(Q) = -J_{\perp} \ , \\
&-2Q_{ik}\left(\frac{\delta\varphi}{\delta Q_{kl}}\right)_{;l} = -J_i \ .
\end{aligned}
\end{equation}
A change of background Hamilton functional $\varphi = \varphi' + \eta$ leads to redefinitions of $J_{\perp}, J_i$~(in analogy to~(\ref{eq:Phi-source-transf})). By defining $S = W-\varphi$, we can rewrite~(\ref{eq:field-HJ}) as
\begin{align}
&-\frac{1}{M}G_{ijlm}\Phi^{ij}\frac{\delta S}{\delta Q_{lm}} = \mathcal{H}^m_{\perp}\left(Q; \frac{\delta S}{\delta q},q\right)-J_{\perp}+\frac{1}{2M}G_{ijlm}\frac{\delta S}{\delta Q_{ij}}\frac{\delta S}{\delta Q_{lm}} \ , \label{eq:field-HJ-S-1}\\
&2Q_{ik}\left(\frac{\delta S}{\delta Q_{kl}}\right)_{;l}=\mathcal{H}^m_{i}\left(Q;\frac{\delta S}{\delta q},q\right)-J_i  \ ,\label{eq:field-HJ-S-2}
\end{align}
where we defined the background momenta $\Phi^{ij}:=\frac{\delta\varphi[Q]}{\delta Q_{ij}(y)}$. Equations~(\ref{eq:field-HJ-S-1}) and~(\ref{eq:field-HJ-S-2}) were also considered in~\cite{Parentani:1998} in the context of a background field approximation, where Parentani emphasised that the arbitrarily chosen sources $J_{\perp}, J_i$ should be compatible with the Bianchi identities.

Given $\varphi$, we define the functionals $X^1(y,Q]$ and $X^r(y,Q]$ ($r = 2,3,4,5,6$) as follows~\cite{Kuchar:1991,Isham:1992}
\begin{align*}
&\frac{\mathcal{N}(y,Q]}{M}G_{ijlm}(y,Q]\frac{\delta\varphi[Q]}{\partial Q_{ij}(y)}\frac{\delta X^1(y',Q]}{\delta Q_{lm}(y)} = \delta(y-y') \ , \\
&G_{ijlm}(y,Q]\frac{\delta\varphi[Q]}{\partial Q_{ij}(y)}\frac{\delta X^r(y',Q]}{\delta Q_{lm}(y)} = 0 \ ,
\end{align*}
where $\mathcal{N}(y,Q]$ is an arbitrary normalisation functional. Now, in analogy to the mechanical case, we define\footnote{Such functional derivatives are formal. The reader is referred to~\cite{Giulini:1994} for a discussion of the consistency of dynamical equations based on time-functionals such as $X^1$ in the presence of momentum constraints.}
\begin{equation}\label{eq:coord-transf-Qx-app-fields}
\begin{aligned}
\frac{\delta}{\delta X^1} &:= \frac{\mathcal{N}(y,Q]}{M}G_{ijlm}(y,Q]\Phi^{lm}(y,Q]\frac{\delta}{\delta Q_{ij}(y)}\equiv B_{1|ij}(y,Q]\frac{\delta}{\delta Q_{ij}(y)} \ , \\
\frac{\delta}{\delta X^r} &:= B_{r|ij}(y,Q]\frac{\delta}{\delta Q_{ij}(y)} \ ,
\end{aligned}
\end{equation}
where the $B$ functionals obey\footnote{As in the mechanical case, we assume for simplicity that $\tilde{G}_{1r} = 0$. In the most general case, $\tilde{G}_{1r} \neq 0$ and there will be additional contributions from such terms to the formulae here presented.} (cf.~(\ref{eq:field-varphi}))
\begin{equation}
\begin{aligned}
\tilde{G}_{11} &:= G^{ijlm}B_{1|ij}B_{1|lm} = -2\mathcal{N}^2\left(\frac{J_{\perp}}{M}+V\right)\ , \\
\tilde{G}_{1r} &:= G^{ijlm}B_{1|ij}B_{r|lm} = 0  \ , \\
\tilde{G}_{rs} &:= G^{ijlm}B_{r|ij}B_{s|lm} \equiv g_{rs} \ .
\end{aligned}
\end{equation}
The inverse metric tensor has components $\tilde{G}^{11} =\left(\tilde{G}_{11}\right)^{-1}$, $\tilde{G}^{1r} = 0$ and $g^{rs} = \tilde{G}^{rs}$, such that $g^{rs}(y,Q]g_{su}(y,Q] = \delta^r_u$. In particular, by contracting both sides of the equation $\tilde{G}_{AB} = G^{ijlm}B_{A|ij}B_{B|lm}$ with $\tilde{G}^{AC}B_{C|nk}$, we find $\tilde{G}^{AC}B_{A|ij}B_{C|nk} = G_{ijnk}$. Thus, the inverse of $B_{A|ij}(y,Q]$ is
\begin{equation}
\left(B^{-1}\right)^{A|ij}(y,Q] = \tilde{G}^{AB}(y,Q]G^{ijlm}(y,Q]B_{B|lm}(y,Q] \ .
\end{equation}
We can thus write
\begin{equation}\label{eq:vectors-Qx-app-fields}
\begin{aligned}
\frac{\delta}{\delta Q_{ij}} &= \left(B^{-1}\right)^{A|ij}\frac{\delta}{\delta X^A}=-\frac{\Phi^{ij}}{2\mathcal{N}\left(J_{\perp}+MV\right)}\frac{\delta}{\delta X^1}+g^{rs}G^{ijlm} B_{r|lm}\frac{\delta}{\delta X^s} \ .
\end{aligned}
\end{equation}
Using~(\ref{eq:coord-transf-Qx-app-fields}) and~(\ref{eq:vectors-Qx-app-fields}), we can rewrite~(\ref{eq:field-HJ-S-1}) as
\begin{equation}\label{eq:field-HJ-S-X}
\begin{aligned}
&-\frac{\delta S}{\delta X^1}\\
&= \mathcal{N}\left[\mathcal{H}^m_{\perp}\left(Q; \frac{\delta S}{\delta q},q\right)-J_{\perp}\right]+\frac{\mathcal{N}}{2M}g^{rs}\frac{\delta S}{\delta X^r}\frac{\delta S}{\delta X^s}-\frac{1}{4\mathcal{N}\left(J_{\perp}+MV\right)}\left(\frac{\delta S}{\delta X^1}\right)^2 \ , 
\end{aligned}
\end{equation}
Equation~(\ref{eq:field-HJ-S-X}) is analogous to~(\ref{eq:constraint-iterative-HJ}) and can also be solved iteratively. The result is
\begin{equation}\label{eq:field-HJ-S-X-pert}
\begin{aligned}
-\frac{\delta S}{\delta X^1} &= \mathcal{N}\left[\mathcal{H}^m_{\perp}\left(Q; \frac{\delta S}{\delta q},q\right)-J_{\perp}\right]\\
&-\frac{\mathcal{N}}{4MV}\left[\mathcal{H}^m_{\perp}\left(Q; \frac{\delta S}{\delta q},q\right)-J_{\perp}\right]^2+\frac{\mathcal{N}}{2M}g^{rs}\frac{\delta S}{\delta X^r}\frac{\delta S}{\delta X^s}+\mathcal{O}\left(\frac{1}{M^2}\right) \ ,
\end{aligned}
\end{equation}
where we have assumed that $J_{\perp}$ 
can be expanded as in~(\ref{eq:J-expansion}).
If we now fix the arbitrary shift vector $N^i(y,Q]$ by a suitable choice of coordinates $y^i$, we can define the `background time' derivative
\begin{equation}\label{eq:field-background-time}
\begin{aligned}
\frac{\partial}{\partial \tau} &= \int\mathrm{d}^3y\ \left\{\frac{\mathcal{N}(y,Q]}{M}G_{ijlm}(y,Q]\frac{\delta\varphi[Q]}{\delta Q_{ij}(y)}\frac{\delta}{\delta Q_{lm}(y)}+2N_{i;j}(y,Q]\frac{\delta}{\delta Q_{ij}(y)}\right\}\\
&=\int\mathrm{d}^3y\ \left\{\frac{\delta}{\delta X^1(y,Q]}+2N_{i;j}(y,Q]\frac{\delta}{\delta Q_{ij}(y)}\right\} \ .
\end{aligned}
\end{equation}
We can then combine~(\ref{eq:field-HJ-S-2}) and~(\ref{eq:field-HJ-S-X-pert}) to obtain
\begin{equation}\label{eq:field-HJ-BO}
\begin{aligned}
&-\frac{\partial S}{\partial\tau}\\
&= \int\mathrm{d}^3y\ \left\{\mathcal{N}\left[\mathcal{H}^m_{\perp}\left(Q; \frac{\delta S}{\delta q},q\right)-J_{\perp}\right]+N^i\left[\mathcal{H}^m_{i}\left(Q;\frac{\delta S}{\delta q},q\right)-J_i\right]\right\}\\
&+\int\mathrm{d}^3y\ \left\{-\frac{\mathcal{N}}{4MV}\left[\mathcal{H}^m_{\perp}\left(Q; \frac{\delta S}{\delta q},q\right)-J_{\perp}\right]^2+\frac{\mathcal{N}}{2M}g^{rs}\frac{\delta S}{\delta X^r}\frac{\delta S}{\delta X^s}\right\}+\mathcal{O}\left(\frac{1}{M^2}\right) \ ,
\end{aligned}
\end{equation}
which is the field-theoretic analogue of~(\ref{eq:HJ-BO}). As we have argued in the mechanical case, equation~(\ref{eq:field-HJ-BO}) is most appropriately interpreted as an approximation to the Hamilton-Jacobi equation for the reduced gauge-fixed system comprised of \emph{both} gravitational and matter degrees of freedom. In this way, the solution $S$ of~(\ref{eq:field-HJ-BO}) is not the ``corrected'' Hamilton principal functional of a system composed of the matter fields alone.

To see the how this corresponds to a particular gauge fixing, we define a field $\mathcal{T}(y,Q]$ as follows
\begin{equation}\label{eq:definition-big-tau-field}
\frac{\mathcal{N}(y,Q]}{M}G_{ijlm}(y,Q]\frac{\delta\varphi[Q]}{\delta Q_{ij}(y)}\frac{\delta \mathcal{T}(y',Q]}{\delta Q_{lm}(y)}+2N_{i;j}(y,Q]\frac{\delta\mathcal{T}(y',Q]}{\delta Q_{ij}(y)} = \delta(y-y') \ ,
\end{equation}
which coincides with $X^1(y,Q]$ only if $N_{i;j} = 0$. Using~(\ref{eq:field-background-time}), we find that the field $\mathcal{T}$ obeys the `background' equation of motion $\frac{\partial\mathcal{T}}{\partial\tau} = 1$. Thus, $\mathcal{T}$ serves as canonical definition of the `background time' functional. In analogy to the derivation of~(\ref{eq:vectors-Qx-app-fields}), we find that a solution to~(\ref{eq:definition-big-tau-field}) obeys
\begin{equation}\label{eq:field-gradient-big-tau}
\frac{\delta\mathcal{T}(z,Q]}{\delta Q_{ij}(y)} = \delta(z-y)\left[\frac{\frac{\mathcal{N}}{M}\Phi^{ij}+2G^{ijlm}N_{l;m}}{-2\mathcal{N}^2\left(\frac{J_{\perp}}{M}+V\right)+\frac{4\mathcal{N}}{M}\Phi^{lm}N_{l;m}+4G^{ablm}N_{a;b}N_{l;m}}\right]_y \ , 
\end{equation}
where we have used the first of equations~(\ref{eq:field-varphi}).

We now fix the canonical gauge condition $\mathcal{T}(y,Q(t)] = t$. For simplicity, we will restrict ourselves to the lowest order of the perturbative regime in order to compute the gauge-fixed lapse. Given a suitable \emph{choice} of the shift vector $N^i$, we obtain the relation between the arbitrary `background' lapse $\mathcal{N}$ and the gauge-fixed lapse $N$ as follows. The chosen gauge condition implies that $\frac{\mathrm{d}\mathcal{T}}{\mathrm{d}t}(y,Q(t)] = 1 = \frac{\partial\mathcal{T}}{\partial\tau}(y,Q(t)]$. Thus,
\begin{align*}
0 &= \frac{\mathrm{d}\mathcal{T}}{\mathrm{d}t}(y,Q(t)]- \frac{\partial\mathcal{T}}{\partial\tau}(y,Q(t)]\\
&=\int\mathrm{d}^3z\ \left\{\frac{N(z,Q]}{M}G_{ijlm}(z,Q]\frac{\delta W[Q]}{\delta Q_{ij}(z)}\frac{\delta \mathcal{T}(y,Q(t)]}{\delta Q_{lm}(z)}+2N_{i;j}(z,Q]\frac{\delta \mathcal{T}(y,Q(t)]}{\delta Q_{ij}(z)}\right\}- \frac{\partial\mathcal{T}}{\partial\tau}(y,Q(t)]\\
&=\frac{1}{M}\int\mathrm{d}^3z\ G_{ijlm}(z,Q]\frac{\delta \mathcal{T}(y,Q(t)]}{\delta Q_{lm}(z)}\left(N(z,Q]\frac{\delta W[Q]}{\delta Q_{ij}(z)}-\mathcal{N}(z,Q]\frac{\delta\varphi[Q]}{\delta Q_{ij}(z)}\right)\ .
\end{align*}
Using~(\ref{eq:field-gradient-big-tau}), we can rewrite the above equation as
\begin{align*}
0 &= \frac{1}{M}\int\mathrm{d}^3z\ \delta(z-y) G_{ijlm}(z,Q]\left[\left(\frac{\mathcal{N}}{M}\Phi^{lm}+2G^{lmnk}N_{n;k}\right)\left(N\frac{\delta W}{\delta Q_{ij}}-\mathcal{N}\frac{\delta\varphi}{\delta Q_{ij}}\right)\right]_z\\
&=\frac{1}{M} G_{ijlm}(y,Q]\left[\left(\frac{\mathcal{N}}{M}\Phi^{lm}+2G^{lmnk}N_{n;k}\right)\left(N\frac{\delta W}{\delta Q_{ij}}-\mathcal{N}\frac{\delta\varphi}{\delta Q_{ij}}\right)\right]_y \ .
\end{align*}
Using $W = S+\varphi$, we obtain
\begin{align*}
0 = \frac{1}{M}\left[\left(\frac{N}{M}-\frac{\mathcal{N}}{M}\right)G_{ijlm}\Phi^{ij}\Phi^{lm}+\frac{N}{M}G_{ijlm}\Phi^{ij}\frac{\delta S}{\delta Q_{lm}}+\frac{2}{\mathcal{N}}(N-\mathcal{N})N_{i;j}\Phi^{ij}+\frac{2N}{\mathcal{N}}N_{i;j}\frac{\delta S}{\delta Q_{ij}}\right] \ .
\end{align*}
Now, using the first of equations~(\ref{eq:field-varphi}) and the first of equations~(\ref{eq:coord-transf-Qx-app-fields}), the above equation becomes
\begin{align*}
N\left[-2\left(\frac{J_{\perp}}{M}+V\right)+\frac{2N_{i;j}\Phi^{ij}}{M\mathcal{N}}+\frac{1}{M\mathcal{N}}\left(\frac{\delta S}{\delta X^1}+2N_{i;j}\frac{\delta S}{\delta Q_{ij}}\right)\right] = \mathcal{N}\left[-2\left(\frac{J_{\perp}}{M}+V\right)+\frac{2N_{i;j}\Phi^{ij}}{M\mathcal{N}}\right] \ .
\end{align*}
Finally, using~(\ref{eq:phi-expansion}) and assuming that $N_i, S$ can be expanded as in~(\ref{eq:J-expansion}), the above equation implies
\begin{equation}
N(y,Q] = \mathcal{N}(y,Q]+ \mathcal{O}\left(\frac{1}{M}\right) \ ,
\end{equation}
which agrees with~(\ref{eq:lowest-order-lapse}) at lowest order for the choice $\kappa = +\mathrm{sgn}(\mathcal{N}V)$, which is the branch of solutions obtained via the iterative procedure as we saw in section~\ref{sec:iterative-procedure}. The terms of higher order in $\frac{1}{M}$ in the expansion of $N(y,Q]$ also depend on the chosen value of $N_i(y,Q]$. If $N_{i;j} = 0$, it is possible to show that the field-theoretic analogue of the next order in~(\ref{eq:lowest-order-lapse}) is also recovered (with $H_m\to\mathcal{H}_{\perp}^m$ and $J\to J_{\perp}$) by using~(\ref{eq:field-HJ-S-X-pert}).

The quantum theory can be constructed in analogy to what was done in section~\ref{sec:quantum}. The formal Laplace-Beltrami-ordered quantum constraint equations read
\begin{equation}\label{eq:field-WDW}
\begin{aligned}
&-\frac{1}{2M\sqrt{|Gh|}}\frac{\delta}{\delta Q_{ij}}\left(\sqrt{|Gh|}G_{ijlm}\frac{\delta\Psi}{\delta Q_{lm}}\right)+MV(Q)\Psi+\hat{\mathcal{H}}^m_{\perp}\left(Q; -\I\frac{\delta}{\delta q},q\right)\Psi = 0 \ , \\
&2\I Q_{ik}\left(\frac{\delta\Psi}{\delta Q_{kl}}\right)_{;l}+\hat{\mathcal{H}}^m_{i}\left(Q;-\I\frac{\delta}{\delta q},q\right)\Psi = 0 \ ,
\end{aligned}
\end{equation}
where $h$ is the determinant of the matter-sector field-space metric. Equations~(\ref{eq:field-WDW}) are the quantum analogues of~(\ref{eq:field-HJ}). Given a \emph{choice} of background Hamilton functional $\varphi[Q]$ (cf.~(\ref{eq:field-varphi})), we perform the phase transformation $\Psi = e^{\I\varphi}\Psi_{\varphi}$ to obtain the constraint equations
\begin{equation}\label{eq:field-WDW-2}
\begin{aligned}
&\frac{\I}{M}G_{ijlm}\Phi^{ij}\frac{\delta\Psi_{\varphi}}{\delta Q_{lm}}=\left(\hat{\mathcal{H}}^m_{\perp}-\mathfrak{J}_{\perp}\right)\Psi_{\varphi}-\frac{1}{2M\sqrt{|Gh|}}\frac{\delta}{\delta Q_{ij}}\left(\sqrt{|Gh|}G_{ijlm}\frac{\delta\Psi_{\varphi}}{\delta Q_{lm}}\right)  \ , \\
&2\I Q_{ik}\left(\frac{\delta\Psi_{\varphi}}{\delta Q_{kl}}\right)_{;l}+\left(\hat{\mathcal{H}}^m_{i}-J_i\right)\Psi_{\varphi} = 0 \ ,
\end{aligned}
\end{equation}
where we have neglected the covariant derivatives $\varphi_{;l}$ and $\left(\Psi_{\varphi}\right)_{;l}$, and we defined $\mathfrak{J}_{\perp} := J_{\perp}+\frac{\I}{2M\sqrt{|Gh|}}\frac{\delta}{\delta Q_{ij}}\left(\sqrt{|Gh|}G_{ijlm}\Phi^{lm}\right)$. Equations~(\ref{eq:field-WDW-2}) are the quantum versions of~(\ref{eq:field-HJ-S-1}) and~(\ref{eq:field-HJ-S-2}) and the field-theoretic analogues of~(\ref{eq:pre-schrodinger}).

Using~(\ref{eq:vectors-Qx-app-fields}) and~(\ref{eq:field-background-time}), we can combine both equations given in~(\ref{eq:field-WDW-2}) into the approximate Schr\"{o}dinger equation
\begin{equation}
\begin{aligned}
&\I\frac{\partial\Psi_{\varphi}}{\partial\tau}= \int\mathrm{d}^3y\ \left[\mathcal{N}\left(\hat{\mathcal{H}}^m_{\perp}-\mathfrak{J}_{\perp}\right)\Psi_{\varphi}+N^i\left(\hat{\mathcal{H}}^m_{i}-J_i\right)\Psi_{\varphi}\right] +\mathcal{O}\left(\frac{1}{M}\right)\ ,
\end{aligned}
\end{equation}
 which was derived in~\cite{LapRuba:1979,Banks:1984,BFS:1984,Kiefer:1991,Kiefer:1993-1,Kiefer:1993-2} without the $\mathfrak{J}_{\perp}, J_i$ terms.
We refrain from computing the corrections of order $\frac{1}{M}$ to the above equation. They should be found in formal analogy to what was done in section~\ref{sec:pt2} for the mechanical case.

\end{appendix}


\begin{thebibliography}{33}
\addcontentsline{toc}{section}{References}\label{sec:ref}

\bibitem{ADM:1962}
  R.~L.~Arnowitt, S.~Deser, and C.~W.~Misner,
  Gen. Rel. Grav. \href{https://doi.org/10.1007/s10714-008-0661-1}{{\bf 40}, 1997} (2008).

\bibitem{Bergmann:1972}
P.~G.~Bergmann and A.~Komar,
 Int.\ J.\ Theor.\ Phys.\  \href{https://doi.org/10.1007/BF00671650}{{\bf 5}, 15} (1972).

\bibitem{Pons:2010}
  J.~M.~Pons, D.~C.~Salisbury, and K.~A.~Sundermeyer,
  J.\ Phys.\ Conf.\ Ser.\  \href{https://doi.org/10.1088/1742-6596/222/1/012018}{{\bf 222}, 012018} (2010).

\bibitem{Kuchar:1991}
  K.~V.~Kucha\v{r},
  Int.\ J.\ Mod.\ Phys.\ D \href{https://doi.org/10.1142/S0218271811019347}{{\bf 20}, 3} (2011). ;

\bibitem{Isham:1992}
  C.~J.~Isham,
  Canonical Quantum Gravity and the Problem of Time
  19th Int. Colloquium on Group Theoretical Methods in Physics, Salamanca, Spain 1992
  \href{https://arxiv.org/abs/gr-qc/9210011}{(arXiv:gr-qc/9210011)}.

\bibitem{Anderson:book}
  E.~Anderson, in: \href{https://doi.org/10.1007/978-3-319-58848-3}{Fundamental \ Theories of \ Physics\ Vol 190},
  Springer International Publishing, Cham, Switzerland 2017.

\bibitem{Gerlach:1969}
  U.~H.~Gerlach,
  Phys.\ Rev.\  \href{https://doi.org/10.1103/PhysRev.177.1929}{{\bf 177}, 1929} (1969).

\bibitem{LapRuba:1979}
  V.~G.~Lapchinsky and V.~A.~Rubakov,
  Acta Phys.\ Polon.\ B \href{http://inspirehep.net/record/148275/files/v10p1041.pdf}{{\bf 10}, 1041} (1979).

\bibitem{Banks:1984}
  T.~Banks,
  Nucl.\ Phys.\ B \href{https://doi.org/10.1016/0550-3213(85)90020-3}{{\bf 249}, 332} (1985).

\bibitem{BFS:1984}
  T.~Banks, W.~Fischler, and L.~Susskind,
  Nucl.\ Phys.\ B \href{https://doi.org/10.1016/0550-3213(85)90070-7}{{\bf 262}, 159} (1985).

\bibitem{Singh:1990}
  T.~P.~Singh,
  Class.\ Quant.\ Grav.\  \href{https://doi.org/10.1088/0264-9381/7/7/006}{{\bf 7}, L149} (1990).

\bibitem{Kiefer:1991}
  C.~Kiefer and T.~P.~Singh,
  Phys.\ Rev.\ D \href{https://doi.org/10.1103/PhysRevD.44.1067}{{\bf 44}, 1067} (1991).

\bibitem{Kim:1995-1}
  S.~P.~Kim,
  Phys.\ Rev.\ D \href{https://doi.org/10.1103/PhysRevD.52.3382}{{\bf 52}, 3382} (1995).

\bibitem{Barvinsky:1997}
  A.~O.~Barvinsky and C.~Kiefer,
  Nucl.\ Phys.\ B \href{https://doi.org/10.1016/S0550-3213(98)00349-6}{{\bf 526}, 509} (1998).

\bibitem{DeWitt:1967}
  B.~S.~DeWitt,
  Phys.\ Rev.\ \href{https://doi.org/10.1103/PhysRev.160.1113}{{\bf 160}, 1113} (1967).

\bibitem{Halliwell:1984}
  J.~J.~Halliwell and S.~W.~Hawking,
  Phys.\ Rev.\ D \href{https://doi.org/10.1103/PhysRevD.31.1777}{{\bf 31} 1777} (1985)
  [Adv.\ Ser.\ Astrophys.\ Cosmol.\  {\bf 3}, 277 (1987)].

\bibitem{Brout:1987-1}
  R.~Brout,
  Found.\ Phys.\  \href{https://doi.org/10.1007/BF01882790}{{\bf 17}, 603} (1987).

\bibitem{Brout:1987-2}
  R.~Brout, G.~Horwitz, and D.~Weil,
  Phys.\ Lett.\ B \href{https://doi.org/10.1016/0370-2693(87)90114-6}{{\bf 192}, 318} (1987).

\bibitem{Brout:1987-3}
  R.~Brout,
  Z. Phys. B Con. Mat. \href{https://doi.org/10.1007/BF01304250}{{\bf 68}, 339} (1987).

\bibitem{Brout:1987-4}
  R.~Brout and G.~Venturi,
  Phys.\ Rev.\ D \href{https://doi.org/10.1103/PhysRevD.39.2436}{{\bf 39}, 2436} (1989).

\bibitem{Vilenkin:1988}
  A.~Vilenkin,
  Phys.\ Rev.\ D \href{https://doi.org/10.1103/PhysRevD.39.1116}{{\bf 39}, 1116} (1989).

\bibitem{PadSingh:1990-1}
  T.~P.~Singh and T.~Padmanabhan,
  Ann. Phys.\  \href{https://doi.org/10.1016/0003-4916(89)90180-2}{{\bf 196}, 296} (1989).

\bibitem{PadSingh:1990-2}
  T.~Padmanabhan and T.~P.~Singh,
  Class.\ Quant.\ Grav.\  \href{https://doi.org/10.1088/0264-9381/7/3/015}{{\bf 7}, 411} (1990).

\bibitem{Kiefer:1993-1}
  C.~Kiefer,
  Report Freiburg THEP-94/4, Contribution for the Lanczos Conference Proceedings,
  \href{https://arxiv.org/abs/gr-qc/9405039}{arXiv:gr-qc/9405039} (1994).

\bibitem{Kiefer:1993-2}
  C.~Kiefer,
  \href{https://doi.org/10.1007/3-540-58339-4_19}{The Semiclassical Approximation to Quantum Gravity}
  Canonical Gravity: From Classical to Quantum (Lecture Notes in Physics vol 434)
  (Eds. J.~Ehlers, H.~Friedrich), Springer, Berlin 1994.

\bibitem{Kiefer:2009T}
C.~Kiefer,
  \href{https://doi.org/10.1007/978-1-4939-3210-8_10}{Does Time Exist in Quantum Gravity?}
  Towards a Theory of Spacetime Theories (Einstein Studies vol 13)
  (Eds. D.~Lehmkuhl, G.~Schiemann, E.~Scholz), Birkhäuser, New York, NY 2017
  [\href{https://arxiv.org/abs/0909.3767}{arXiv:0909.3767 [gr-qc]}]

\bibitem{Englert:1989}
  F.~Englert,
  Phys.\ Lett.\ B \href{https://doi.org/10.1016/0370-2693(89)90534-0}{{\bf 228}, 111} (1989).

\bibitem{Briggs:2000-1}
  J.~S.~Briggs and J.~M.~Rost,
  Eur.\ Phys.\ J.\ D \href{https://doi.org/10.1007/s100530050554}{{\bf 10}, 311} (2000).

\bibitem{Briggs:2000-2}
  J.~S.~Briggs and J.~M.~Rost,
  Found. Phys. \href{https://doi.org/10.1023/A:1017525227832}{{\bf 31}, 693} (2001).

\bibitem{Hehl:2013}
  \href{https://www.worldscientific.com/worldscibooks/10.1142/p781}{Gauge Theories of Gravitation}
  (Eds. M.~Blagojevi\'{c}, F.~W.~Hehl), Imperial College Press 2013
  [\href{https://arxiv.org/abs/1210.3775}{arXiv:1210.3775 [gr-qc]}].

\bibitem{Parentani:1998}
  R.~Parentani,
  Class.\ Quant.\ Grav.\  \href{https://doi.org/10.1088/0264-9381/17/6/314}{{\bf 17}, 1527} (2000).

\bibitem{Briggs:2015}
  J.~S.~Briggs,
  Phys.\ Rev.\ A \href{https://doi.org/10.1103/PhysRevA.91.052119}{{\bf 91}, 052119} (2015).

\bibitem{Parentani:1997-1}
  R.~Parentani,
  Phys.\ Rev.\ D \href{https://doi.org/10.1103/PhysRevD.56.4618}{{\bf 56}, 4618} (1997).

\bibitem{Parentani:1997-2}
  R.~Brout and R.~Parentani
  Int.\ J.\ Mod.\ Phys.\ D \href{https://doi.org/10.1142/S0218271899000031}{{\bf 8}, 1} (1999).

\bibitem{Marolf:1995-1}
  D.~Marolf,
  \href{https://arxiv.org/abs/gr-qc/9508015}{arXiv:gr-qc/9508015} (1995).
  
\bibitem{Marolf:1995-2}
  J.~B.~Hartle and D.~Marolf,
  Phys.\ Rev.\ D \href{https://doi.org/10.1103/PhysRevD.56.6247}{{\bf 56}, 6247} (1997).

\bibitem{Kamenshchik:2013}
  A.~Y.~Kamenshchik, A.~Tronconi, and G.~Venturi,
  Phys.\ Lett.\ B \href{https://doi.org/10.1016/j.physletb.2013.08.067}{{\bf 726}, 518} (2013).

\bibitem{Kamenshchik:2014}
  A.~Y.~Kamenshchik, A.~Tronconi, and G.~Venturi,
  Phys.\ Lett.\ B \href{https://doi.org/10.1016/j.physletb.2014.05.028}{{\bf 734}, 72} (2014).

\bibitem{Kamenshchik:2017}
  A.~Y.~Kamenshchik, A.~Tronconi, and G.~Venturi,
  Class.\ Quant.\ Grav.\  \href{https://doi.org/10.1088/1361-6382/aa8fb3}{{\bf 35}, 015012} (2018).

\bibitem{Kamenshchik:2018}
  A.~Y.~Kamenshchik, A.~Tronconi, T.~Vardanyan, and G.~Venturi,
  Int.\ J.\ Mod.\ Phys.\ D \href{https://doi.org/10.1142/S0218271819500731}{{\bf 28}, 1950073} (2019).

\bibitem{Balbinot:1990}
  R.~Balbinot, A.~Barletta, and G.~Venturi,
  Phys.\ Rev.\ D \href{https://doi.org/10.1103/PhysRevD.41.1848}{{\bf 41}, 1848} (1990).

\bibitem{Anderson:2006vs-1}
  E.~Anderson,
  Class.\ Quant.\ Grav.\  \href{https://doi.org/10.1088/0264-9381/24/11/011}{{\bf 24}, 2935} (2007).

\bibitem{Anderson:2006vs-2}
  E.~Anderson,
  Class.\ Quant.\ Grav.\  \href{https://doi.org/10.1088/0264-9381/24/11/012}{{\bf 24}, 2979} (2007).

\bibitem{Anderson:2011wq}
  E.~Anderson,
  Class.\ Quant.\ Grav.\  \href{https://doi.org/10.1088/0264-9381/28/18/185008}{{\bf 28}, 185008} (2011).

\bibitem{Anderson:2013vaa}
  E.~Anderson,
  Class.\ Quant.\ Grav.\  \href{https://doi.org/10.1088/0264-9381/31/2/025006}{{\bf 31}, 025006} (2014).

\bibitem{Anderson:2013yza}
  E.~Anderson,
  Gen.\ Rel.\ Grav.\  \href{https://doi.org/10.1007/s10714-014-1708-0}{{\bf 46}, 1708} (2014).

\bibitem{BO:1927}
 M.~Born and R.~Oppenheimer,
 Ann. der Phys. \href{https://doi.org/10.1002/andp.19273892002}{{\bf 389}, 457} (1927).

\bibitem{Cederbaum:2008}
L.~S.~Cederbaum,
  J. Chem. Phys. \href{https://doi.org/10.1063/1.2895043}{{\bf 128}, 124101} (2008).

\bibitem{Mott:1929}
 N.~F.~Mott,
 Proc. R. Soc. Lond. A \href{https://doi.org/10.1098/rspa.1929.0205}{{\bf126}, 79} (1929).

\bibitem{Mott:1931}
 N.~F.~Mott,
 Math. Proc. Cambridge \href{https://doi.org/10.1017/S0305004100009816}{{\bf27}, 553} (1931).

 \bibitem{Barbour:1994-2}
  J.~B.~Barbour,
  Class.\ Quant.\ Grav.\  \href{https://doi.org/10.1088/0264-9381/11/12/006}{{\bf 11}, 2875} (1994).

\bibitem{Halliwell:2000}
  J.~J.~Halliwell,
  Phys.\ Rev.\ D \href{https://doi.org/10.1103/PhysRevD.64.044008}{{\bf 64}, 044008} (2001).

\bibitem{Zeh:1987}
  H.~D.~Zeh,
  Phys.\ Lett.\ A \href{https://doi.org/10.1016/0375-9601(88)90842-0}{{\bf 126}, 311} (1988).

\bibitem{Arce:2012}
  J.~C.~Arce,
  Phys.\ Rev.\ A \href{https://doi.org/10.1103/PhysRevA.85.042108}{{\bf 85}, 042108} (2012).

\bibitem{MeadBerry:1979-1}
 C.~A.~Mead and D.~G.~Truhlar,
 J. Chem. Phys. \href{https://doi.org/10.1063/1.437734}{{\bf 70}, 2284} (1979).

\bibitem{MeadBerry:1979-2}
 M.~V.~Berry,
 Proc. R. Soc. Lond. A \href{https://doi.org/10.1098/rspa.1984.0023}{{\bf 392}, 45} (1984).

\bibitem{Abedi:2010}
  A.~Abedi, N.~T.~Maitra, and E.~K.~U.~Gross,
  Phys. Rev. Lett. \href{https://doi.org/10.1103/PhysRevLett.105.123002}{{\bf 105}, 123002} (2010).

\bibitem{Abedi:2012-1}
  A.~Abedi, N.~T.~Maitra, and E.~K.~U.~Gross,
  J. Chem. Phys. \href{https://doi.org/10.1063/1.4745836}{{\bf 137}, 22A530} (2012).
  
\bibitem{Abedi:2012-2}
  J.~L.~Alonso, J.~Clemente-Gallardo, P.~Echenique-Robba, and J.~A.~Jover-Galtier,
  J. Chem. Phys. \href{https://doi.org/10.1063/1.4818521}{{\bf 139}, 087101} (2012).

\bibitem{Bertoni:1996}
  C.~Bertoni, F.~Finelli, and G.~Venturi,
  Class.\ Quant.\ Grav.\  \href{https://doi.org/10.1088/0264-9381/13/9/005}{{\bf 13}, 2375} (1996).

\bibitem{Kiefer:2018}
  C.~Kiefer and D.~Wichmann,
  Gen.\ Rel.\ Grav.\  \href{https://doi.org/10.1007/s10714-018-2390-4}{{\bf 50}, 66} (2018).

\bibitem{Hartle:1987}
  J.~B.~Hartle,
  \href{https://doi.org/10.1007/978-1-4613-1897-2_12}{ASI Series (Series B: Physics) vol 156}
  (Eds B.~Carter, J.~B.~Hartle), Springer, Boston, MA 1987.

\bibitem{Halliwell:1987}
  J.~J.~Halliwell,
  Phys.\ Rev.\ D \href{https://doi.org/10.1103/PhysRevD.36.3626}{{\bf 36}, 3626} (1987).

\bibitem{DEath:1986}
  P.~D.~D'Eath and J.~J.~Halliwell,
  Phys.\ Rev.\ D \href{https://doi.org/10.1103/PhysRevD.35.1100}{{\bf 35}, 1100} (1987).

\bibitem{Padmanabhan:1988}
  T.~Padmanabhan,
  Class.\ Quant.\ Grav.\  \href{https://doi.org/10.1088/0264-9381/6/4/012}{{\bf 6}, 533} (1989).

\bibitem{Kiefer:book}
C.~Kiefer,
Quantum Gravity
(International Series of Monographs on Physics), 3rd ed.,
Oxford University Press, Oxford 2012.

\bibitem{Hunter:1975}
 G.~Hunter,
 Int. J. Quantum Chem. \href{https://onlinelibrary.wiley.com/doi/abs/10.1002/qua.560090205}{{\bf9}, 237} (1975).

\bibitem{Cederbaum:2013}
  L.~S.~Cederbaum,
  J. Chem. Phys. \href{https://doi.org/10.1063/1.4807115}{{\bf 138}, 224110} (2013).

\bibitem{Anderson:2013}
  E.~Anderson, in:
  \href{https://doi.org/10.1142/9789814578745_0024}{XXIX-th International Workshop on High Energy Physics: New Results and Actual Problems in Particle \& Astroparticle Physics and Cosmology}. World Scientific Publishing Co. Pte. Ltd, Singapore 2014, p. 182 (\href{https://arxiv.org/abs/1306.5812}{arXiv:1306.5812 [gr-qc]}).

\bibitem{Schild:2018}
  A.~Schild,
  Phys.\ Rev.\ A \href{https://doi.org/10.1103/PhysRevA.98.052113}{{\bf 98}, 052113} (2018).

\bibitem{Greensite:1989-1}
  J.~Greensite,
  Nucl.\ Phys.\ B \href{https://doi.org/10.1016/0550-3213(90)90196-K}{{\bf 342}, 409} (1990).

\bibitem{Greensite:1989-2}
  T.~Padmanabhan,
  Pramana \href{https://doi.org/10.1007/BF02875295}{{\bf 35}, L199} (1990).

\bibitem{Greensite:1989-3}
  J.~Greensite,
  Nucl.\ Phys.\ B \href{https://doi.org/10.1016/S0550-3213(05)80043-4}{{\bf 351}, 749} (1991).

\bibitem{Greensite:1989-4}
  T.~Brotz and C.~Kiefer,
  Nucl.\ Phys.\ B \href{https://doi.org/10.1016/0550-3213(96)00304-5}{{\bf 475}, 339} (1996).

\bibitem{Nelson:2018}
  N.~Pinto-Neto and W.~Struyve,
  \href{https://arxiv.org/abs/1801.03353}{arXiv:1801.03353 [gr-qc]} (2018).

\bibitem{Venturi:1990}
  G.~Venturi,
  Class.\ Quant.\ Grav.\  \href{https://doi.org/10.1088/0264-9381/7/6/014}{{\bf 7}, 1075} (1990).

\bibitem{Kim:1995-2}
  S.~P.~Kim,
  Phys.\ Lett.\ A \href{https://doi.org/10.1016/0375-9601(95)00584-P}{{\bf 205}, 359} (1995).

\bibitem{Massar:1998}
  S.~Massar and R.~Parentani
  Phys.\ Rev.\ D \href{https://doi.org/10.1103/PhysRevD.59.123519}{{\bf 59}, 123519} (1999).

\bibitem{Kiefer:1993-0}
  C.~Kiefer,
  Phys.\ Rev.\ D \href{https://doi.org/10.1103/PhysRevD.47.5414}{{\bf 47}, 5414} (1993).

\bibitem{Kiefer:1987}
  C.~Kiefer,
  Class.\ Quant.\ Grav.\  \href{https://doi.org/10.1088/0264-9381/4/5/031}{{\bf 4}, 1369} (1987).

\bibitem{Halliwell:1989}
  J.~J.~Halliwell,
  Phys.\ Rev.\ D \href{https://doi.org/10.1103/PhysRevD.39.2912}{{\bf 39}, 2912} (1989).

\bibitem{Barbour:1993}
  J.~B.~Barbour,
  Phys.\ Rev.\ D \href{https://doi.org/10.1103/PhysRevD.47.5422}{{\bf 47}, 5422} (1993).

\bibitem{DeWitt:1957}
  B.~S.~DeWitt,
  Rev.\ Mod.\ Phys.\  \href{https://doi.org/10.1103/RevModPhys.29.377}{{\bf 29}, 377} (1957).

\bibitem{Lammerzahl:1995}
  C.~L\"{a}mmerzahl,
  Phys.\ Lett.\ A \href{https://doi.org/10.1016/0375-9601(95)00345-4}{{\bf 203}, 12} (1995).

\bibitem{Kiefer:2011-1}
C.~Kiefer and M.~Kr\"{a}mer,
 Phys.\ Rev.\ Lett.\  \href{https://doi.org/10.1103/PhysRevLett.108.021301}{{\bf 108}, 021301} (2012).

\bibitem{Kiefer:2011-2}
  C.~Kiefer,
  J.\ Phys.\ Conf.\ Ser.\  \href{https://doi.org/10.1088/1742-6596/442/1/012025}{{\bf 442}, 012025} (2013).

\bibitem{Bini:2013}
  D.~Bini, G.~Esposito, C.~Kiefer, M.~Kr\"{a}mer, and F.~Pessina,
  Phys.\ Rev.\ D \href{https://doi.org/10.1103/PhysRevD.87.104008}{{\bf 87}, 104008} (2013).

\bibitem{Brizuela:2015-1}
  D.~Brizuela, C.~Kiefer, and M.~Kr\"{a}mer,
  Phys.\ Rev.\ D \href{https://doi.org/10.1103/PhysRevD.93.104035}{{\bf 93}, 104035} (2016).

\bibitem{Brizuela:2015-2}
  D.~Brizuela, C.~Kiefer, and M.~Kr\"{a}mer,
  Phys.\ Rev.\ D \href{https://doi.org/10.1103/PhysRevD.94.123527}{{\bf 94}, 123527} (2016).

\bibitem{Brizuela:2015-3}
  D.~Brizuela and M.~Kr\"{a}mer,
  Galaxies \href{https://doi.org/10.3390/galaxies6010006}{{\bf 6}, 6} (2018).

\bibitem{Kamenshchik:2016}
  A.~Y.~Kamenshchik, A.~Tronconi, and G.~Venturi,
  Phys.\ Rev.\ D \href{https://doi.org/10.1103/PhysRevD.94.123524}{{\bf 94}, 123524} (2016).

\bibitem{Kamenshchik:2018rpw}
  A.~Y.~Kamenshchik, A.~Tronconi, T.~Vardanyan, and G.~Venturi,
  Phys.\ Rev.\ D \href{https://doi.org/10.1103/PhysRevD.97.123517}{{\bf 97}, 123517} (2018).

\bibitem{Giulini:1994}
  D.~Giulini and C.~Kiefer,
  Class.\ Quant.\ Grav.\  \href{https://doi.org/10.1088/0264-9381/12/2/009}{{\bf 12}, 403} (1995).

%
%

\end{thebibliography}
\end{document}